\newif\ifsubmode
\submodetrue

\newif\ifprintfig
\printfigtrue
\newif\iffullout
\fullouttrue
\ifsubmode
  \documentclass[12pt,preprint]{aastex}
  \usepackage{epsf}
  \received{} 
  \revised{}
  \accepted{}
  \ccc{}
  \cpright{}{}
\else
  \documentclass[preprint]{aastex}
  \usepackage{epsf}
  \slugcomment{{\it draft \today, do not distribute}}
\fi

\newcommand{\kms}{${\rm km~s^{-1}}$}

\shorttitle{Emission line gas in FR-I Galaxies}
\shortauthors{Jacob Noel-Storr, et al.}

\begin{document}

\title{STIS spectroscopy of the emission line gas in the nuclei of nearby FR-I galaxies\altaffilmark{1}}

\author{Jacob Noel-Storr\altaffilmark{2,3}, Stefi A.~Baum\altaffilmark{3}, Gijs Verdoes Kleijn\altaffilmark{4}, Roeland P.~van der Marel\altaffilmark{3}, Christopher P.~O'Dea\altaffilmark{3}, P.~Tim de Zeeuw\altaffilmark{5}, and C.~Marcella Carollo\altaffilmark{6}}

\altaffiltext{1}{Based on observations with the NASA/ESA {\em Hubble Space Telescope}, obtained at the Space Telescope Science Institute, which is operated by the Association of Universities for Research in Astronomy, Inc., under NASA contract NAS 5-26555}
\altaffiltext{2}{Columbia University, Astronomy Department, New York, NY 10027 {\tt jake@astro.columbia.edu}}
\altaffiltext{3}{Space Telescope Science Institute, 3700 San Martin Drive, Baltimore, MD 21218}
\altaffiltext{4}{European Southern Observatory, Karl-Schwarzschild Str. 2, D-85748 Garching, Germany}
\altaffiltext{5}{Sterrewacht Leiden, Postbus 9513, 2300 RA Leiden, The Netherlands}
\altaffiltext{6}{Institute of Astronomy, ETH Zentrum, CH-8092 Zurich, Switzerland}

\begin{abstract}
We present the results of the analysis of a set of medium resolution spectra, obtained by the Space Telescope Imaging Spectrograph on board the {\em Hubble Space Telescope}, of the emission line gas present in the nuclei of a complete sample of 21 nearby, early-type galaxies with radio jets (the UGC FR-I Sample). For each galaxy nucleus we present spectroscopic data in the region of H$\alpha$ and the dervived kinematics.

We find that in 67\% of the nuclei the gas appears to be rotating and, with one exception, the cases where rotation is not seen are either face on or have complex central morphologies. We find that in 62\% of the nuclei the fit to the central spectrum is improved by the inclusion of a broad component. The broad components have a mean velocity dispersion of $1349 \pm 345$~\kms\ and are redshifted from the narrow line components (assuming an origin in H$\alpha$) by $486 \pm 443$~\kms .
\end{abstract}

\keywords{galaxies: active --- galaxies: elliptical and lenticular --- galaxies: individual (NGC~193, NGC~315, NGC~383, NGC~541, NGC~741, UGC~1841, NGC~2329, NGC~2892, NGC~3801, NGC~3862, UGC~7115, NGC~4261, NGC~4335, M84, M87, NGC~5127, NGC~5141, NGC~5490, NGC~7052, UGC~12064, NGC~7626) --- galaxies: kinematics and dynamics --- galaxies: nuclei}

\section{Introduction}
A nearby radio galaxy can produce a radio jet that reaches up to $\sim 1$~Mpc, emerging from a central engine that resides in an active region $\sim 10^{-3}$~pc in radius in its nucleus. The energy that powers radio jets and active nuclei is typically believed to be produced by the accretion of material onto a central supermassive black hole \citep{rees84}. Central emission line gas is detected in virtually all nearby radio galaxies that harbor kiloparsec-scale radio jets, and this gas along with nuclear dust presumably provide sources of fuel for the central black hole. However, the relationship between normal early-type galaxies and early-type galaxies that are radio-loud remains one of the mysteries in active galaxy research. 

Current evidence suggests that all galaxies may have central black holes and that the size of the black hole is a strong function of the mass of the bulge \citep[for example,][]{kr95} and and even stronger function of the central stellar velocity dispersion \citep{FM00,geb00,MF01,trem02}. It is not clear what drives these relationships or whether they are the same for active and quiescent galaxies. Since black hole growth and nuclear activity are causally related, scatter in these relationship can, in principle, put limits on the frequency and duration of nuclear activity in galaxies.

Typical findings are that 60\%\ of quiescent early-type galaxies have detections of emission line gas \citep[e.g.][]{philips86, goud94} and about 40\%\ have nuclear dust \citep[e.g.][]{vdf95, tran01}. Despite the apparent presence of fuel, these galaxies fail to produce radio jets or any significant nuclear activity.

Nearby early-type galaxies with radio jets provide an opportunity to gain an understanding of the conditions in a galaxy which lead to the formation of a radio-active nucleus and of the physics of the regions which harbor the black hole and jet formation regimes. To this end, we are undertaking a coordinated, multiwavelength, study of a complete sample of the nearest radio galaxies (see \S 2).

Based on our imaging with the {\em Hubble Space Telescope} (HST) Wide Field and Planetary Camera - 2 (WFPC2) we found that the galaxies of our sample show ubiquitous nuclear dust and emission line gas \citep{gijs99}. An understanding of the gas kinematics in the central region can enable us to determine the black hole masses \citep{harms94, FFJ96, macheto97, vdm98, ff99, gvk00, sarzi01, barth01} and the ionization mechanisms at work in the nuclear regions \citep[e.g.][]{dopita97} and relate these to other properties of the activity and of the galaxy as a whole. In order to achieve this goal we have obtained HST Space Telescope Imaging Spectrograph (STIS) medium resolution long slit spectra of the nuclear regions of our sample galaxies. We present those data here. In future work we will go on to model the kinematics and other properties of the emission line gas.

The paper is organized as follows. In section 2 we describe the sample and in section 3 we describe the STIS spectroscopic observations and the data reduction. In section 4 we describe our analysis procedures. We present our initial interpretations in section 5 and draw conclusions in section 6. We use a Hubble constant of $H_0 = 50 {\rm km~s^{-1}~Mpc^{-1}}$ throughout. 

\section{Sample selection and Properties}
Our galaxy sample (the UGC~FR-I sample) contains all 21 nearby ($v_{\rm r} < 7000\
\rm{km~s^{-1}}$), elliptical or S0 galaxies in the declination range
$-5^\circ < \delta < 70^\circ$ in the UGC~catalog \citep[limits
magnitude $m_B < 14\fm 6$ and angular size $\theta_p > 1\farcm
0$]{nilson73} that are extended radio-loud sources (larger than
10\arcsec\ at $3\sigma$ on VLA A-Array maps and brighter than 150~mJy
from single dish flux density measurements at 1400 MHz)\footnote{We have slightly adjusted the sample definition since our earlier work, however the galaxies that the sample contains remain the same. The sample is complete within the definition given here as of the time of writing according to up to date data drawn from the NED database.}. The source information is shown in Table \ref{sample}. 

This complete sample was drawn from a catalog of 176 radio-loud galaxies constructed by \citet{CB88}, by position coincidence of radio identifications in the Green Bank 1400~MHz sky maps and galaxies in the UGC catalog.

All of the galaxies fall into the \citet{FR74} Type-I (FR-I) radio
classification \citep[see][for a description of the radio properties
of our sample]{xu2000}; i.e. they are low luminosity radio galaxies, with jets
that are brightest nearest to the nucleus. (In contrast, FR-II
galaxies are more powerful radio sources, and have bright spots
at the far edges of their radio lobes.)

The primary energy source of each galaxy in the sample falls into
`monster' rather than `starburst' classification of \citet{CB88} based on their infra-red to radio flux ratios \citep[for example, see][]{heckman}

\begin{equation}
u \equiv {\rm log}~\left({{S_{60{\rm \mu m}}} \over {S_{1400{\rm MHz}}}}\right) \ll 1.6,
\end{equation}

and infra-red spectral gradients 

\begin{equation}
\alpha_{\rm IR} \equiv {{{\rm log}~(S_{60{\rm \mu m}}/S_{25{\rm \mu m}})} \over {{\rm log}~(60 / 25)}} < +1.25.
\end{equation}

A photometric analysis of the nuclei of these galaxies was performed by \citet{gijs99}, based on observations made using the WFPC2 instrument (using the {\tt F555W} and {\tt F814W} filters) on board the HST \citep[the photometric analyses of UGC~7115 and UGC~12064 are presented in Appendix A of][]{gvknoc}.

\section{STIS Observations}
In program 8236 we used HST/STIS \citep[see][]{kimble1998} to take spectra of the 19 sample members not previously or concurrently observed by others. Observations were carried out at both
medium (for 19/19 galaxies) and low (for 4/19 galaxies)
resolution, with the {\tt G750M} and the {\tt G430L} and {\tt G750L} gratings respectively. Our medium resolution observation log is shown in
Table~\ref{obsets}. We will present the low resolution spectra in a
future paper. We include in our analysis similar medium resolution data obtained by
R. Green and collaborators for the nucleus of M84 \citep[program 7124, see][]{bower98} and H. Ford and collaborators for the nucleus of M87 (program 8666) to complete the data set for UGC FR-I galaxies.

In our program, we observed each galaxy in three parallel,
adjacent slit positions. For each slit position we obtained two
exposures with a shift of 0\farcs202800 (4 unbinned STIS Pixels) along
the slit direction to enable us to more efficiently remove detector
effects (bad pixels, etc.). In the case of UGC~7115 we observed in only
one slit position - we sacrificed STIS observing time to make WFPC2 observations of this
galaxy as it had not been included in our earlier WFPC2 program.

The data for M87 (NGC~4486) were obtained in a similar manner to our own. In the case of M84 (NGC~4374) the observation pairs were not shifted along the slit direction, thus some detector effects may remain, though will be much less significant thanks to the far greater signal to noise.

We list the instrumental properties of the STIS configurations used in Table~\ref{stisets}. Panel (a) of Figures~\ref{sgplimage} to \ref{sgpi21} (see key in Figure \ref{key}) shows the location of the STIS slits on each galaxy observed, along with the position angles of the galaxy major axes and radio-jet axes. For the majority of cases the STIS slits were aligned within $10^\circ$ of the galaxy major axes (Table~\ref{pas}), with the exceptions that we note below.

In the cases of NGC~741 [$\Delta PA(galaxy,STIS) = 19.4^\circ$] and NGC~2892 [$\Delta PA(galaxy,STIS) = 22^\circ$] more freedom in orientation was allowed to provide reasonable observing windows. 

In NGC~3862 and UGC~7115 the position angles of the galaxy major axes are hard to determine, we decided to position the slits approximately perpendicular to the radio jets [$\Delta PA(radio,STIS) = 92.3^\circ\ \&\ 110.4^\circ$, respectively]. The central major axis is also hard to determine in NGC~541, where the central isophotes rotate considerably. In this case we chose to use a slit position somewhere around the mean of the isophotal position angles with considerable leeway given to allow for reasonable observing windows. 

For UGC~12064 the slits were aligned along the major axis of the prominent dust disk, which is offset from the galaxy major axis by $\sim 50^\circ$. 

The slits in M84 were positioned approximately perpendicular to the radio jet, which lies close to the major axis of the nuclear gas. For NGC~4486 (M87) the slits were positioned to follow certain morphological structures across the nuclear regions.

\subsection{Data Reduction}
We used the standard STIS calibration pipeline \citep[{\tt calstis}, see][]{stis_hb} to
perform bias, dark and flat-field corrections using the best available
reference files. We used {\tt calstis} version 2.13 (26-April-2002) throughout the data reduction\footnote{We found that significant variations in measured parameters could be introduced by using different {\tt calstis} versions. Versions 2.4 and 2.13 produce consistent results, while intermediate versions do not.}. 

We shifted the rows of alternating observations by 4 pixels, so that they were properly aligned with their counterparts and combined them using the STSDAS routine {\tt ocrreject}. We cleaned co-incident cosmic rays and negative bad pixels, which would not be caught by {\tt ocrreject}, using the NOAO/IRAF task {\tt cosmicrays}. At each step we carefully investigated the effects of varying task parameters to insure we were not damaging valid data while removing most cosmic rays. 

We made use of the STIS calibration pipeline tasks {\tt wavecal} and {\tt x2d} to perform wavelength calibration and image rectification respectively. The error introduced by rectifying after shifting one of the images is $\la 0.05$~pixels (which is $\la 1.3$~\kms\ at 6750\AA).

Panel (b) of Figures \ref{sgplimage} to \ref{sgpi21} (see key in Figure \ref{key}) shows (i) the central strip of the reduced spectrum for each slit position observed on each galaxy, along with (ii) Gaussian line fits to the same (see \S 4.1).

\section{Analysis}
In this section we describe the line fitting that we carried out on each spectral row of each reduced CCD spectral image, firstly with a single Gaussian per spectral line (\S 4.1) and secondly with the inclusion of an additional free component (\S 4.2). In \S 4.3 we discuss the sizes of the errors on quoted parameters from various sources. In \S 4.4 we describe each of the UGC~FR-I sample members in turn. For each galaxy we define the central spectrum as the row with the greatest integrated flux after the data reduction. We list the row numbers in the final {\tt x2d} image corresponding to the central spectrum in Table \ref{obsets}.

\subsection{Single Gaussian line fitting}
In the {\tt G750M} spectra we expect to find the emission lines in the vicinity of H$\alpha$ that are listed in Table \ref{medlinelist}. We used wavelengths from the recent measurements of \citet{wallerstein01} and converted from air to vacuum wavelengths using the IAU standard formula

\begin{equation}
{{\lambda_{\rm vac} - \lambda_{\rm air}} \over {\lambda_{\rm air}}} = 6.4328\times 10^{-5} + {{2.94981\times 10^{-2}} \over {146 - \left( {10^{4}/\lambda_{\rm air}}\right) ^{2}}} + {{2.5540\times 10^{-4}} \over {41 - \left( {10^{4}/\lambda_{\rm air}}\right) ^{2}}}.
\end{equation}

Using one Gaussian to represent each of these five lines we obtain a set of 7
free parameters to fit: the continuum flux level, velocity ($v_{\rm r}$), velocity dispersion ($\sigma$), and the fluxes of each line. The flux of [\ion{N}{2}]$_{6550}$ was fixed in a ratio of 1:3 with the flux of [\ion{N}{2}]$_{6585}$ based on the transition probabilities derived from atomic physics \citep{osterbrock}.

We used a $\chi^2$ minimization routine \citep[using Levenberg-Marquardt iterations, see][]{press92} to fit the Gaussian template to the observed spectra. The application of this fitting technique and development of this routine are described by \citet{vdm98}. Formal errors are drawn from the covariance matrix of the fit. As we do not expect the noise in each spectrum to be normally-distributed after the steps of wavelength calibration and two-dimensional rectification, these error values should be treated strictly as the formal fit errors under the understanding that the size of the real errors may be somewhat different (see \S 4.3 below).

From this point on we only consider data points where the formal errors from the fits meet the following criteria, allowing us to exclude unreliable data points originating from poorly constrained fits:

\begin{eqnarray}
\Delta\sigma & < & 50~{\rm km~s^{-1}}\\
{\Delta F(H\alpha)} \over {F(H\alpha)} & < & 0.75
\end{eqnarray}

where $\Delta\sigma$ and $\Delta F(H\alpha)$ are the errors in velocity dispersion and line flux respectively. The main constraint arises from the limit on the velocity dispersion error. The very large flux error allowed is in place to remove only the few remaining bad data points where the profile very precisely fits the noise.

In Panel (e) of Figures~\ref{sgplimage} to \ref{sgpi21} (see key in Figure \ref{key}) we show profiles of (i) radial velocity, (ii) velocity dispersion, (iii) [\ion{N}{2}]$_{6585}$ line flux and (iv) [\ion{N}{2}] / H$\alpha$ ratios
resulting from this fitting procedure for each of our sample galaxies. These profiles are combined and visualized in 2D for each galaxy in Panel (c) of Figures \ref{sgplimage} to \ref{sgpi21}. 

We present the fit data in Tables \ref{ngc193_p1} to \ref{ngc7626_p1}, where the errors given are the formal errors from the fit. In these tables Column (1) is the row number of the portion of the spectrum fitted. Column (2) shows the offset along the slit direction in arcseconds from the row with the greatest integrated flux. Columns (3) and (4) give the radial velocities ($v_{\rm r}$) and gas velocity dispersions ($\sigma_{gas}$) respectively. Column (5) gives the line flux of the H$\alpha$ line and columns (6) and (7) show its ratio against the fluxes of the [\ion{N}{2}]$_{6585}$ and [\ion{S}{2}]$_{total}$ (the total flux of the two [\ion{S}{2}] lines) respectively. Column (8) gives the reduced $\chi^2$ ($R^2$) value of the resulting fit.

We repeated the fit for the central row of the galaxy NGC~4335 varying the set of free parameters in order to estimate the reliability of the fits that we had used. We found that the fit was stable to within one formal error on all quoted parameters when the velocities, velocity dispersions and fluxes of all parameters were fit independently. The signal to noise falls off rapidly outside of the very nuclear regions so it is not possible to consistently run fits with a large number of free parameters. The results for the nucleus of NGC~4335 satisfy us that we are justified in fixing the parameters in the manner that we have chosen, without adding any obvious biases to our results.

\subsection{Fits with an additional free component}
In many cases, as the very central pixels are reached the fit begins to do a poorer job of matching the observed profile. In an attempt to improve the fit to the narrow centers of the lines we tested a fit for the central spectrum of each galaxy including an additional fit component with independent velocity, velocity dispersion and flux, along with the original set of five Gaussians. The fits to the central spectra are shown in Panel (d) of Figures \ref{sgplimage} to \ref{sgpi21} (see key in Figure \ref{key}), (i) excluding and (ii) including the additional component.

We assessed the effectiveness of including this component in each galaxy based on (1) an improvement in the mean of the absolute value of the residuals from the fit $\ge 5\%$ (2) an improvement in the reduced $\chi^2$ value of the fit such that $(R^2_1 - R^2_2)/R^2_1 \ge 0.15$ and (3) an improvement judged by eye in the fit compared to the data. We assigned a score to each galaxy, with one point available for each of the three categories. We consider scores of 2 or 3 to be indicative of the presence of a broad component, a score of 1 indicates the possibility of a broad component, while we treat a score of 0 as a none detection. The three parameters and scores are listed in Table~\ref{blrtab}. 

We find an additional free component improves the fit in 62\% (N = 13) of the sample galaxies. In the cases of NGC~2329 and NGC~3862 the component appears to represent a non-flat continuum. The kinematic parameters for each galaxy including the additional free component are listed in Table \ref{ffbblfits} and the flux parameters in Table \ref{ffbblflux}. We present further interpretation of the nature and origin of the features fit by the additional free component in \S 5.3.

\subsection{Quantifying error sources \label{rser}}
The STIS data handbook \citep{stis_hb} gives the following absolute and relative accuracies applicable to this work: A wavelength absolute calibration error ($\Delta \lambda$ offset) of 0.1 to 0.3 pixels (2.6 to 7.7~\kms\ at 6500\AA ) within an exposure, and from 0.2 to 0.5 pixels (5.1 to 12.8~\kms\ at 6500\AA ) between exposures. An absolute photometry error of 5\% and a relative photometry error of 2\% within a single exposure assuming a wide slit observation. 5~${\rm \mu m}$ variations in slit width along the slit lengths could result in variations of up to 20\% in flux along the 0\farcs 1 slit. 

In \citet{gvknoc} H$\alpha$ + [\ion{N}{2}] fluxes were presented for each nucleus in the sample. The values presented there agree well with the values we find here, certainly given our limited ability to extract comparable apertures and within the 20\% potential flux errors noted above.

In Section 1 we indicated that a 1.3~\kms\ error could be incorporated into the final data as a result of shifting the spectra for image combination and cosmic ray rejection. This shift is insignificant compared to other error sources.

In Table \ref{hafree} we showed that by allowing different free parameters within the single-Gaussian-per-line fit produced changes in the measurement in velocity of $\sim 8~{\rm km~s^{-1}}$ and of $\sim 16~{\rm km~s^{-1}}$ in velocity dispersion for the nucleus of NGC~4335.

In Table \ref{vsmodels} we show the effect on the measured velocities and velocity dispersions of the various components for each of the models described in the previous section, again for the case of NGC~4335. This illustrates that the measured velocities of the narrow lines may vary by up to $\sim 20~{\rm km~s^{-1}}$ (and velocity dispersions by as much as $\sim 110~{\rm km~s^{-1}}$) when additional components in the line shape are taken into account.

In the nuclei of NGC~383 and NGC~4335 (representing cases with blended and less blended lines respectively) we repeated the narrow line fit to the central spectrum with 286 different combinations of input velocity and velocity dispersion; varying the velocities over a range of 2000 \kms\ and the velocity dispersions over a range of 6000 \kms. In the case of NGC~4335 we found that the velocity varied by $\pm 10.91~{\rm km~s^{-1}}$ and the velocity dispersion by just $\pm 0.02~{\rm km~s^{-1}}$. In the case of NGC~383 we found that the velocity varied by $\pm 13.04~{\rm km~s^{-1}}$ and the velocity dispersion by just $\pm 3.76~{\rm km~s^{-1}}$. There was a systematic effect relating input and output velocities in both cases.

We conclude that reasonable estimates of the genuine errors on each of our measured parameters are: 5\% - 10\% on fluxes (dominated by the effects of variations along the narrow slits and the STIS absolute calibration); and $\sim 20$~\kms\ on velocities and velocity dispersions (dominated by the fit model dependency of the results).

\subsection{Individual source descriptions}
Below, we give descriptions of each member of the UGC~FR-I sample in turn. The galaxy classifications are taken from the NASA Extragalactic Database, which lists references in which the terms used are described. Descriptions of dust properties and radio sources are as presented by \citet{gijs99} and \citet{xu2000} respectively.

\paragraph{NGC~193 (UGC~408)} This S0 galaxy has a complex gas morphology with two lanes apparent in the central regions (the most clearly defined lane has a $width:length = 0.18$). It has a core-jet radio morphology on VLA and VLBA scales. The STIS slits were aligned parallel to the galaxy major axis. The central kinematic and flux properties are listed in Table \ref{ngc193_p1}; the gas does not exhibit a regular rotation curve, though it does appear dominated by systematic rather than random motions. The fit to the central spectrum is improved by the addition of a broad component. Data for this galaxy are shown in Figure \ref{sgplimage} (see key in Figure \ref{key} for an explanation of these plots).

\paragraph{NGC~315 (UGC~597)} This elliptical galaxy has a nuclear dust disk $(b/a = 0.23)$. It has a core-jet radio morphology on VLA and VLBA scales. The STIS slits were aligned parallel to the galaxy major axis. The central kinematic and flux properties are listed in Table \ref{ngc315_p1}; the gas appears to be in organized motion, possibly regular rotation. The fit to the central spectrum is improved by the addition of a broad component. Data for this galaxy are shown in Figure \ref{sgpi2} (see key in Figure \ref{key} for an explanation of these plots).

\paragraph{NGC~383 (UGC~689)} This S0 galaxy has a nuclear dust disk $(b/a = 0.77)$. It has a core-jet radio morphology on VLBA scales, and a twin-jet morphology on VLA scales. The STIS slits were aligned parallel to the galaxy major axis. The central kinematic and flux properties are listed in Table \ref{ngc383_p1}; the gas exhibits a regular rotation profile. In the negative offset side slit there is a dip in the velocity dispersion profile at a position close to the nucleus. The fit to the central spectrum is improved by the addition of a broad component. Data for this galaxy are shown in Figure \ref{sgpi3} (see key in Figure \ref{key} for an explanation of these plots).

\paragraph{NGC~541 (UGC~1004)} This cD S0 galaxy has a nuclear dust disk $(b/a = 0.91)$. It has a radio core on VLBA scales and a core-jet morphology on VLA scales. The STIS slits were aligned to a mean of the position angles of the central isophotes measured from our WFPC/2 images, which vary considerably. We allowed considerable flexibility in position angle to enable reasonable observing windows. The central kinematic and flux properties are listed in Table \ref{ngc541_p1}; the gas does not exhibit a regular rotation profile. The fit to the central spectrum is not significantly improved by the addition of a broad component, though the fit improves somewhat when judged by eye. Data for this galaxy are shown in Figure \ref{sgpi4} (see key in Figure \ref{key} for an explanation of these plots).

\paragraph{NGC~741 (UGC~1413)} This E0 galaxy has no apparent nuclear dust. It has a radio core on VLBA scales and a core-jet morphology on VLA scales. The STIS slits were aligned approximately parallel to the galaxy major axis, however a certain degree of freedom was allowed in slit placement to allow reasonable observing windows. The central kinematic and flux properties are listed in Table \ref{ngc741_p1}. Very few points had sufficient signal to noise to obtain good fits in these data, it has not been included in further analysis of global kinematic properties. The fit to the central spectrum is not improved by the addition of a broad component. Data for this galaxy are shown in Figure \ref{sgpi5} (see key in Figure \ref{key} for an explanation of these plots).

\paragraph{UGC~1841} This elliptical galaxy has a nuclear dust disk $(b/a \sim 0.98)$. It has a core-jet radio morphology on VLBA and VLA scales. The STIS slits were aligned parallel to the galaxy major axis. The central kinematic and flux properties are listed in Table \ref{ugc01841_p1}; the gas does not exhibit a regular rotation profile. The fit to the central spectrum is improved by the addition of a broad component. Data for this galaxy are shown in Figure \ref{sgpi6} (see key in Figure \ref{key} for an explanation of these plots).

\paragraph{NGC~2329 (UGC~3695)} This S0 galaxy has a nuclear dust disk $(b/a = 0.68)$. It has a core-jet radio morphology on VLBA and VLA scales.  The STIS slits were aligned parallel to the galaxy major axis. The central kinematic and flux properties are listed in Table \ref{ngc2329_p1}; the gas does not exhibit a regular rotation profile. The fit to the central spectrum is improved by the addition of a broad component which appears to represent a non-flat continuum in this case. Data for this galaxy are shown in Figure \ref{sgpi7} (see key in Figure \ref{key} for an explanation of these plots).

\paragraph{NGC~2892 (UGC~5073)} This elliptical galaxy has no apparent nuclear dust. It has a radio core on VLBA scales and a twin-jet morphology on VLA scales. The STIS slits were aligned approximately parallel to the galaxy major axis, however a certain degree of freedom was allowed in slit placement to allow reasonable observing windows. The central kinematic and flux properties are listed in Table \ref{ngc2892_p1}; the gas does not exhibit a regular rotation profile. The fit to the central spectrum is not significantly improved by the addition of a broad component. Data for this galaxy are shown in Figure \ref{sgpi8} (see key in Figure \ref{key} for an explanation of these plots).

\paragraph{NGC~3801 (UGC~6635)} This S0/a galaxy has a complex nuclear dust morphology with a large scale dust lane $(width:length = 0.12)$. It has a twin-jet radio morphology on VLA scales. The STIS slits were aligned parallel to the galaxy major axis. The central kinematic and flux properties are listed in Table \ref{ngc3801_p1}; the gas does not exhibit a regular rotation profile. The fit to the central spectrum is not significantly improved by the addition of a broad component. Data for this galaxy are shown in Figure \ref{sgpi9} (see key in Figure \ref{key} for an explanation of these plots).

\paragraph{NGC~3862 (UGC~6723)} This elliptical galaxy has a nuclear dust disk $(b/a \sim 0.99)$. It has a core-jet radio morphology on VLBA and VLA scales. The STIS slits were aligned approximately perpendicular to the radio jet as the nuclear isophotal position angles are poorly constrained. The central kinematic and flux properties are listed in Table \ref{ngc3862_p1}; the gas does not exhibit a regular rotation profile. The fit to the central spectrum is improved by the addition of a broad component which appears to represent a non-flat continuum in this case. Data for this galaxy are shown in Figure \ref{sgpi10} (see key in Figure \ref{key} for an explanation of these plots).

\paragraph{UGC~7115} This elliptical galaxy has a nuclear dust disk $(b/a \sim 0.95)$. It has a core-jet radio morphology on VLA scales. The STIS slit were aligned approximately perpendicular to the radio jet as the nuclear isophotal position angles are poorly constrained, a certain degree of freedom was allowed in slit placement to allow reasonable observing windows. This galaxy was observed in only one slit position, as we also required WFPC2 observations of this target in order to measure the central photometric properties \citep[see][]{gvknoc}. The central kinematic and flux properties are listed in Table \ref{ugc7115-cor_p1}; the gas exhibits a regular rotation profile. The fit to the central spectrum is not significantly improved by the addition of a broad component. Data for this galaxy are shown in Figure \ref{sgpi11} (see key in Figure \ref{key} for an explanation of these plots).

\paragraph{NGC~4261 (UGC~7360)} This E2-3 galaxy has a nuclear dust disk $(b/a = 0.46)$. It has a twin-jet radio morphology on VLBA and VLA scales. The STIS slits were aligned parallel to the galaxy major axis. The central kinematic and flux properties are listed in Table \ref{ngc4261_p1}. The nucleus of this galaxy lies closer to one of the side slits (slit one) than the central position, however it is still possible to see a clear rotation curve along that slit. The fit to the central spectrum is improved by the addition of a broad component. Data for this galaxy are shown in Figure \ref{sgpi12} (see key in Figure \ref{key} for an explanation of these plots).

\paragraph{NGC~4335 (UGC~7455)} This elliptical galaxy has a nuclear dust disk $(b/a = 0.41)$. It has a radio core on VLBA scales and a twin-jet morphology on VLA scales. The STIS slits were aligned parallel to the galaxy major axis. The central kinematic and flux properties are listed in Table \ref{ngc4335_p1}; the gas exhibits a regular rotation profile. In the positive offset side slit there is a dip in the velocity dispersion profile at the position closest to the nucleus. See also \citet{gvk4335}. The fit to the central spectrum is improved by the addition of a broad component. Data for this galaxy are shown in Figure \ref{sgpi13} (see key in Figure \ref{key} for an explanation of these plots).

\paragraph{NGC~4374 (M84; UGC~7494)} This E1 galaxy has a nuclear dust lane $(width:length = 0.15)$. It has a core-jet radio morphology on VLBA scales, and a twin-jet morphology on VLA scales. The STIS slits were aligned approximately perpendicular to the radio jets, which lies close to the major axis of the emission line gas. The central kinematic and flux properties are listed in Table \ref{m84_p1}; the gas exhibits a regular rotation profile. See also \citet{bower98}. The fit to the central spectrum is not significantly improved the addition of a broad component. Data for this galaxy are shown in Figure \ref{sgpi14} (see key in Figure \ref{key} for an explanation of these plots).

\paragraph{NGC~4486 (M87; UGC~7654)} This elliptical galaxy has an irregular nuclear dust morphology. It has a core-jet radio morphology on VLBA and VLA scales. The STIS slits were aligned to trace morphological features in the emission line gas across the nuclear region of this galaxy. The central kinematic and flux properties are listed in Table \ref{ngc4486_p1}; the gas exhibits a regular rotation profile. The fit to the central spectrum is improved by the addition of a broad component. Data for this galaxy are shown in Figure \ref{sgpi15} (see key in Figure \ref{key} for an explanation of these plots).

\paragraph{NGC~5127 (UGC~8419)} This elliptical peculiar galaxy has a nuclear dust lane $(width:length = 0.25)$. It has a radio core on VLBA scales and a twin-jet morphology on VLA scales. The STIS slits were aligned parallel to the galaxy major axis. The central kinematic and flux properties are listed in Table \ref{ngc5127_p1}; the gas exhibits a regular rotation profile. The fit to the central spectrum is not significantly improved by the addition of a broad component. Data for this galaxy are shown in Figure \ref{sgpi16} (see key in Figure \ref{key} for an explanation of these plots).

\paragraph{NGC~5141 (UGC~8433)} This S0 galaxy has a nuclear dust lane $(width:length = 0.25)$. It has a core-jet radio morphology on VLBA scales and a twin-jet morphology on VLA scales. The STIS slits were aligned parallel to the galaxy major axis. The central kinematic and flux properties are listed in Table \ref{ngc5141_p1}; the gas exhibits a regular rotation profile. The fit to the central spectrum is improved by the addition of a broad component. Data for this galaxy are shown in Figure \ref{sgpi17} (see key in Figure \ref{key} for an explanation of these plots).

\paragraph{NGC~5490 (UGC~9058)} This elliptical galaxy has a nuclear dust lane $(width:length = 0.35)$. It has a core-jet radio morphology on VLBA scales and a twin-jet morphology on VLA scales. The STIS slits were aligned parallel to the galaxy major axis. The central kinematic and flux properties are listed in Table \ref{ngc5490_p1}; the gas does not exhibit a regular rotation profile. The fit to the central spectrum is improved by the addition of a broad component. Data for this galaxy are shown in Figure \ref{sgpi18} (see key in Figure \ref{key} for an explanation of these plots).

\paragraph{NGC~7052 (UGC~11718)} This elliptical galaxy has a nuclear dust disk $(b/a =0.30)$. It has a twin-jet radio morphology on VLBA scales and a core-jet morphology on VLA scales. The STIS slits were aligned parallel to the galaxy major axis. The central kinematic and flux properties are listed in Table \ref{ngc7052_p1}; the gas exhibits a regular rotation profile. The fit to the central spectrum is not significantly improved by the addition of a broad component. Data for this galaxy are shown in Figure \ref{sgpi19} (see key in Figure \ref{key} for an explanation of these plots).

\paragraph{UGC~12064} This S0 galaxy has a nuclear dust disk $(b/a = 0.54)$. It has a twin-jet radio morphology on VLA scales. The STIS slits were aligned parallel to the dust disk major axis. The central kinematic and flux properties are listed in Table \ref{ugc12064_p1}; the gas exhibits a regular rotation profile. The fit to the central spectrum is improved by the addition of a broad component. Data for this galaxy are shown in Figure \ref{sgpi20} (see key in Figure \ref{key} for an explanation of these plots).

\paragraph{NGC~7626 (UGC~12531)} This elliptical peculiar galaxy has a nuclear dust lane $(width:length = 0.17)$. It has a core-jet radio morphology on VLBA scales and a twin-jet morphology on VLA scales. The STIS slits were aligned parallel to the galaxy major axis. The central kinematic and flux properties are listed in Table \ref{ngc7626_p1}; the gas exhibits a regular rotation profile. The fit to the central spectrum is improved by the addition of a broad component. Data for this galaxy are shown in Figure \ref{sgpi21} (see key in Figure \ref{key} for an explanation of these plots).

\section{Interpretation and Discussion}
In our initial interpretation we have focussed on understanding the general parameters of the data set. We will undertake more detailed analyses in future work that we outline in \S 6 below. Here, we first describe the categorization of sources as rotating and non-rotating systems based on the observed kinematics (\S 5.1). We then discuss the ionization states of the nuclear regions (\S 5.2). We go on to discuss the presence of broad components in these nuclei and a more detailed analysis of the line shapes (\S 5.3). 

\subsection{Rotators and non-rotators}
By inspecting maps of the central kinematics and the velocity profiles along each slit (as presented above in figures \ref{sgplimage} to \ref{sgpi21}), we have classified, by eye, the galaxies into two classes: rotators and non-rotators. Rotators are systems where we see patterns reminiscent of rotation curves; in non-rotators we find no such patterns - the kinematics seem either irregular or organized in some manner that does not represent regular rotation. We do not include NGC~741 in discussions of kinematics as very few points were well fit during our analysis. We classify 67\% (N = 14/21) of the UGC FR-I galaxies as rotators. 73\% of galaxies with dust disks (N = 8/11), 100\% of galaxies with dust lanes (N = 5/5) and 50\% of galaxies with complex dust or no dust (N = 2/4) are rotators.

We have made use of the mean velocity dispersion

\begin{equation}
{\overline{\sigma_{100{\rm pc}}} = {{1} \over {N}} \displaystyle \sum_{i} \sigma_i : x_i \le 100{\rm pc,} }
\end{equation}

and the difference in mean velocities on each side of the nucleus

\begin{equation}
\Delta_{100{\rm pc}} = \left|{\left( {{1} \over {N_1}} \displaystyle {\sum_{i}v_i} : -100{\rm pc}\le x_i<0\right) - \left( {{1} \over {N_2}} \displaystyle {\sum_{j}v_j} : 0<x_j\le 100{\rm pc}\right)}\right|
\end{equation}

within $100~{\rm pc}$ of the brightest pixel as illustrative of the global kinematic parameters along the central slit\footnote{For NGC~4261 we used the offset slit closest to the nucleus as explained above.}. These parameters are shown in Table \ref{ketab} for each galaxy, along with the mean properties for each class of galaxy. In Figure \ref{rotornot} we show the relationship between the two parameters for each galaxy. It is clear that the non-rotators lie at the bottom of the $\Delta_{100{\rm pc}}$ distribution, so it is conceivable that irregular motions conceal any remaining signs of rotation in these cases. We note that we detect rotation in all cases where the disk is $\ga 25^{\circ}$ from face on, other than those cases where the dust morphology is highly irregular. The only exception is NGC~5490 where the signal to noise was particularly poor.

In Figure \ref{kpar} we show the values of $\Delta_{100{\rm pc}}$ plotted as a function of dust disk axis ratio, or dust lane width to length ratio; if the dust were all in circular disks this would be an indicator of disk inclination. We included lines showing an indication of the projection of several values of $\Delta_{100{\rm pc}}$ at different disk inclinations through the relation

\begin{equation}
\Delta_{obs} \approx \Delta_{int}.{\rm sin} i \approx \Delta_{int} \times \sqrt{1 - \left( {{b} \over {a}} \right)^2},
\end{equation}

Where $\Delta_{obs}$ is the observed $\Delta_{100{\rm pc}}$ parameter at a given inclination, $\Delta_{int}$ is the presumed intrinsic rotation of the disk and b/a is the axis (or width:length) ratio. This is not a rigorously valid means of projecting these values, but it serves our purposes of illustration here. 

It is easy to see that the non-rotators with face on disks would have to be intrinsically rotating very fast for their rotation to register over the random motions in these cases. From the observations of systems at greater inclinations we have no cause to expect any disks to be rotating that fast. In a similar manner we plot the values of $\overline{\sigma_{100{\rm pc}}}$ against dust disk axis ratio, or dust lane width to length ratio in Figure \ref{kpar2}. Here we see no clear trend in velocity dispersion with axis ratio.

We find no significant systematic differences between the typical values $\overline{\sigma_{100{\rm pc}}}$ between the two classes. Using the Kolmogorov-Smirnov Test to assess the two groups we find a probability of 0.93 that the distributions of velocity dispersions were drawn from the same underlying distribution. Applying the same test to the $\Delta_{100{\rm pc}}$ parameter gives only a 0.08 probability that these were drawn from the same distribution.

This evidence suggests to us that rotators and non-rotators represent the same type of kinematic systems, with observational effects such as inclination and dust properties limiting our ability to detect the rotation of those systems where we do not. We conclude that a model of a rotating gas disk with significant random motions is compatible with all of the observations.

\subsection{Flux ratios and ionization}
In Figure \ref{flux_cfx} we show the [\ion{N}{2}] flux as a function of H$\alpha$ narrow line flux for each central spectrum. The values are compatible with values typically found for photo-ionization and shock models \citep[see, for example,][]{dopita97}. The unusually low ratios shown in two nuclei (NGC~383 and M84) are likely a consequence of the high degree of blending of the lines in these observations (see panel (d) of Figures \ref{sgpi3} and \ref{sgpi14}), rather than any underlying physics.

While we are satisfied that our data have not produced any results that are incompatible with reasonable parameters, detailed modeling is required to understand the various ionization mechanisms at work in each individual case. We intend to undertake this type of modeling and present the outcomes in future work.

\subsection{Are we observing broad lines?}
The fit to the central spectrum is improved in 62\% of the galaxies by the inclusion of an additional free Gaussian component (see \S 4.2). Each of these fits resulted in the supplementary component being centered redwards of the H$\alpha$ line with a velocity dispersion between about two and ten times that of the narrow lines; thus we describe this feature as a broad fit component. We identify these broad components in 73\% of galaxies with dust disks (N=8/11), 60\% of galaxies with dust lanes (N=3/5), 67\% of galaxies with irregular dust (N=2/3) and 0\% of galaxies with no dust (N=0/2). 

If we make a flux cut in H$\alpha$ we find broad components in 40\%\ of nuclei with $1\times 10^{-15} \le F(H\alpha ) \le 1\times 10^{-14}$, in 55\%\ of nuclei with $1\times 10^{-14} \le F(H\alpha ) \le 5\times 10^{-14}$ and in 100\%\ of nuclei with $F(H\alpha ) \ge 5\times 10^{-14}$. This detection trend with flux is reflected in Figure \ref{flux_cfx}. This suggests that the detection of nuclear broad components is somewhat flux (and therefore signal to noise) dependent across the sample.

The broad components could originate either as an artifact of attempting to fit Gaussians to non-Gaussian line profiles, or from a physical source -- such as a broad line region or a change in the characteristics of the gas as the inner regions of the disk are approached.

We detect broad components only in the central few pixels, this could be a consequence of either the fall in signal to noise or that the component is an unresolved source. It is important to note as a consequence of this, that fitting single Gaussians to each line samples a different part of the line shape as the central pixels are approached.

In Table \ref{blstats} we show the mean properties of the broad components and compare them to the samples of LINERS by \citet{ho97} and Radio Galaxies summarized by \citet{sulentic00}. These comparisons show that our broad components are compatible with the broad lines seen in LINERs and that the observed offsets from the narrow line components are compatible with those observed in samples of radio loud galaxies.

\subsection{Constraining the line shapes}

In order to better establish the line shape, we investigated the profiles from the nucleus of NGC~4335, which has a relatively good signal to noise and relatively non-blended lines \citep[][]{gvk4335}. 

We first investigated the profile of the two [\ion{S}{2}] lines, as they are less entangled than the H$\alpha$ + [\ion{N}{2}] complex. We fit these two lines with a varying number of Gaussians per line. The minimum in the reduced $\chi^2$ parameter resulting from each of these fits lay between two and three Gaussians per line. We extended the two Gaussian per line model to fit the entire profile of the five emission lines. By then testing various sets of free parameters we were able to discover an optimal set. 

Fitting these models quickly illustrated that the additional broad component is a very strongly favored feature; in any case where it was possible for the parameters to contrive to create a broad component they did so. Table \ref{vsmodels} shows the kinematic parameters and reduced $\chi^2$ values from five combinations of fit parameters that produce distinct fits. These were: 

\begin{enumerate}
\item {\em Narrow lines only:} The original 5 Gaussian model (one Gaussian for each of the following 5 lines: [\ion{N}{2}]$_{6550}$, H$\alpha$, [\ion{N}{2}]$_{6585}$, [\ion{S}{2}]$_{6718}$, [\ion{S}{2}]$_{6733}$), with a single value of velocity and velocity dispersion for all lines.
\item {\em Additional broad component:} The same as the {\em Narrow lines only} model, with an additional broad component, with an independent velocity, velocity dispersion and flux.
\item {\em Flux-constrained Broad Bases (i):} The same as model 2, with an additional set of 5 Gaussians, representing the broader wings (Broad Bases). This new set of lines had a single velocity and velocity dispersion and the velocity was fixed to be the same as for the other lines. The fluxes of each line in the second set of Gaussians was fixed at a constant ratio to its counterpart in the first set, based on the mean ratio measured in the fit of two Gaussians per line to the two [\ion{S}{2}] lines as described above.
\item {\em Flux-constrained Broad Bases (ii):} The same as model 3, however in this case the velocity of the set of broad bases was allowed to vary from that of the narrow lines. 
\item {\em Flux-unconstrained Broad Bases:} the same as model 3, however in this case the lines in both sets were able to vary independently in flux, other than the fixed 1:3 ratio between the two [\ion{N}{2}] lines in each set.
\end{enumerate}

We conclude, as the {\em Flux-constrained Broad Bases (ii)} model not only produces the most satisfying fit judged by eye, but also the lowest reduced $\chi^2$ value, that this model is likely to most closely represent the line profile present in the central regions. The broad bases in this case are able to represent an asymmetric red wing on each line, but they do not reduce the importance of the broad component in the fit. 

\section{Conclusions}
In this paper we presented the medium resolution spectra of the 21 galaxies in our UGC~FR-I sample, obtained by ourselves and others using STIS. Data were obtained for three parallel slit positions on each nucleus (other than UGC~7115, which was observed in only one slit position).

We find that all nuclei are compatible with a single kinematic description: a rotating gas disk, where random motions are always important. We observe patterns reminiscent of rotation in 67\% of the nuclei, in the remainder the non-detection can be accounted for as the systems are either face on, have complex central morphologies (as judged from the nuclear dust distribution) or, in the case of NGC~5490, particularly poor signal to noise.

We find that the inclusion of an additional fit component with unconstrained parameters improves our fit to the nuclear spectrum in 62\% of the galaxies, where it fits in every case as a broad component in the vicinity of H$\alpha$ and [\ion{N}{2}]. The detection of the broad component is related to the line flux (and therefore the signal to noise). The broad components have a mean velocity dispersion of $1349 \pm 345$~\kms\ and are redshifted from the narrow line components (assuming an origin in H$\alpha$) by $486 \pm 443$~\kms.

The broad component could be a consequence of non-Gaussian line profiles with broad wings, which would be biased toward the brightest [\ion{N}{2}] line (redwards of $H\alpha$). However, our more detailed analysis of line shape shows that it is very hard to reproduce this effect by including broad (even asymmetric) wings on each line, suggesting the broad component may indeed have a physical origin.

The measured H$\alpha$ to [\ion{N}{2}] ratios for the narrow components in the central spectra are consistent with standard photo-ionization or shock-ionization models \citep[for example,][]{dopita97} other than two examples where the lines are highly blended and may be leading us to misleading fits in the very center.

As we continue with this research we will next model the galaxies with thin disks and with spherical gas distributions \citep[see, for example,][]{gvk4335} in order to obtain estimates of the black hole masses for each galaxy in this complete sample. We will also extend our investigation into the ionization profiles of each nucleus; investigating the ionization properties of each nucleus and seeking the signs of jet / disk interactions.

\acknowledgments
The authors would like to thank Jacqueline van Gorkom for her very valuable comments on this work.

Support for this work was provided by NASA through grant number HST-GO-08236.01-A from the Space Telescope Science Institute, which is operated by the Association of Universities for Research in Astronomy, Inc., under NASA contract NAS5-26555.

This research has made use of the NASA/IPAC Extragalactic Database (NED) which is operated by the Jet Propulsion Laboratory, California Institute of Technology, under contract with the National Aeronautics and Space Administration.

%%%%%%%%%%%%%%
% FIGURES
%%%%%%%%%%%%%%

%\input{figs}

%%%%%%%%%%%%%%%
% FIGURES
%%%%%%%%%%%%%%%

\ifprintfig

\clearpage
\begin{figure}[t]
\epsscale{0.8}
\plottwo{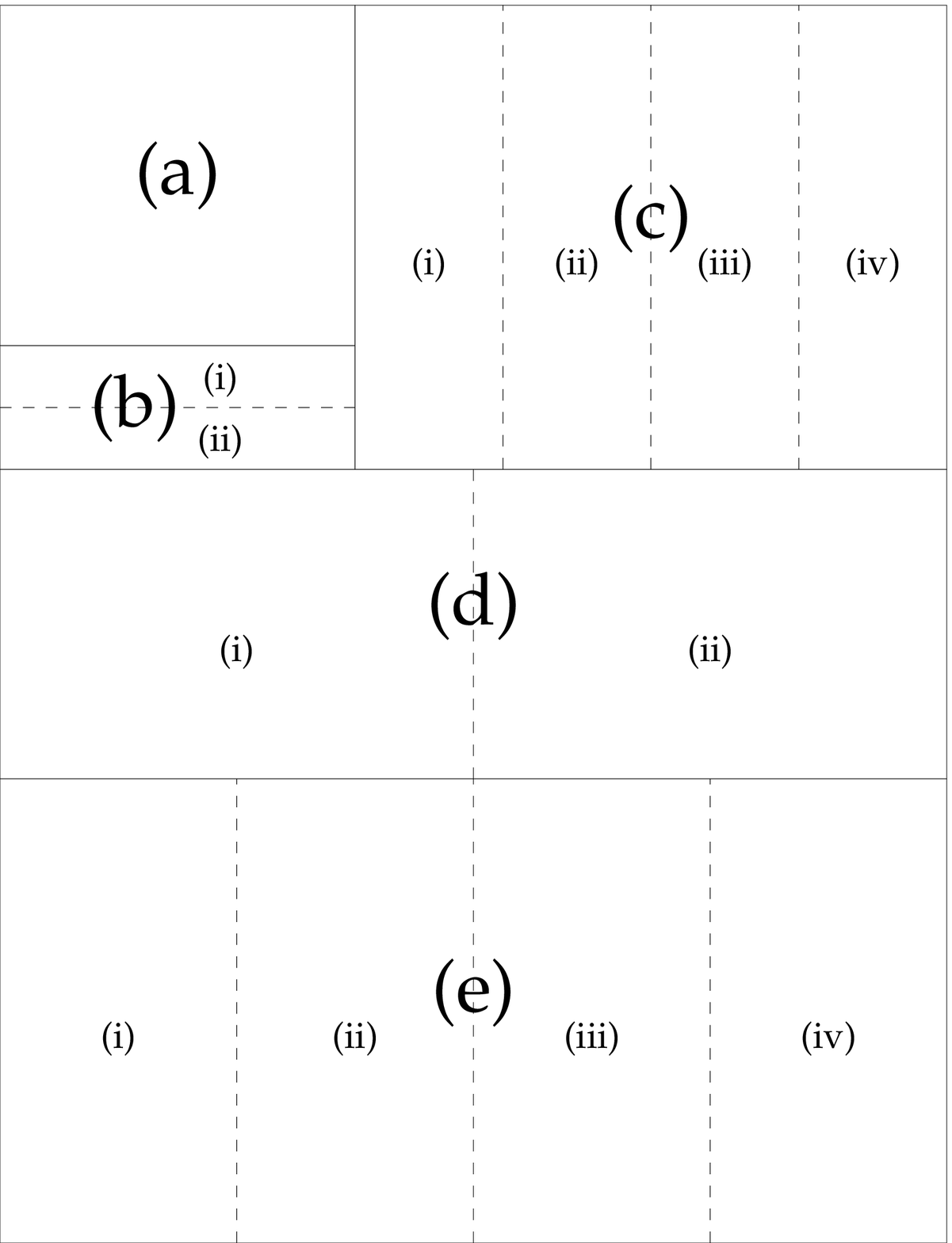}{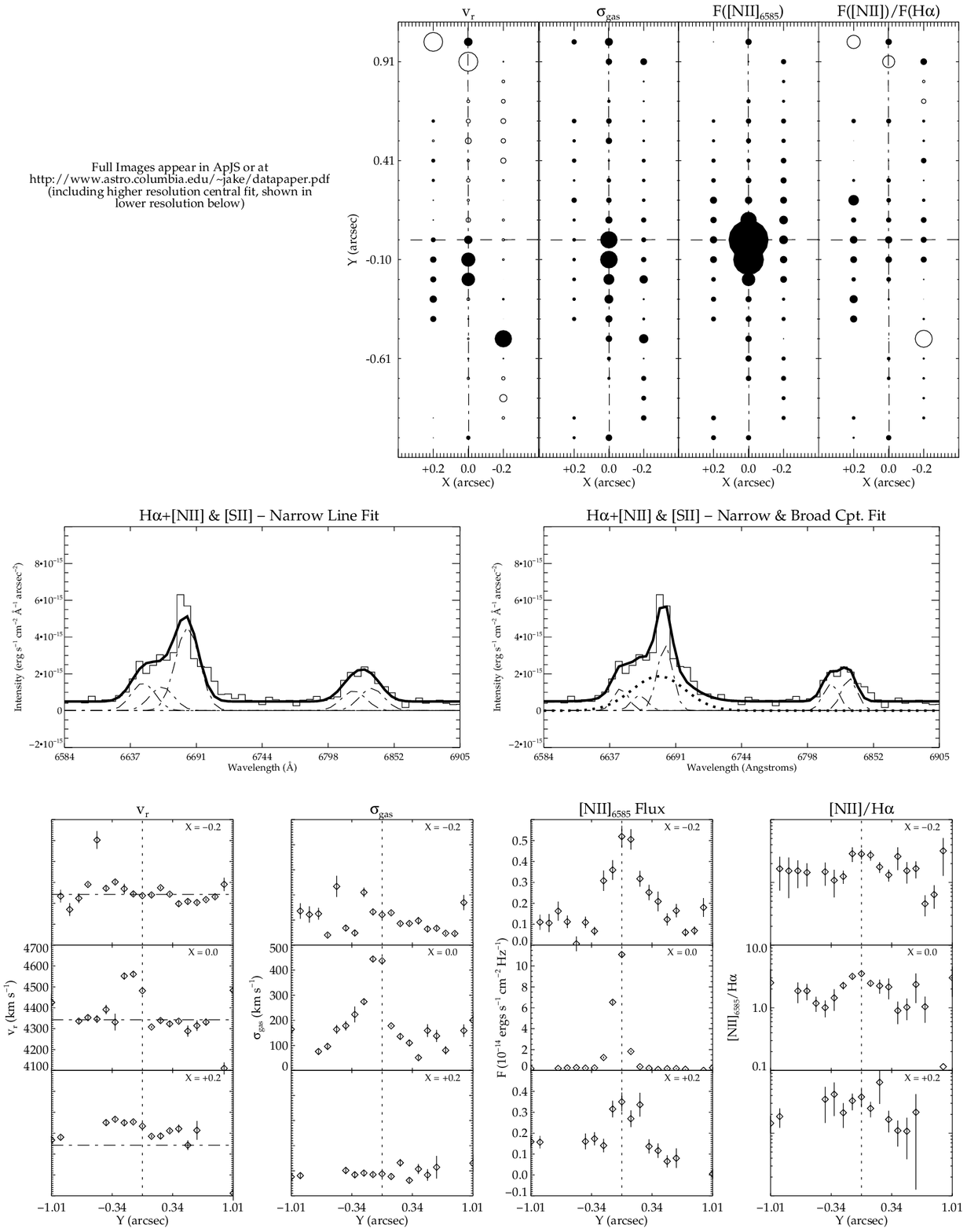}
\figcaption{Key to plots. {\bf Region (a):} the HST acquisition image for each galaxy observed is shown with the positions of the observed long slits overlaid. The relative position angles of the galaxy major axes measured from the central isophotes (dashed lines) and the arcsecond scale radio jets (dotted lines) are shown. These lines cross, and a North-East indicator is drawn, at the location of the central pixel (see text). {\bf Region (b):} (i) the central portion of the reduced 2-D spectrum from the central slit position is shown along with (ii) an image created from fitting a set of 5 Gaussian lines to the same. {\bf Region (c):} plots representing the two-dimensional distributions of parameters measured by fitting a one Gaussian per emission line model to spectra from the nuclear region. The dashed and dash-dot lines indicate the directions of the minor and major axes respectively. The lines cross at the location of the central pixel. We show: (i) radial velocity (filled circles represent velocities greater than the mean, empty circles represent velocities less than the mean and the radius of each point [i,j] is proportional to $|v_{i,j}-v_{mean}|$); (ii) velocity dispersion (the radius of the circle is proportional to $\sigma_{i,j}$); (iii) integrated line flux (the area of each point is proportional to $F_{i,j}$); and (iv) [\ion{N}{2}]$_{6585}$ / H$\alpha$ ratio (The radius of each circle is proportional to \rm log(F([\ion{N}{2}]$_{6585}$)/F($H\alpha$))). {\bf Region (d):} single Gaussian per line fit and residuals without (i) and with (ii) the additional free component described in the text for the central spectrum of each galaxy. {\bf Region (e):} plots of (i) radial velocity, (ii) velocity dispersion, (iii) [\ion{N}{2}] line flux and (iv) [\ion{N}{2}]$_{6585}$ / H$\alpha$ ratio along each slit measured by fitting a one Gaussian per line model to the emission line spectra. The vertical dotted line indicates the Y position of the central row. The dash-dot line in the velocity panel indicates the quoted recession velocity for the galaxy (see Table \ref{sample}).\label{key}}
\end{figure}

% \label{sgplimage}

\clearpage
\begin{figure}
\epsscale{0.9}
\plotone{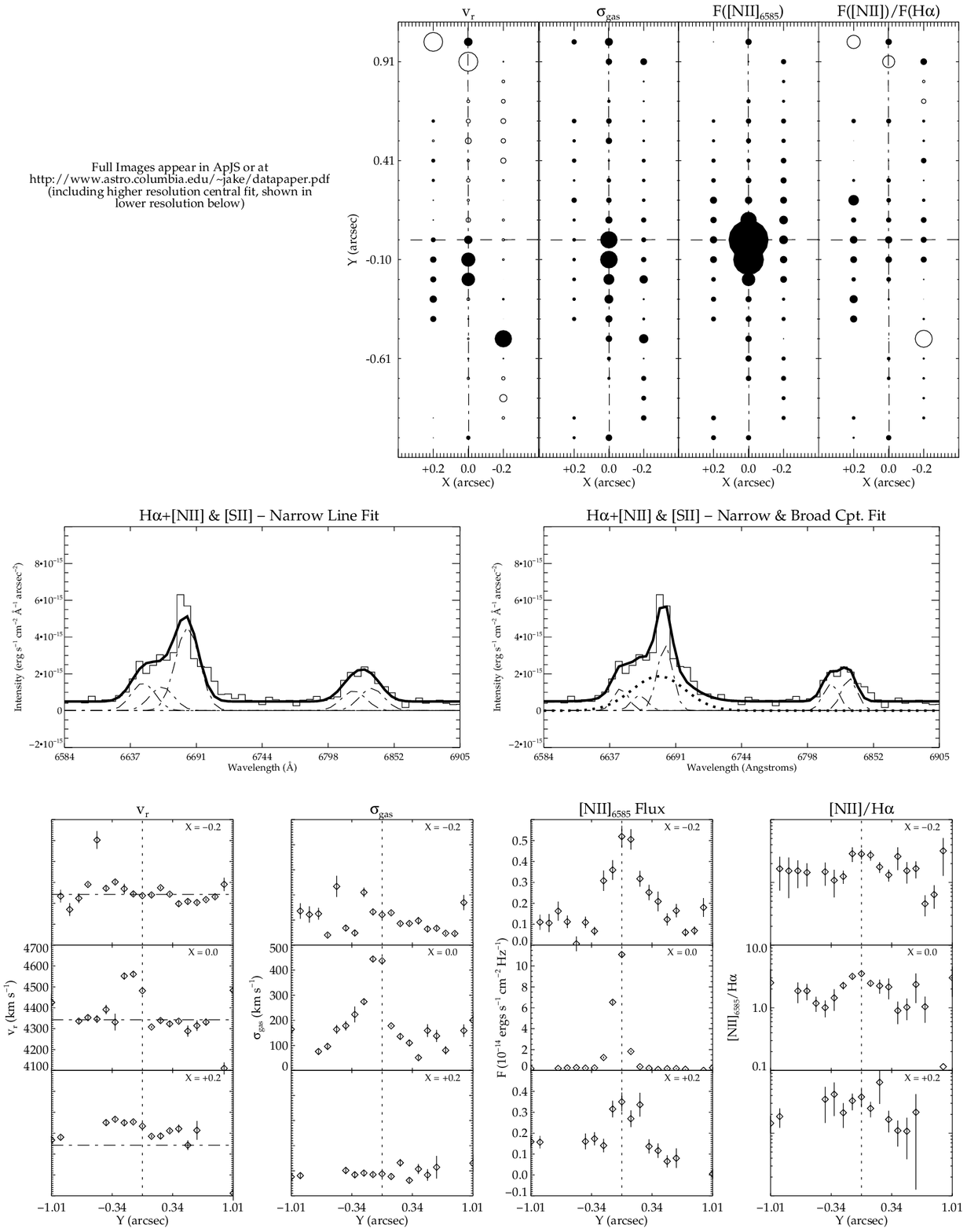}
\figcaption{Observation and fit data for NGC 193, see Figure \ref{key} for description. \label{sgplimage}}
\end{figure}

\iffullout

\clearpage
\begin{figure}
\epsscale{0.9}
\plotone{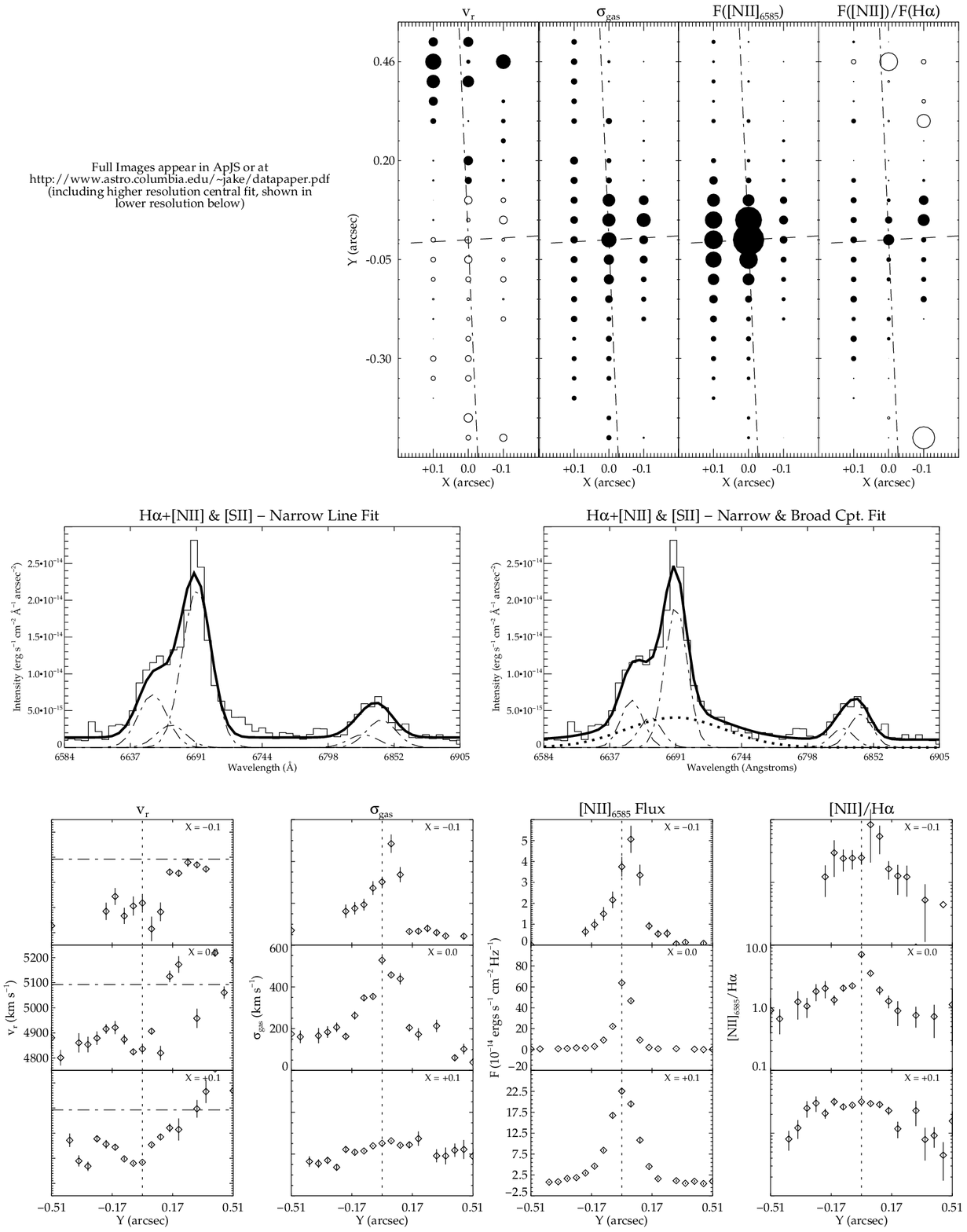}
\figcaption{Observation and fit data for NGC 315, see Figure \ref{key} for description. \label{sgpi2}}
\end{figure}

\clearpage
\begin{figure}
\epsscale{0.9}
\plotone{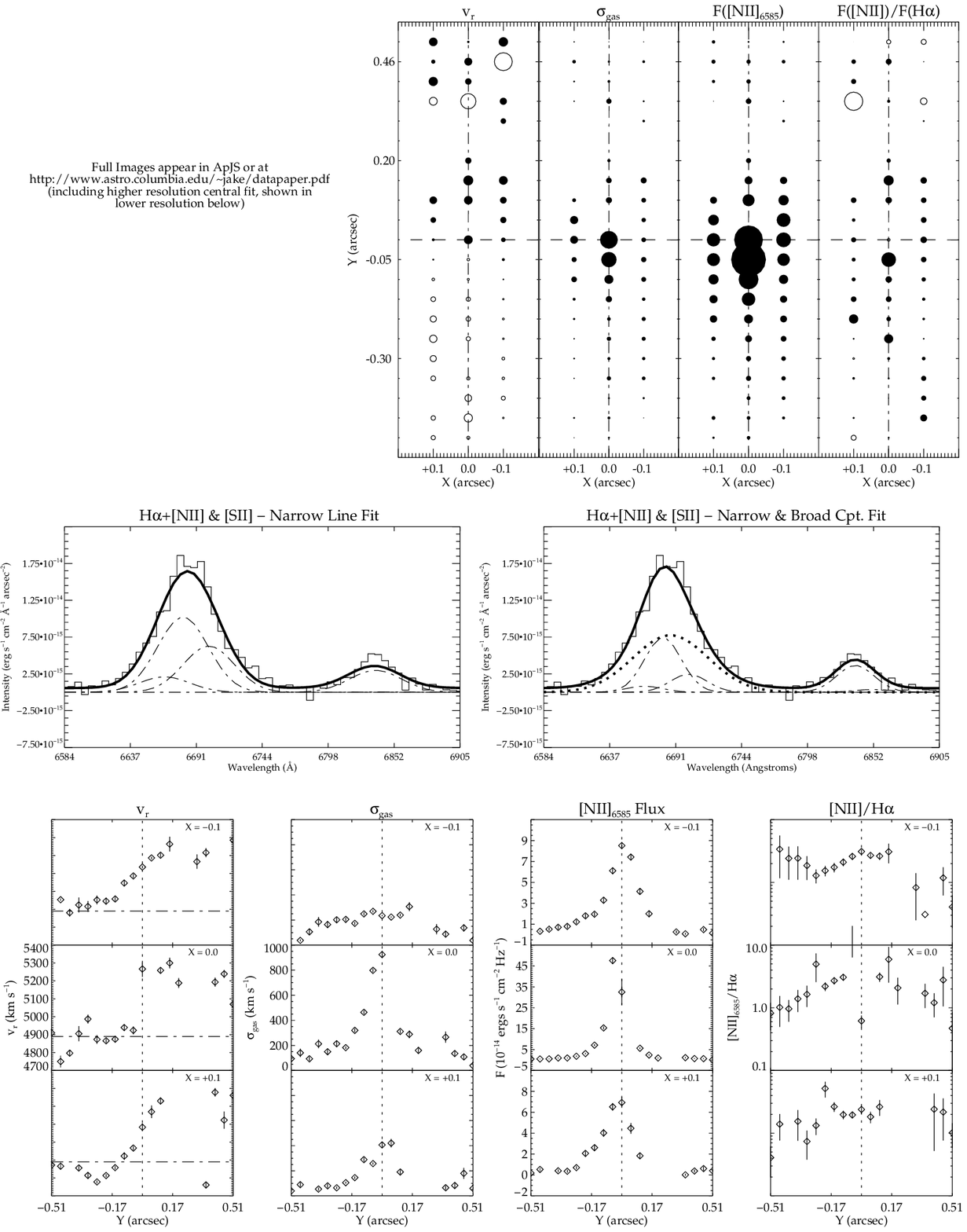}
\figcaption{Observation and fit data for NGC 383, see Figure \ref{key} for description. \label{sgpi3}}
\end{figure}

\clearpage
\begin{figure}
\epsscale{0.9}
\plotone{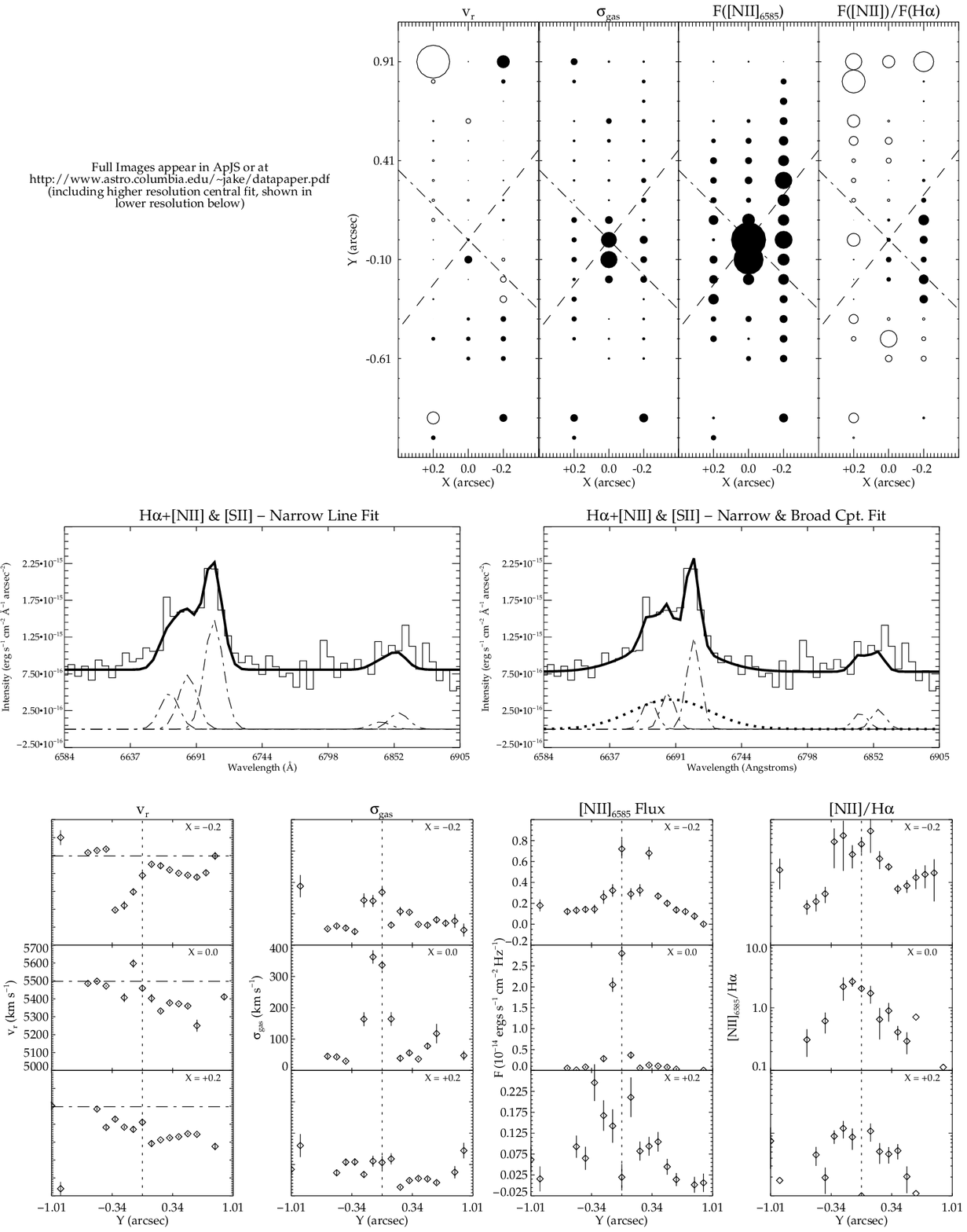}
\figcaption{Observation and fit data for NGC 541, see Figure \ref{key} for description. \label{sgpi4}}
\end{figure}

\clearpage
\begin{figure}
\epsscale{0.9}
\plotone{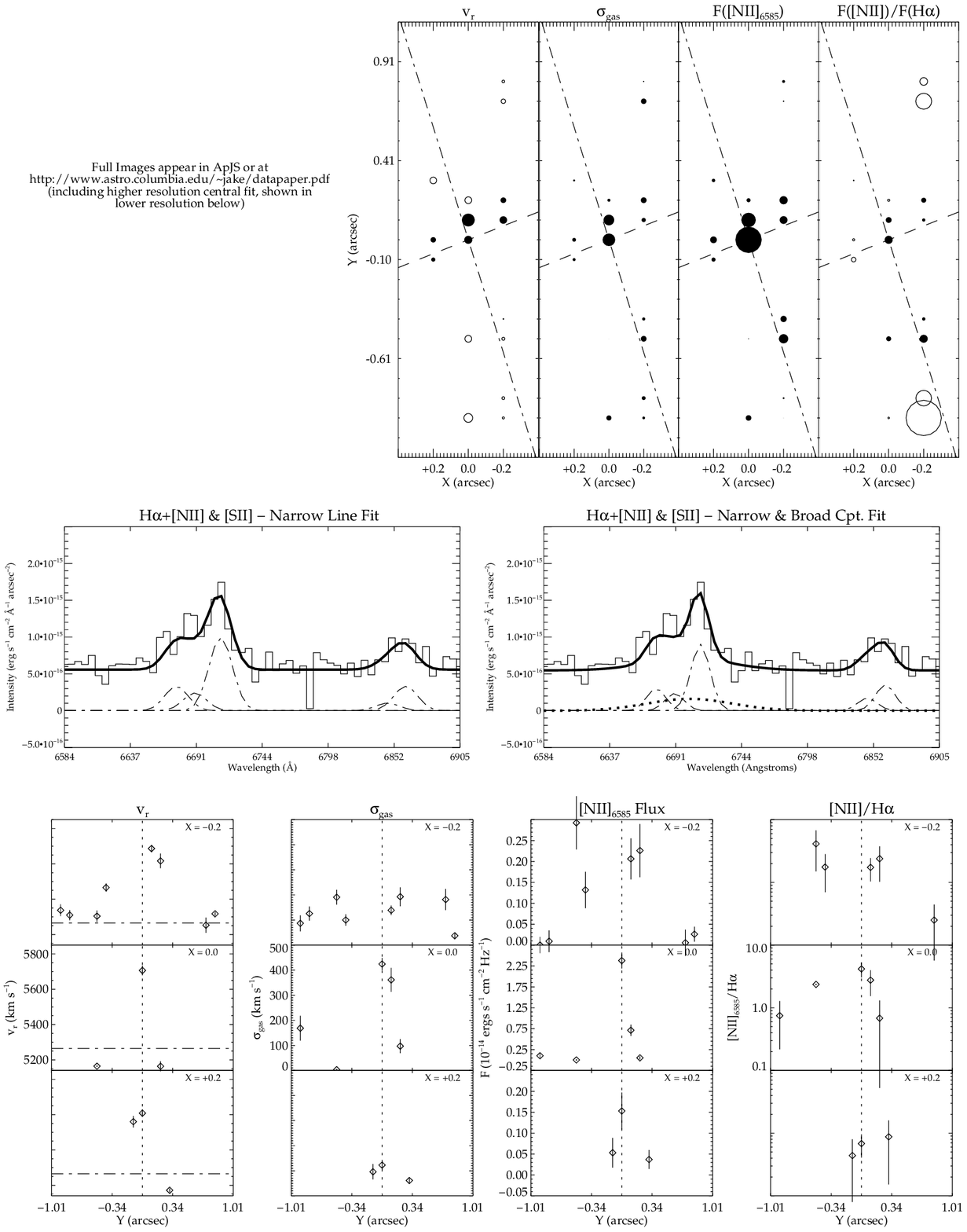}
\figcaption{Observation and fit data for NGC 741, see Figure \ref{key} for description. \label{sgpi5}}
\end{figure}

\clearpage
\begin{figure}
\epsscale{0.9}
\plotone{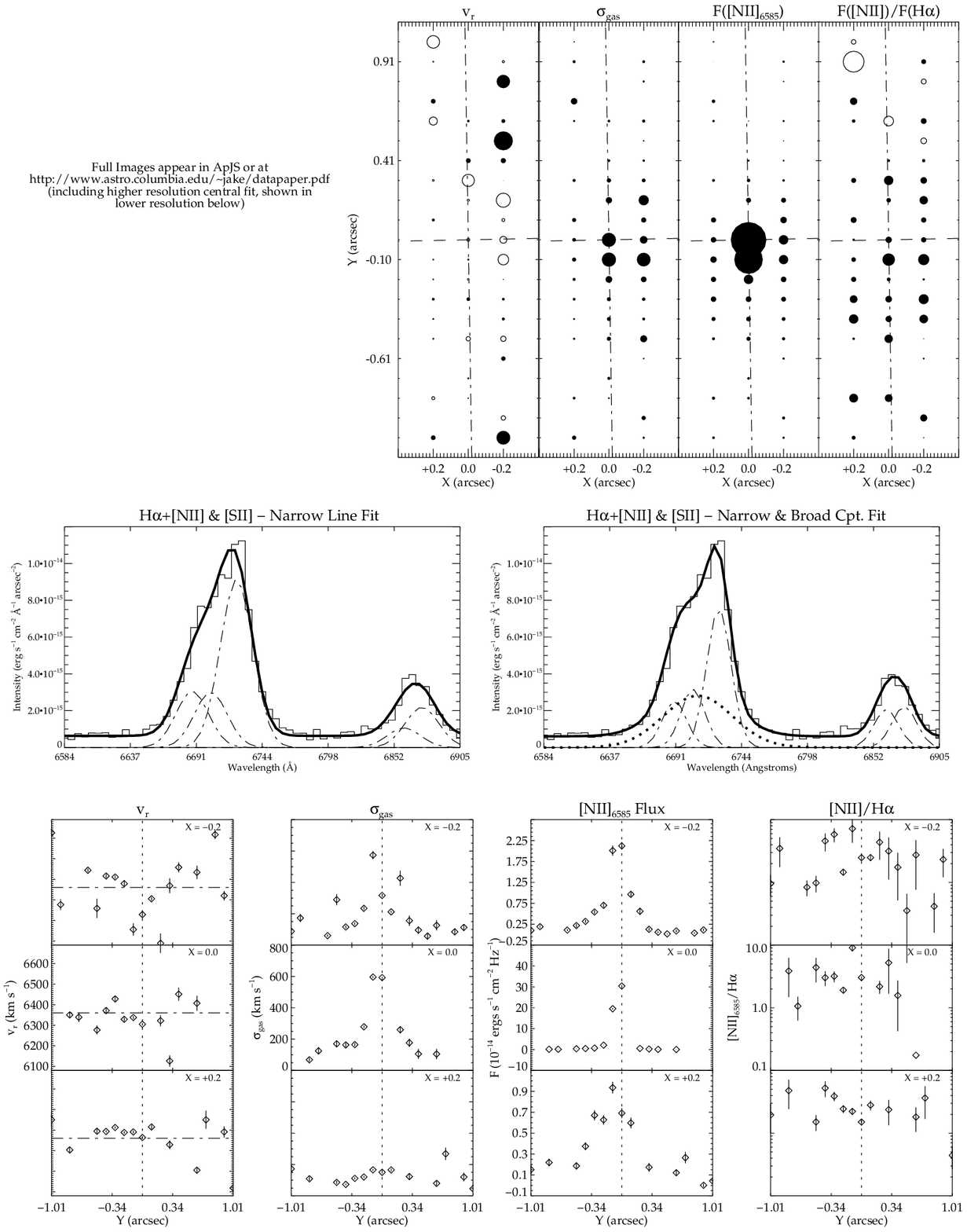}
\figcaption{Observation and fit data for UGC 1841, see Figure \ref{key} for description. \label{sgpi6}}
\end{figure}

\clearpage
\begin{figure}
\epsscale{0.9}
\plotone{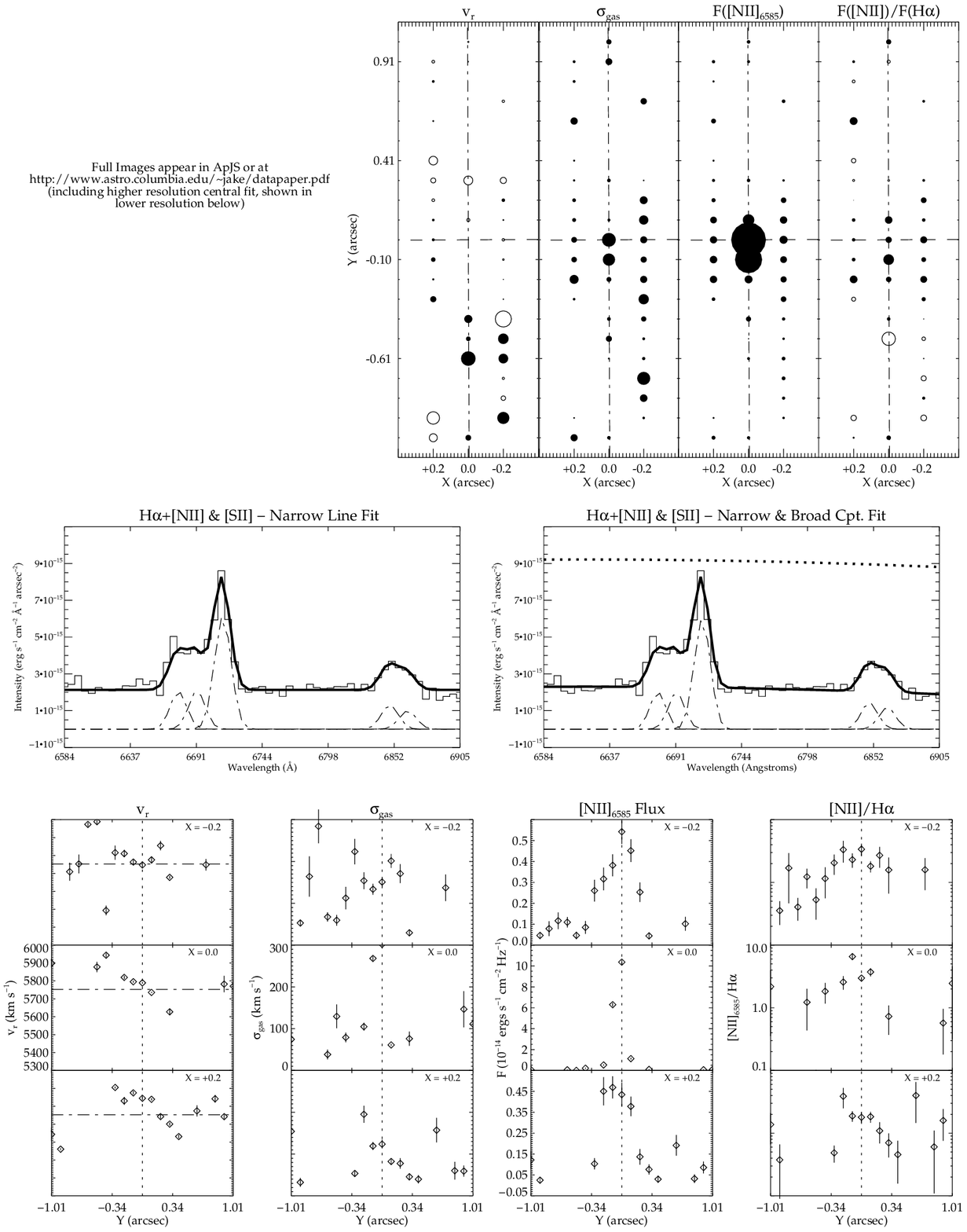}
\figcaption{Observation and fit data for NGC 2329, see Figure \ref{key} for description. \label{sgpi7}}
\end{figure}

\clearpage
\begin{figure}
\epsscale{0.9}
\plotone{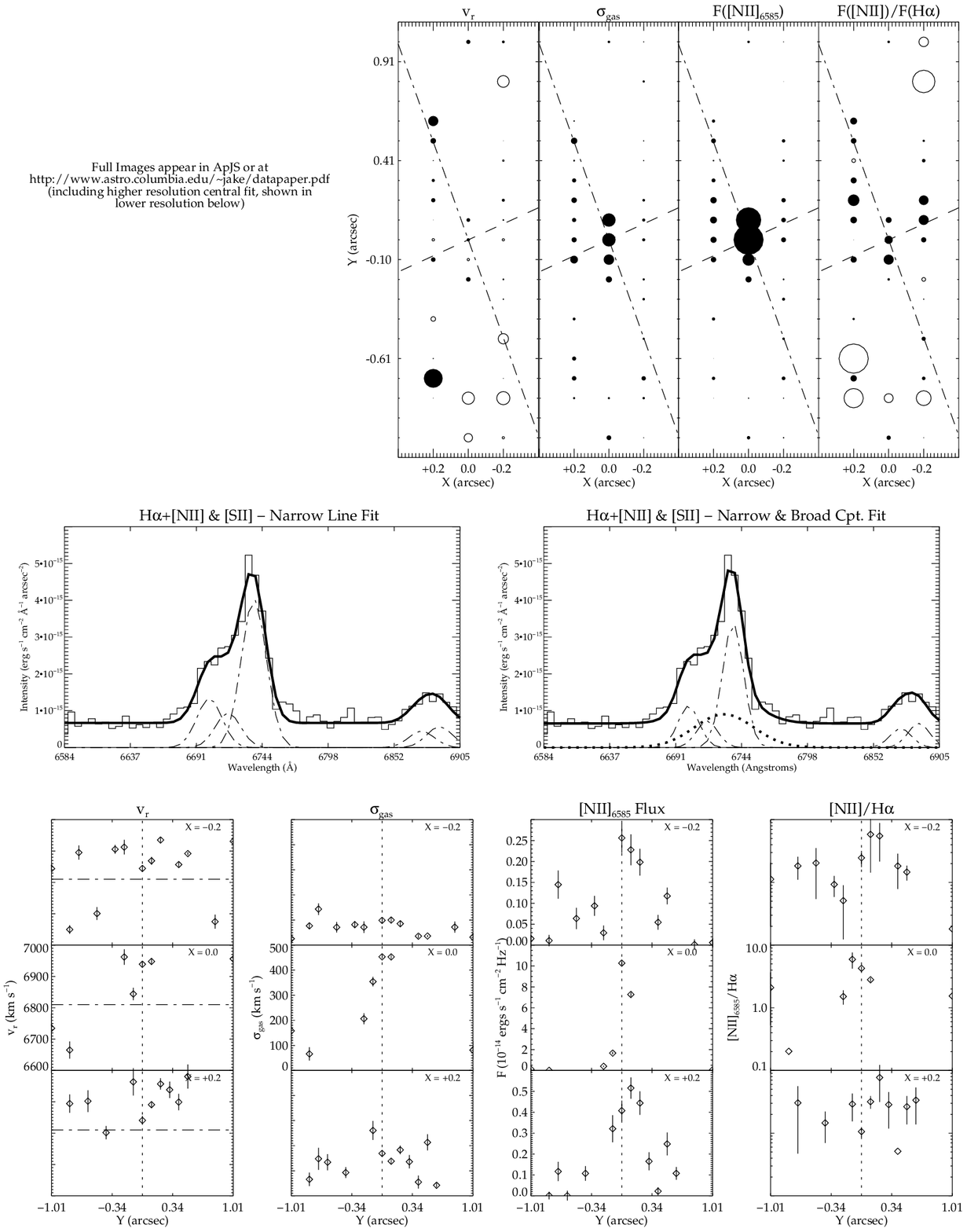}
\figcaption{Observation and fit data for NGC 2892, see Figure \ref{key} for description. \label{sgpi8}}
\end{figure}

\clearpage
\begin{figure}
\epsscale{0.9}
\plotone{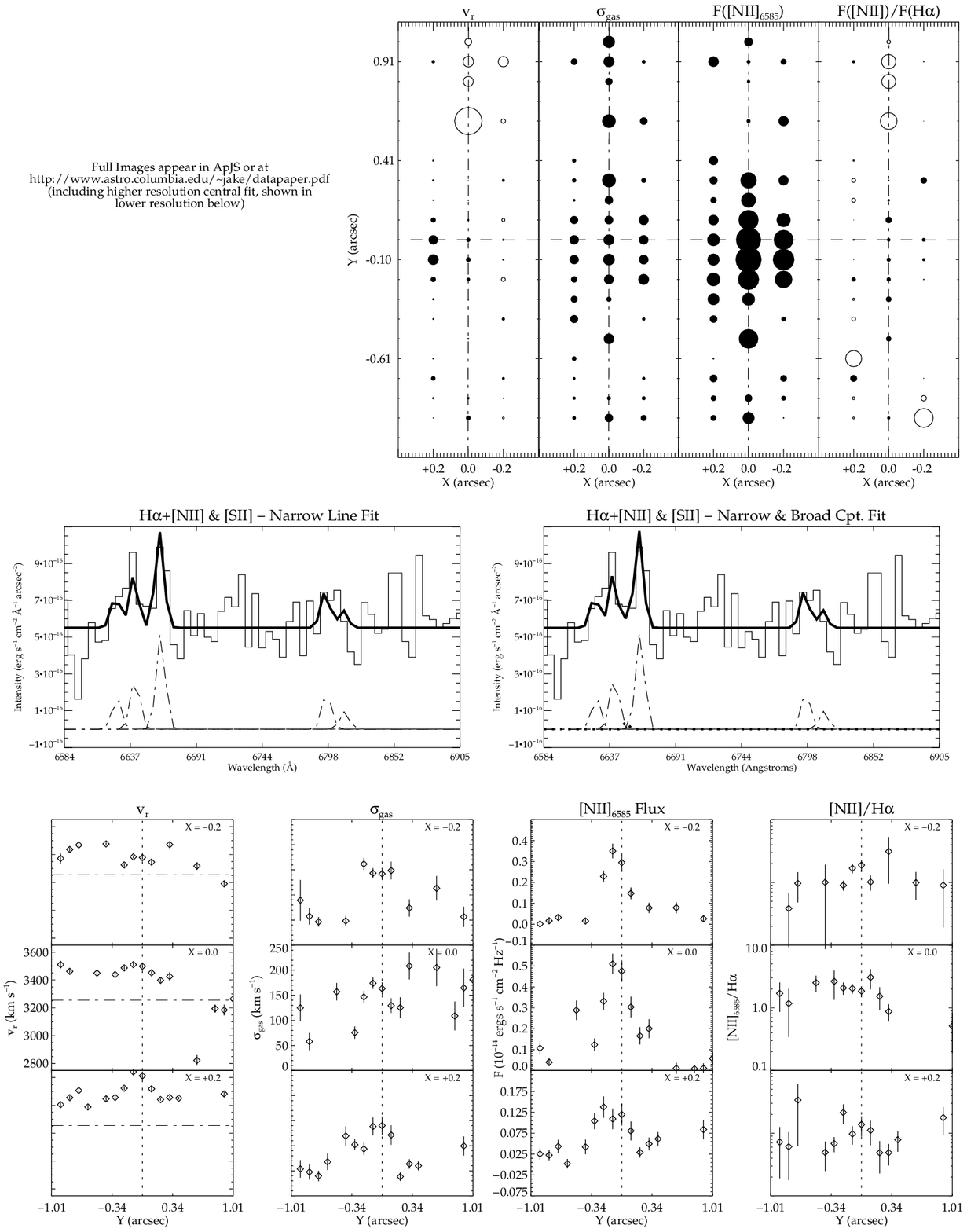}
\figcaption{Observation and fit data for NGC 3801, see Figure \ref{key} for description. \label{sgpi9}}
\end{figure}

\clearpage
\begin{figure}
\epsscale{0.9}
\plotone{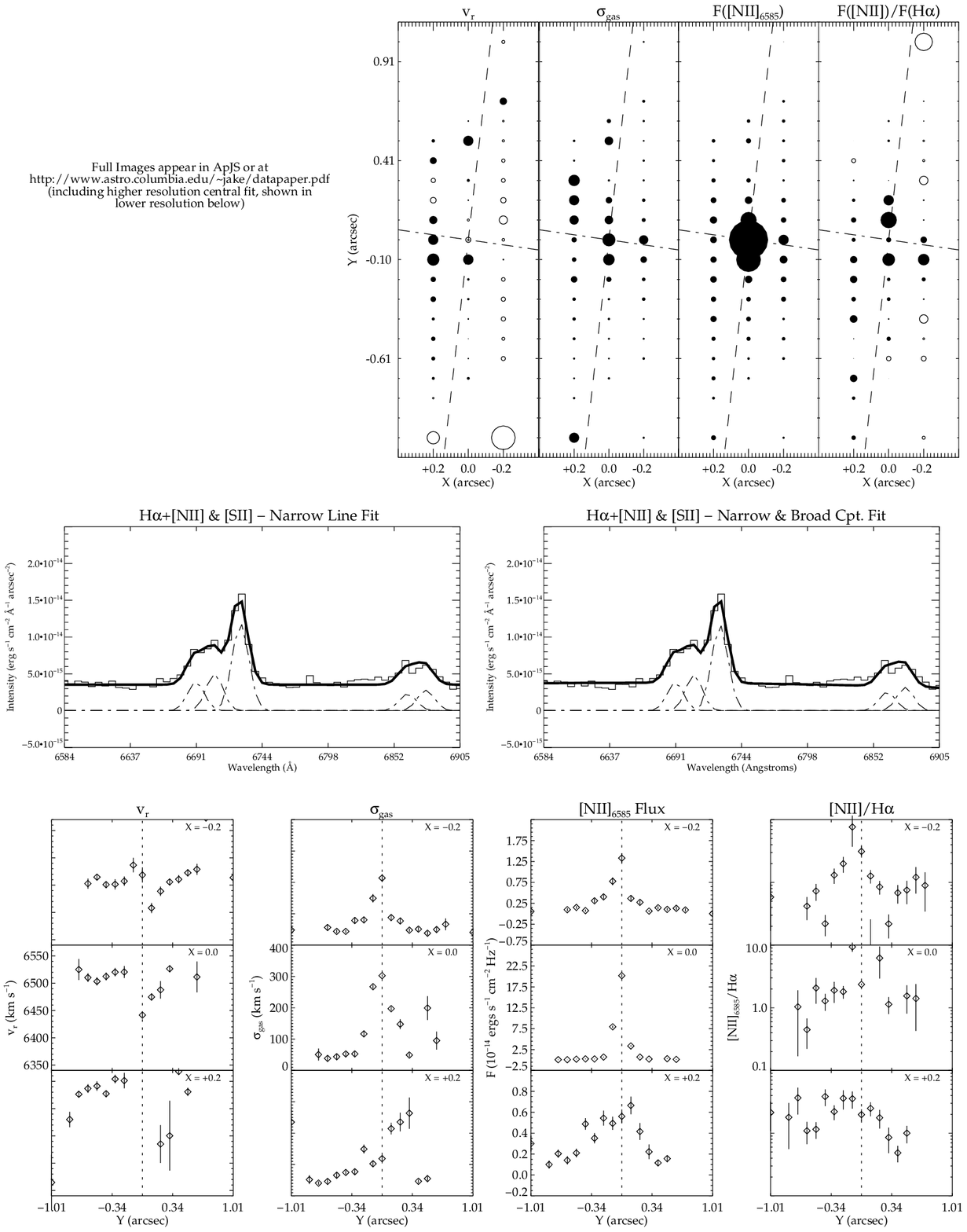}
\figcaption{Observation and fit data for NGC 3862, see Figure \ref{key} for description. \label{sgpi10}}
\end{figure}

\clearpage
\begin{figure}
\epsscale{0.9}
\plotone{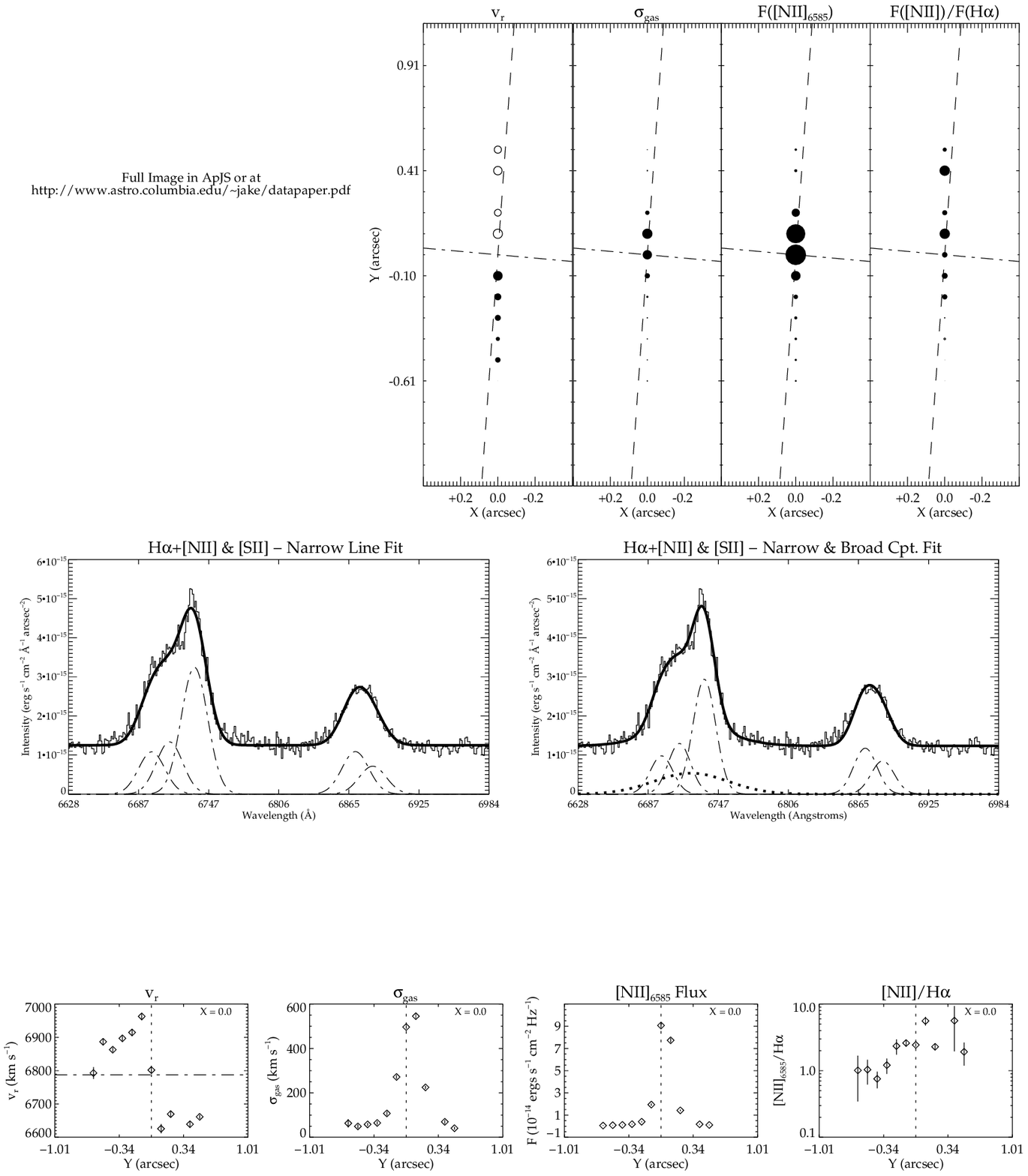}
\figcaption{Observation and fit data for UGC 7115, see Figure \ref{key} for description. \label{sgpi11}}
\end{figure}

\clearpage
\begin{figure}
\epsscale{0.9}
\plotone{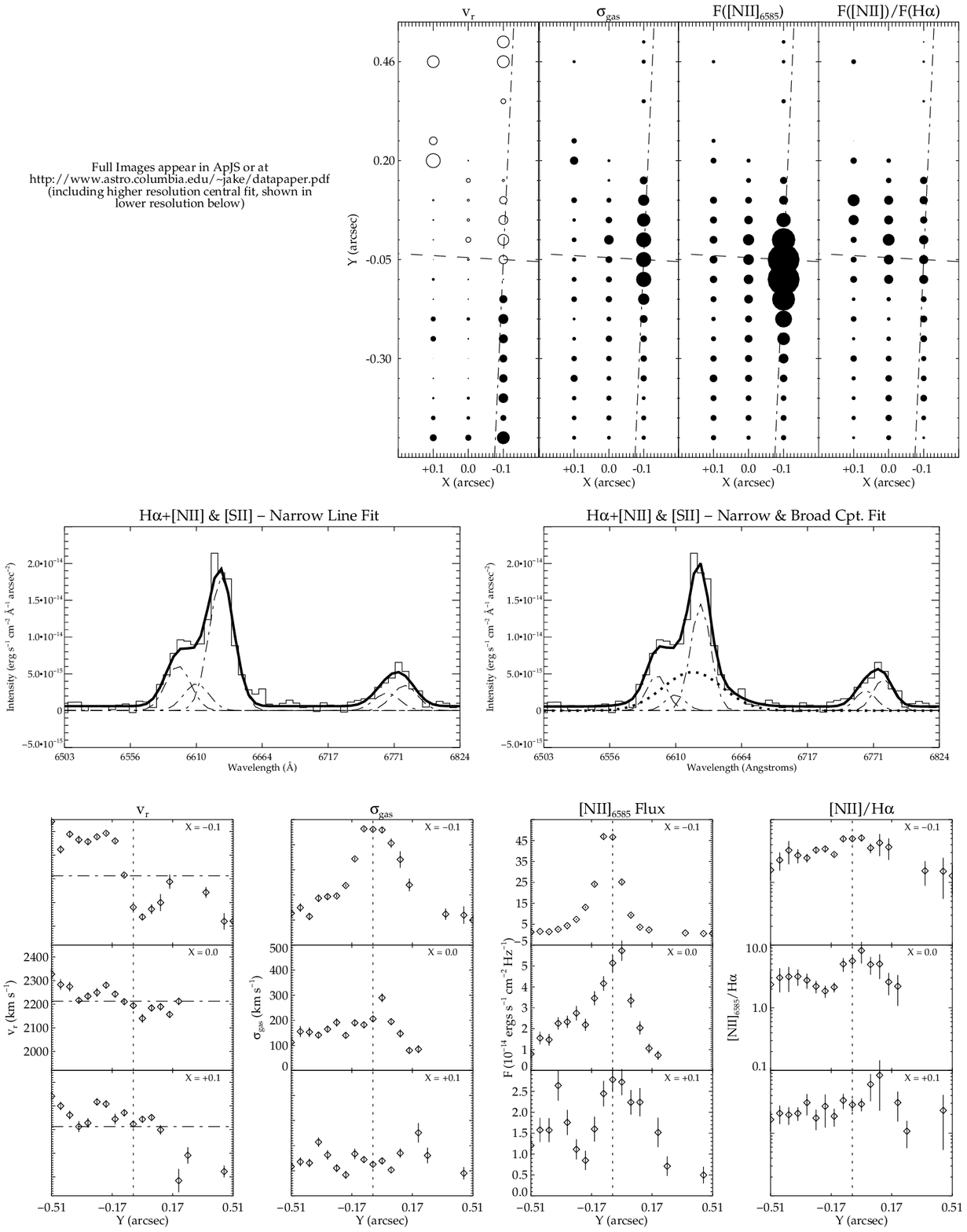}
\figcaption{Observation and fit data for NGC 4261, see Figure \ref{key} for description. \label{sgpi12}}
\end{figure}

\clearpage
\begin{figure}
\epsscale{0.9}
\plotone{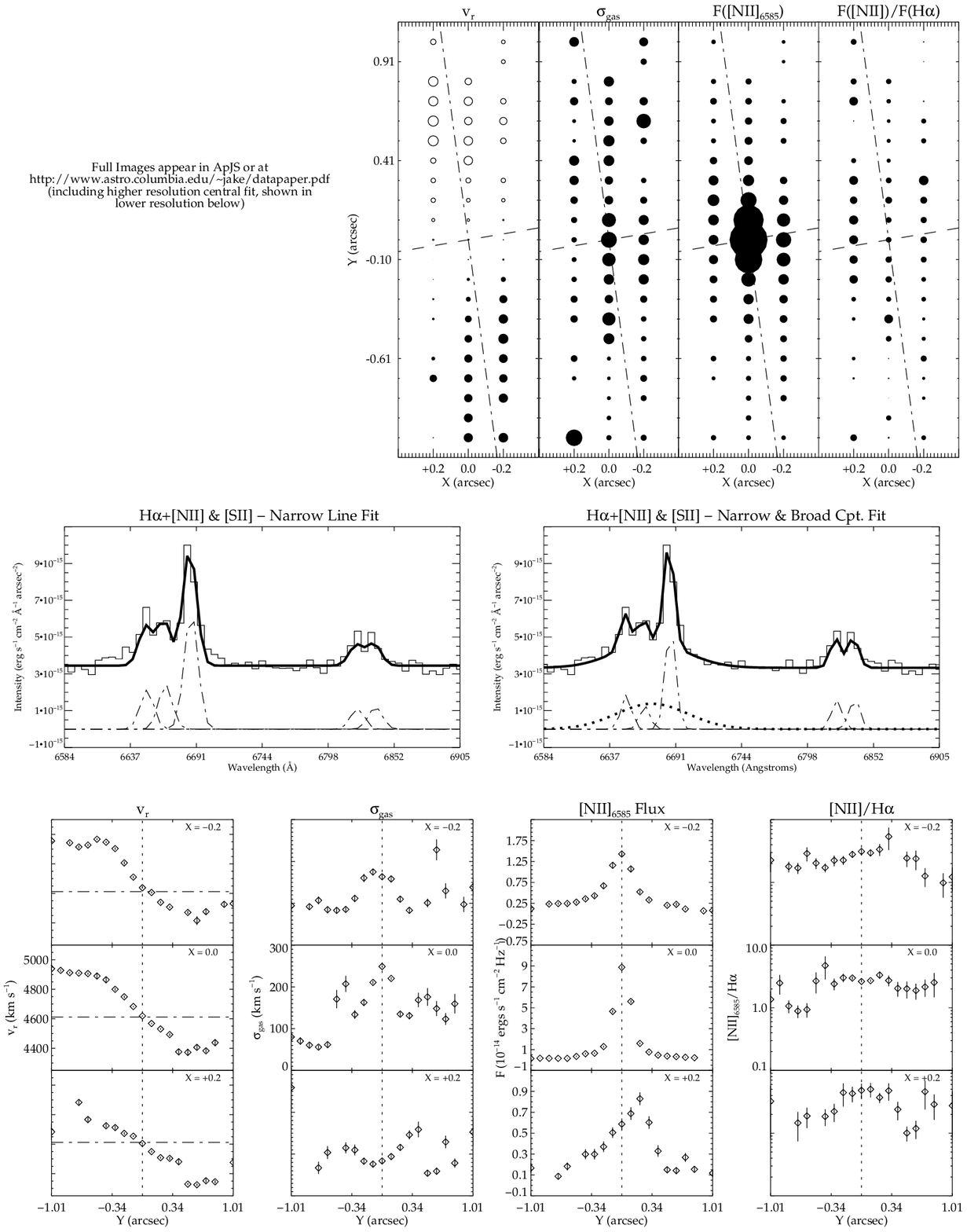}
\figcaption{Observation and fit data for NGC 4335, see Figure \ref{key} for description. \label{sgpi13}}
\end{figure}

\clearpage
\begin{figure}
\epsscale{0.9}
\plotone{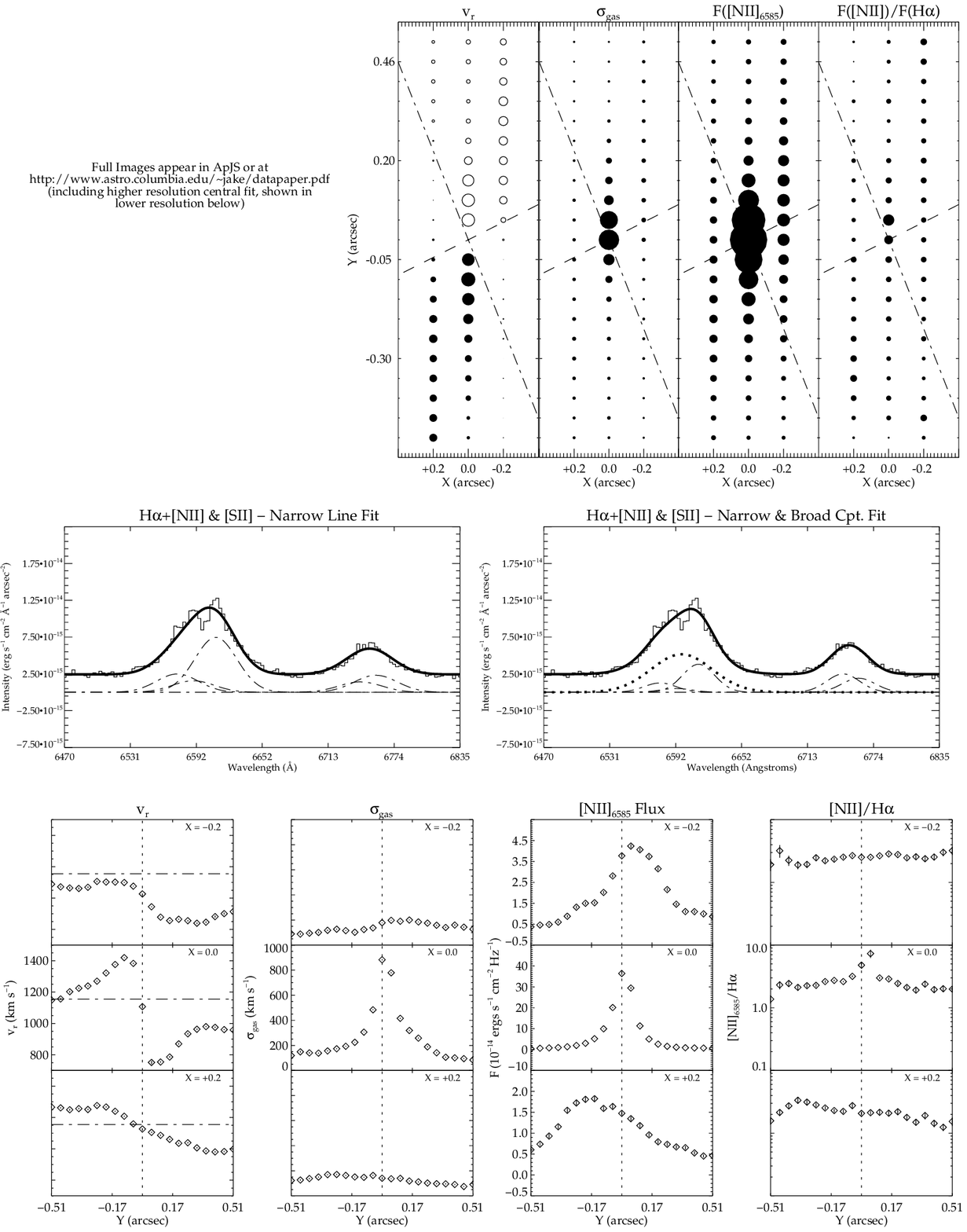}
\figcaption{Observation and fit data for NGC 4374, see Figure \ref{key} for description. \label{sgpi14}}
\end{figure}

\clearpage
\begin{figure}
\epsscale{0.9}
\plotone{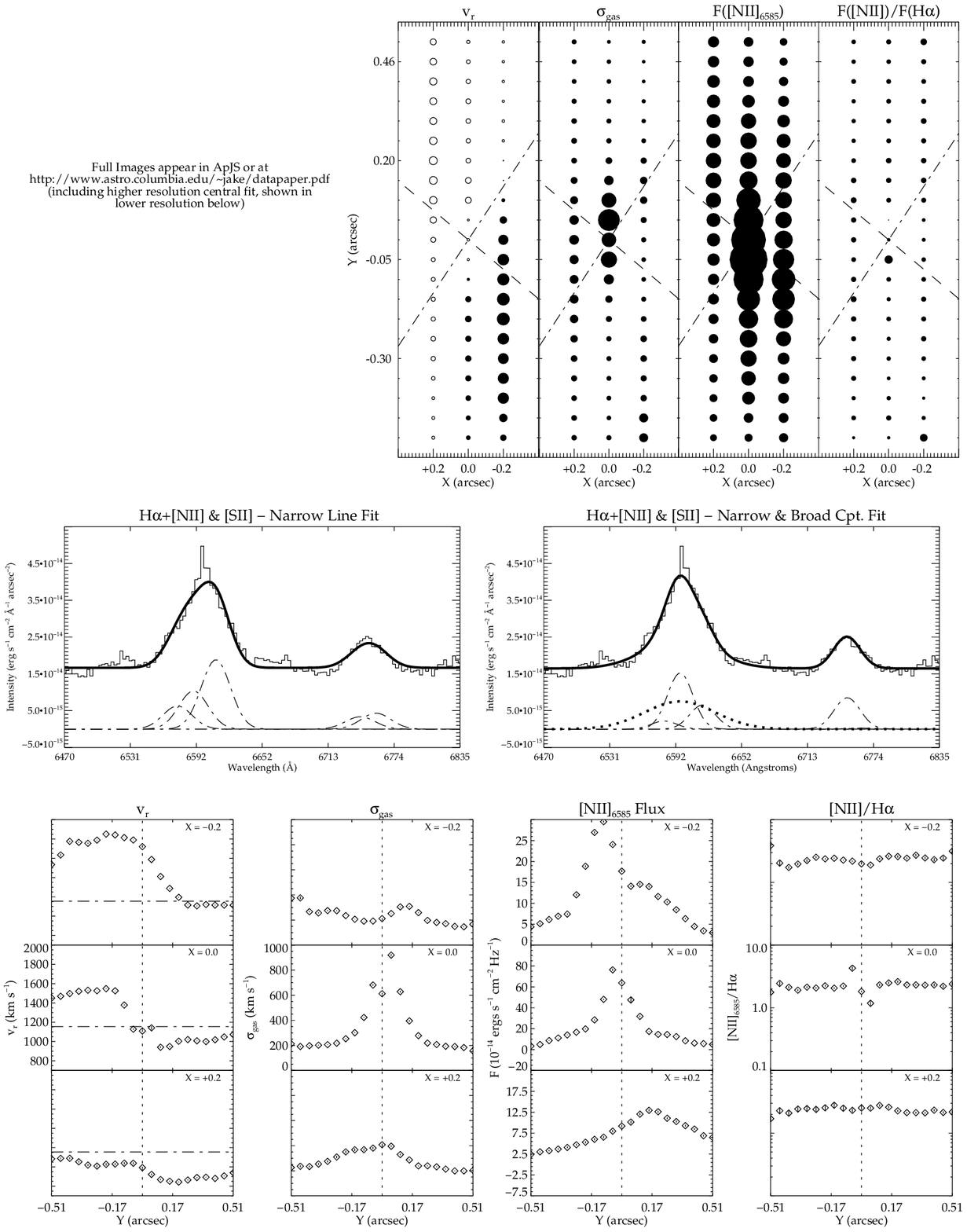}
\figcaption{Observation and fit data for NGC 4486, see Figure \ref{key} for description. \label{sgpi15}}
\end{figure}

\clearpage
\begin{figure}
\epsscale{0.9}
\plotone{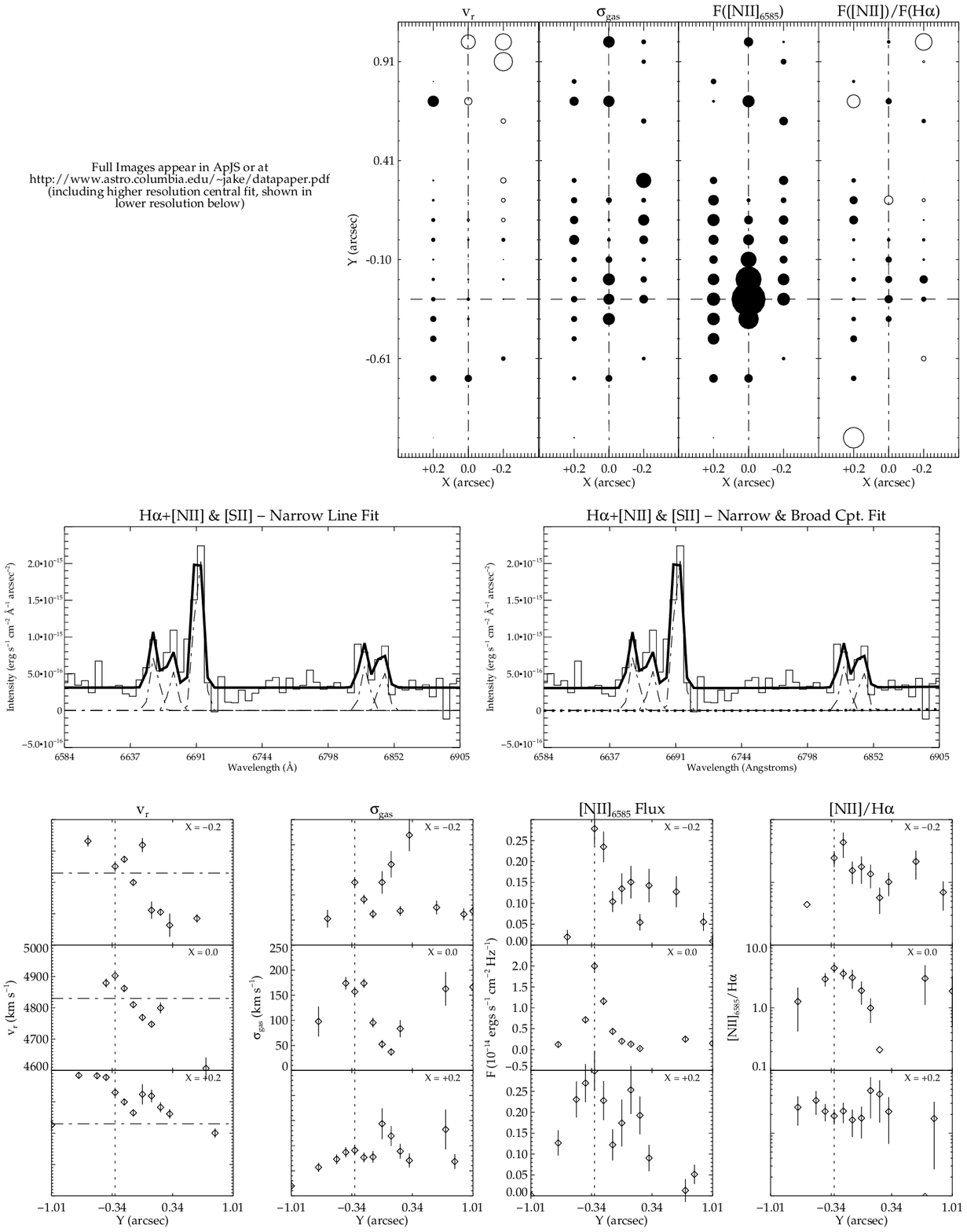}
\figcaption{Observation and fit data for NGC 5127, see Figure \ref{key} for description. \label{sgpi16}}
\end{figure}

\clearpage
\begin{figure}
\epsscale{0.9}
\plotone{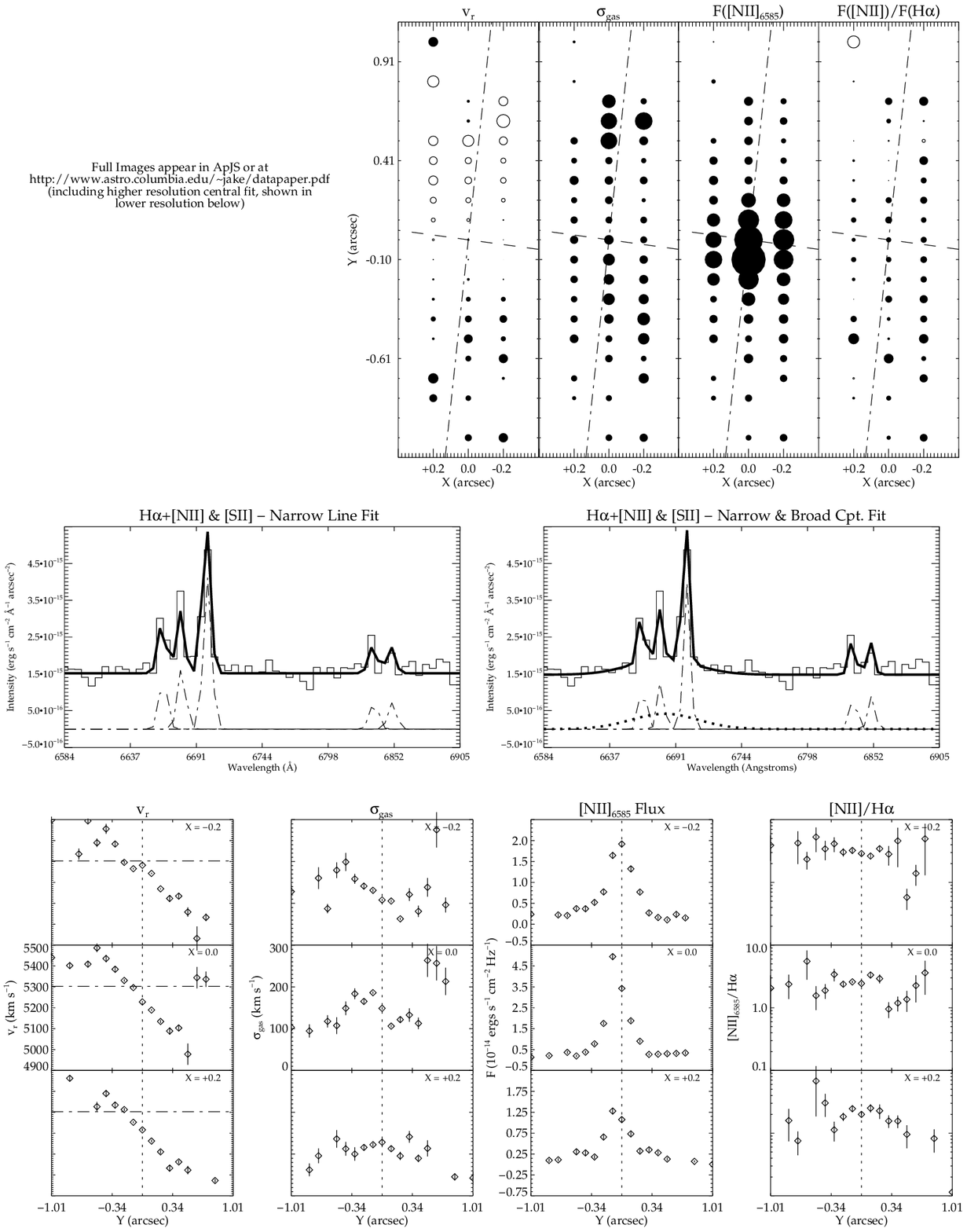}
\figcaption{Observation and fit data for NGC 5141, see Figure \ref{key} for description. \label{sgpi17}}
\end{figure}

\clearpage
\begin{figure}
\epsscale{0.9}
\plotone{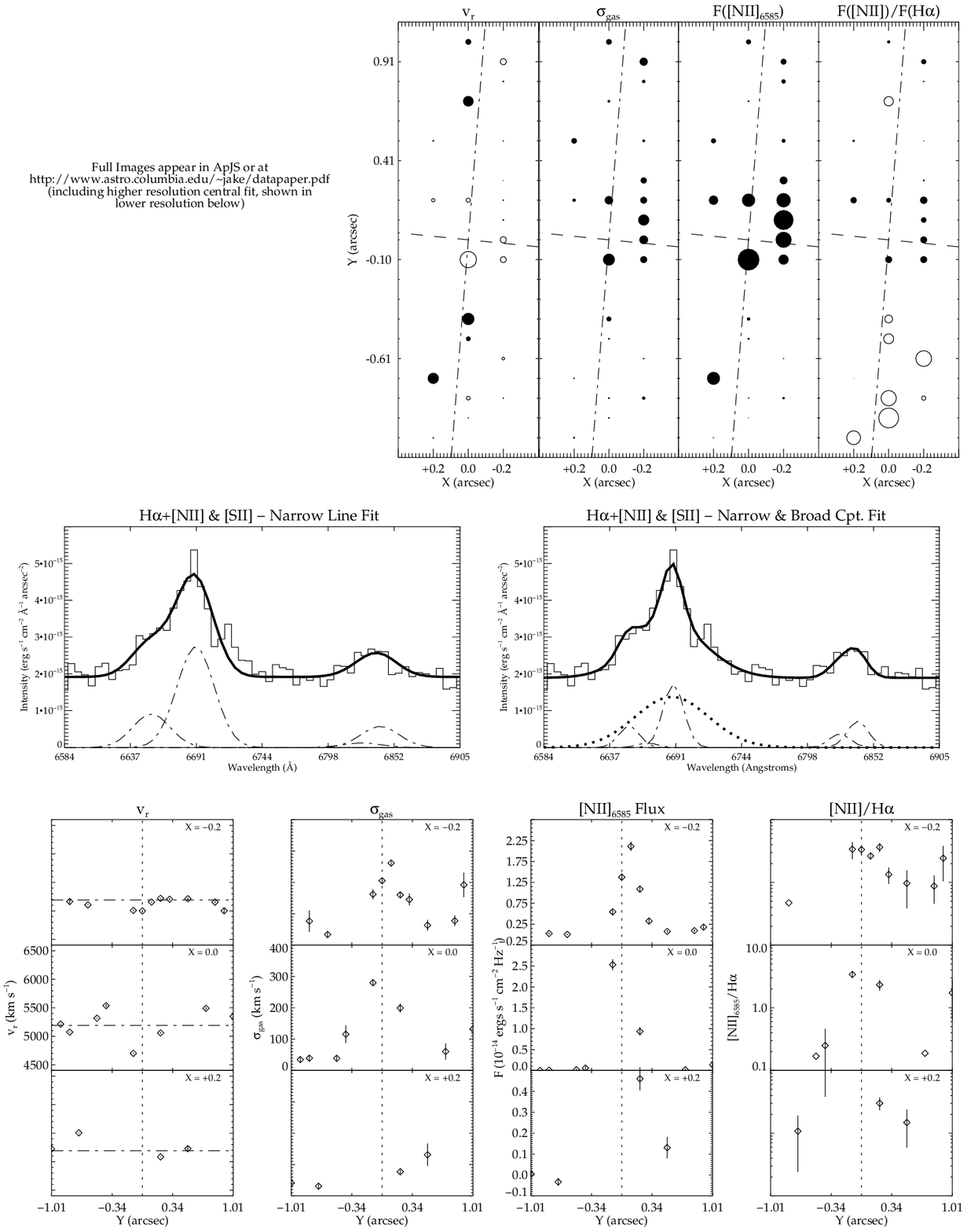}
\figcaption{Observation and fit data for NGC 5490, see Figure \ref{key} for description. \label{sgpi18}}
\end{figure}

\clearpage
\begin{figure}
\epsscale{0.9}
\plotone{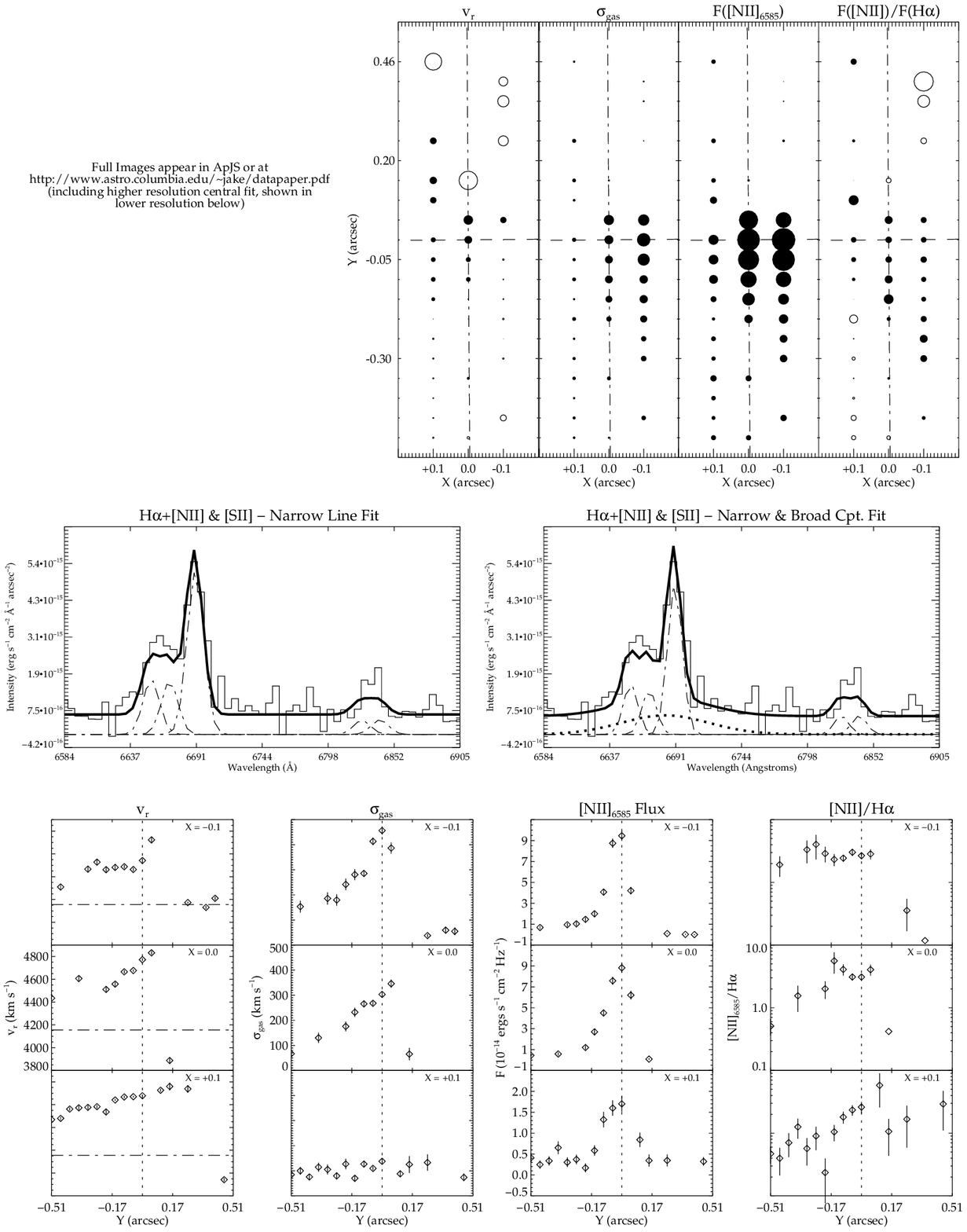}
\figcaption{Observation and fit data for NGC 7052, see Figure \ref{key} for description. \label{sgpi19}}
\end{figure}

\clearpage
\begin{figure}
\epsscale{0.9}
\plotone{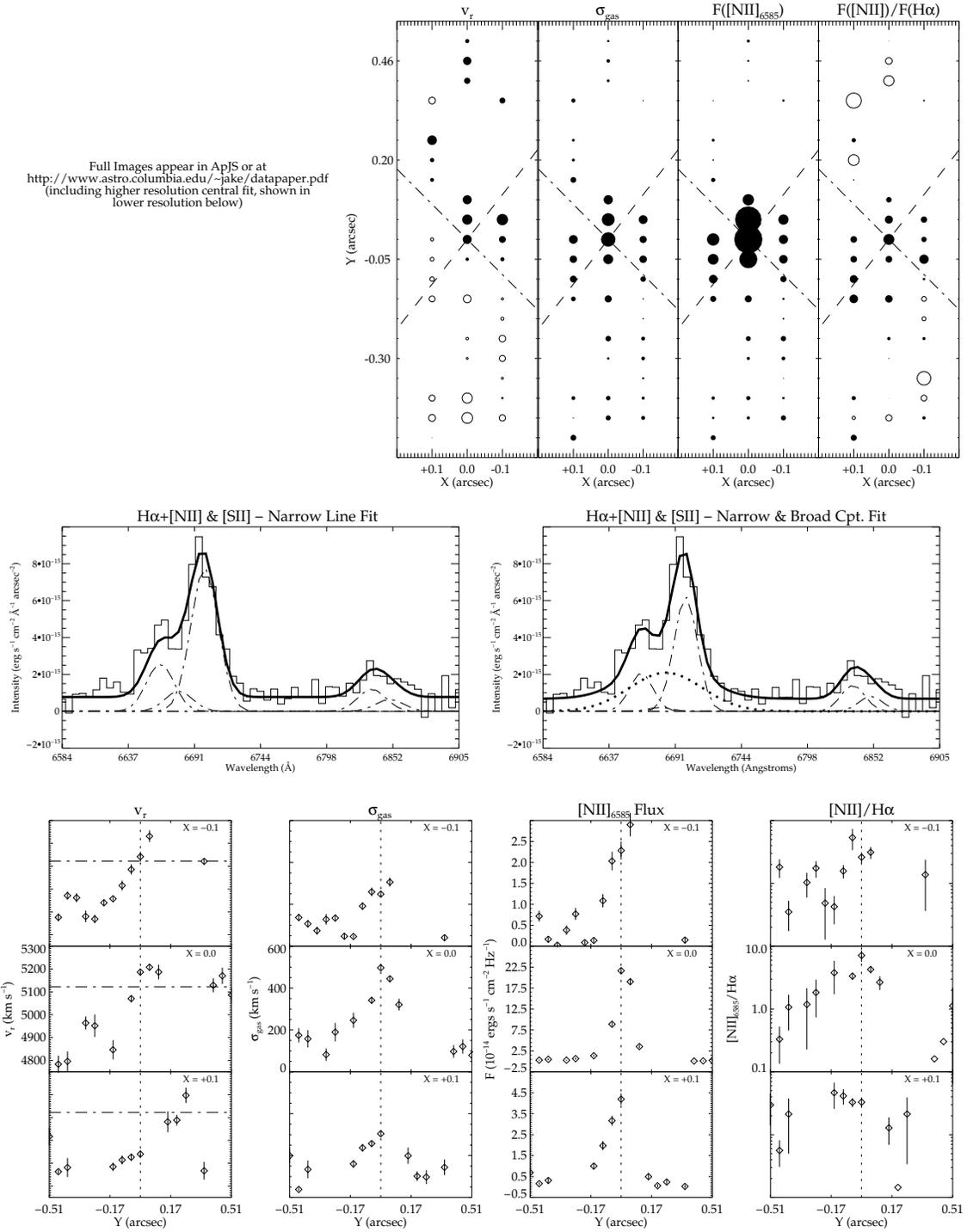}
\figcaption{Observation and fit data for UGC 12064, see Figure \ref{key} for description. \label{sgpi20}}
\end{figure}

\clearpage
\begin{figure}
\epsscale{0.9}
\plotone{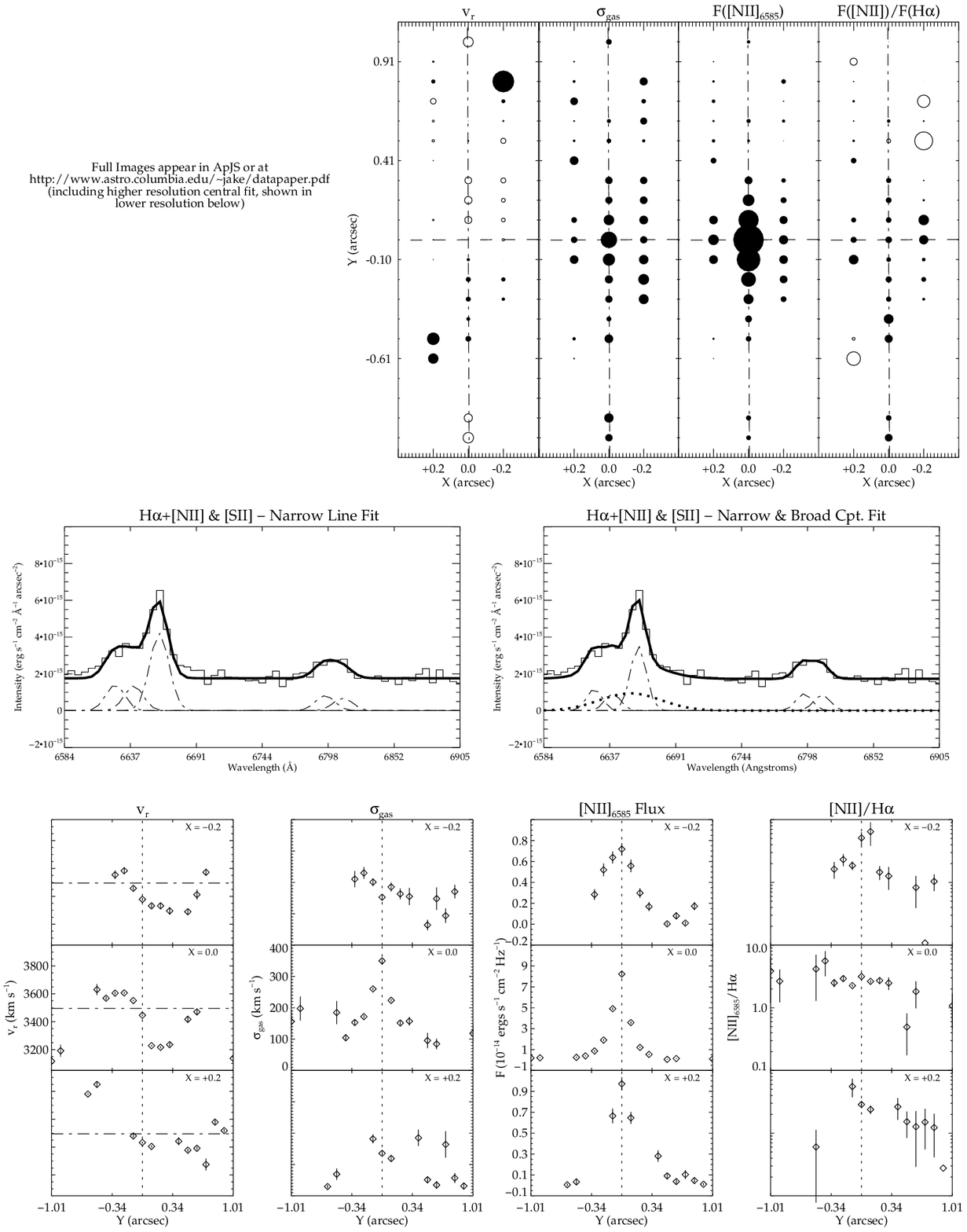}
\figcaption{Observation and fit data for NGC 7626, see Figure \ref{key} for description. \label{sgpi21}}
\end{figure}

\fi

\clearpage
\begin{figure}
\epsscale{1.0}
\plotone{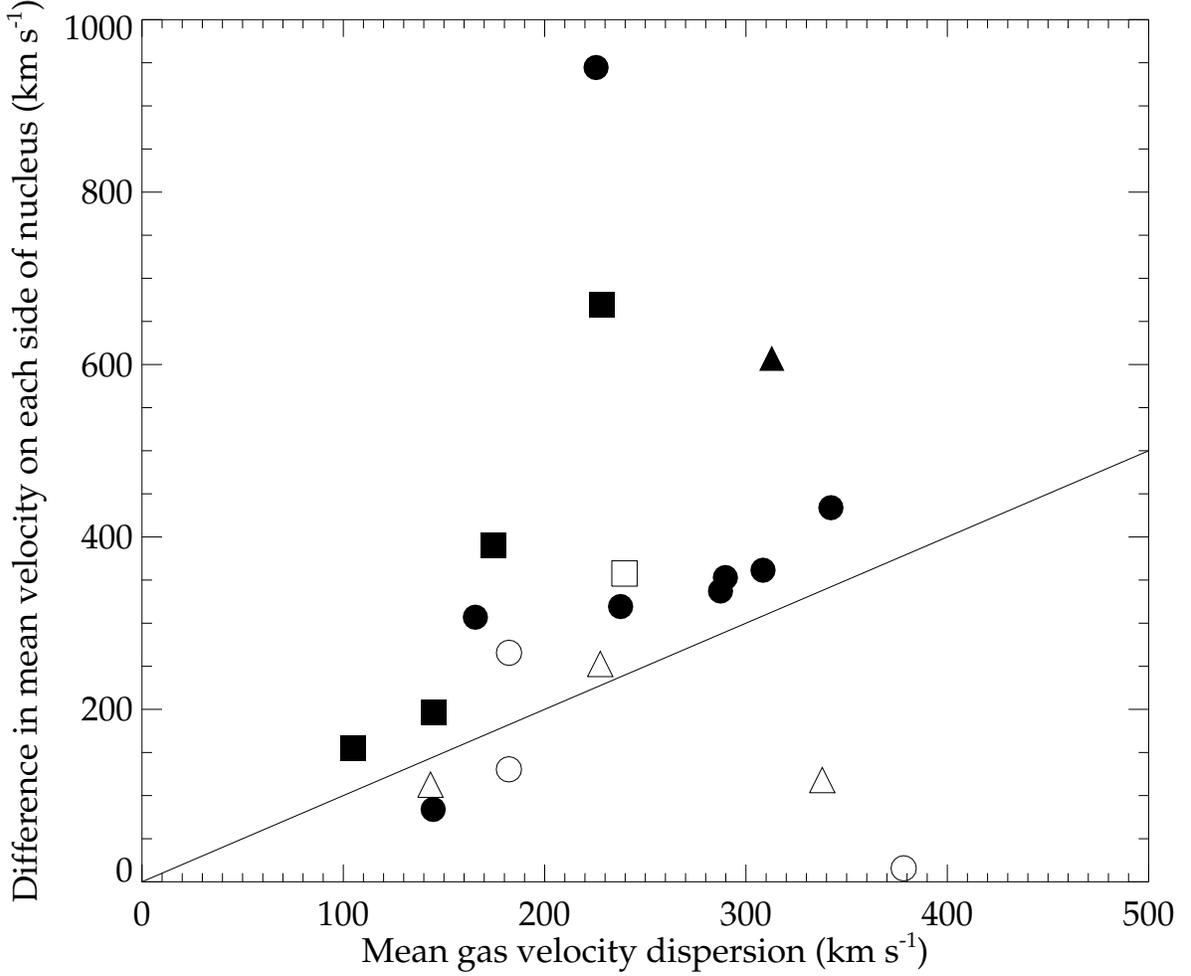}
\figcaption{Difference in mean velocity within 100~pc of each side of the nucleus ($\Delta_{100{\rm pc}}$, see text) as a function of the mean gas velocity dispersion within 100~pc of the nucleus ($\overline{\sigma_{100{\rm pc}}}$). The different kinematic classes and dust morphologies are indicated. Filled symbols represent rotators: with dust disks ({\huge $\bullet$}), with dust lanes ($\blacksquare$) or with irregular dust ($\blacktriangle$); Empty symbols represent non-rotators: with dust disks ({\huge $\circ$}), with dust lanes ($\square$) or with no-dust or irregular dust ($\triangle$). The solid line is the 1:1 ratio between $\Delta_{100{\rm pc}}$ and $\overline{\sigma_{100{\rm pc}}}$, indicating regions where organized motions (above the line) or random motions (below the line) dominate. \label{rotornot}}
\end{figure}

\clearpage
\begin{figure}
\epsscale{1.0}
\plotone{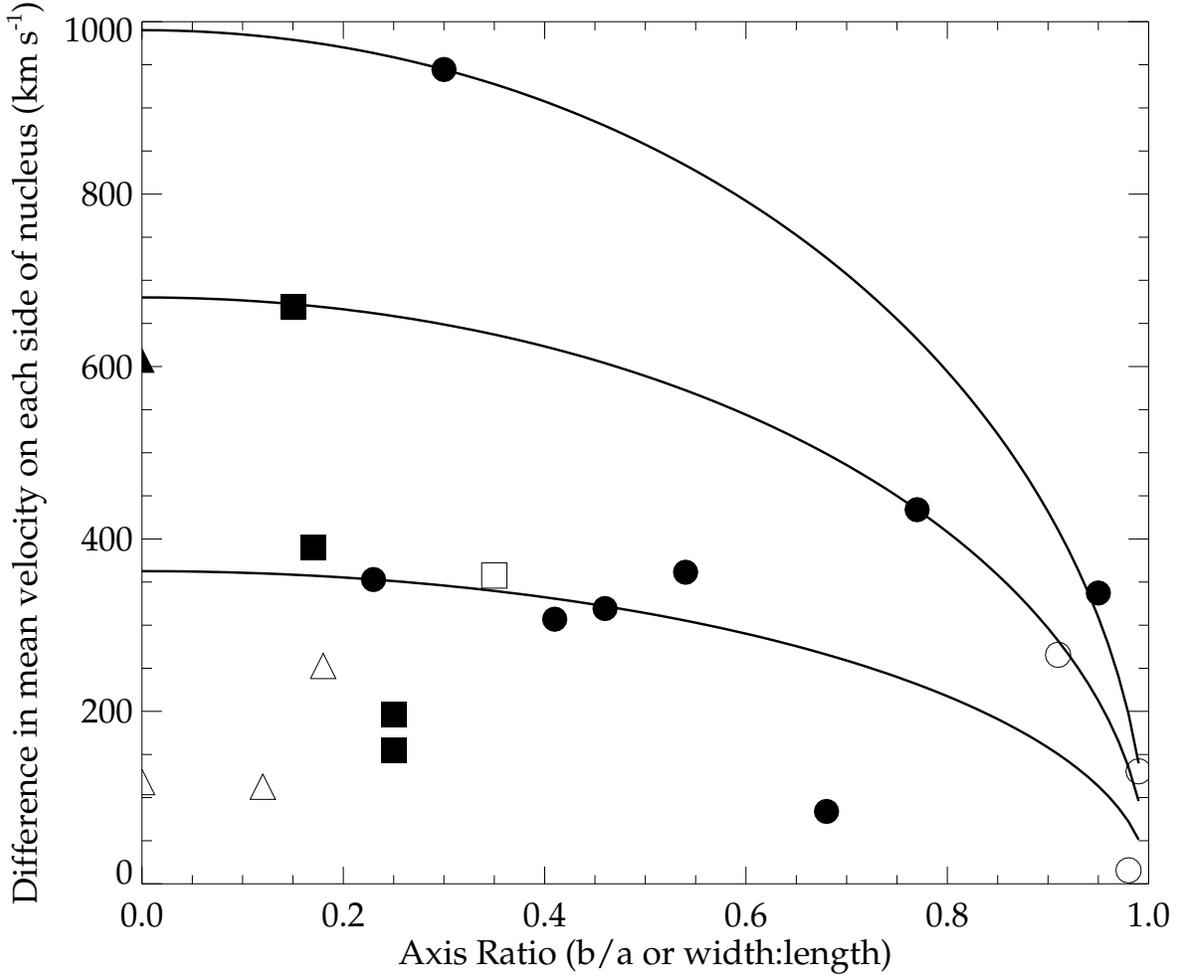}
\figcaption{Difference in mean velocity within 100~pc of each side of the nucleus ($\Delta_{100{\rm pc}}$, see text) as a function of dust disk axis ratio ($b/a$) or dust lane $width:length$ ratio, which is an indicator of inclination. The different kinematic classes and dust morphologies are indicated. Filled symbols represent rotators: with dust disks ({\huge $\bullet$}), with dust lanes ($\blacksquare$) or with irregular dust ($\blacktriangle$); Empty symbols represent non-rotators: with dust disks ({\huge $\circ$}), with dust lanes ($\square$) or with no-dust or irregular dust ($\triangle$). The solid lines are loci for disks with the same intrinsic $\Delta_{100{\rm pc}}$ (990, 680 and 360~${\rm km~s^{-1}}$) viewed at inclinations projected assuming $b/a = \sin i$ (i.e. a circular disk). \label{kpar}}
\end{figure}

\clearpage
\begin{figure}
\epsscale{1.0}
\plotone{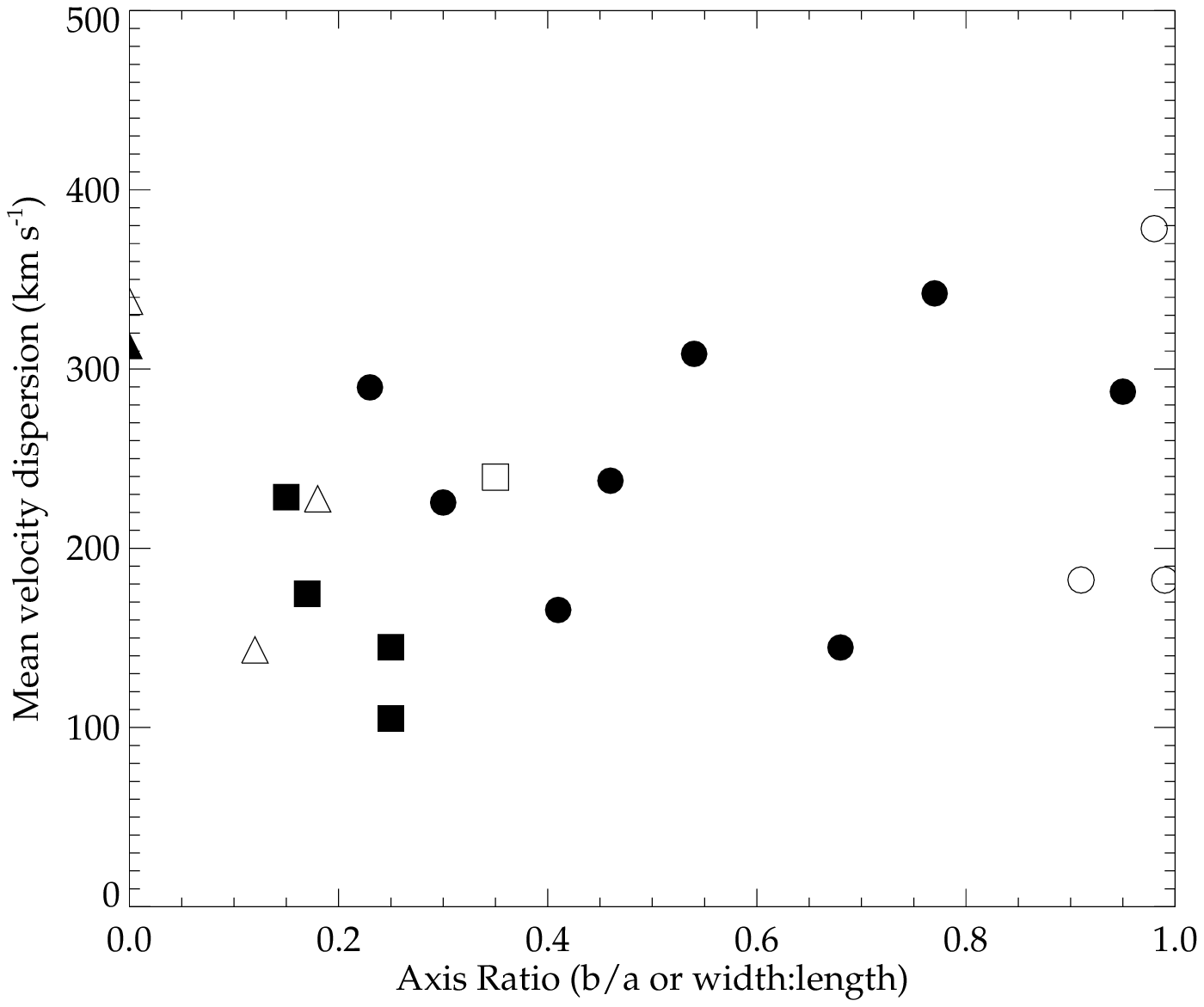}
\figcaption{Mean gas velocity dispersion within 100~pc of the nucleus ($\overline{\sigma_{100{\rm pc}}}$) as a function of dust disk axis ratio ($b/a$) or dust lane $width:length$ ratio, which is an indicator of inclination. The different kinematic classes and dust morphologies are indicated. Filled symbols represent rotators: with dust disks ({\huge $\bullet$}), with dust lanes ($\blacksquare$) or with irregular dust ($\blacktriangle$); Empty symbols represent non-rotators: with dust disks ({\huge $\circ$}), with dust lanes ($\square$) or with no-dust or irregular dust ($\triangle$). \label{kpar2}}
\end{figure}

\clearpage
\begin{figure}
\epsscale{1.0}
\plotone{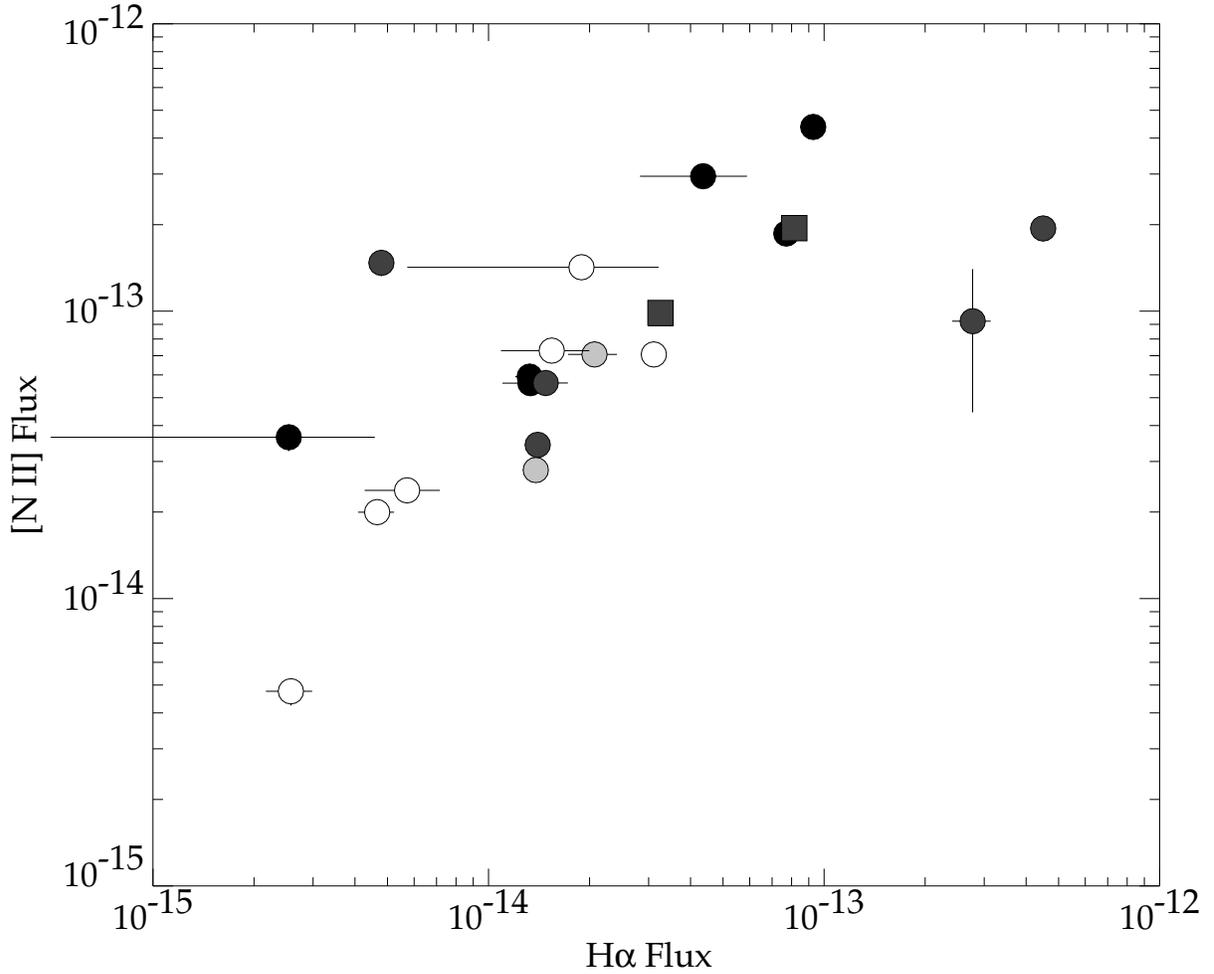}
\figcaption{[\ion{N}{2}] against H$\alpha$ fluxes for the UGC FR-I sample members. Formal errors in the fluxes are shown. The shading of the symbols indicates the score assigned to the broad component in each case (see text). A score of 3 (most confident) is shown by black symbols, through to white symbols for a score of zero (no confidence in a broad line). It is possible to see a trend with flux from less to more confidence in  broad-line detection.\label{flux_cfx}}
\end{figure}

\fi

%%%%%%%%%%%%%%%
% Tables
%%%%%%%%%%%%%%%

\clearpage
\ifsubmode\pagestyle{empty}\fi

\begin{deluxetable}{llllrrrrr}
\tablecolumns{9}
\tablewidth{0pc}
\tablecaption{Properties of the galaxy sample members.\label{sample}}
\tablehead{
\colhead{NGC} & \colhead{UGC} & \colhead{Other Names} & \colhead{Type} & \colhead{v$_{\rm sys}$} & \colhead{STIS Scale} & \colhead{$M_B$} & \colhead{${\rm log}(L_{1400})$} &\colhead{Axis Ratio}\\
\colhead{} &\colhead{} &\colhead{} &\colhead{} & \colhead{(km s$^{-1}$)} & \colhead{(${\rm pc / pixel}$)} & \colhead{(mag)} & \colhead{(W Hz$^{-1}$)} &\colhead{b/a}\\
\colhead{(1)}&\colhead{(2)}&\colhead{(3)}&\colhead{(4)}&\colhead{(5)}&\colhead{(6)}&\colhead{(7)}&\colhead{(8)}&\colhead{(9)}
}
\startdata
193 & 408 & & E-S0 & 4342.5 & 15.0 &-21.0&23.93&(0.18)\\
315 & 597 & & E & 5092.5 &17.6&-22.6&24.10&0.23\\
383 & 689 & 3C 31& E-S0 & 4890.0 &16.9&-22.2&24.51&0.77\\
541 & 1004 & & E & 5497.5 &19.0&-21.7&23.94&0.91\\
741 & 1413 & & E & 5265.0 &18.2&-22.6&23.85&\nodata\\
 & 1841 & 3C 66B & E & 6360.0 &22.0&-22.5&24.94&$\sim 0.98$\\
2329 & 3695 & & E-S0 & 5725.0 &19.8&-21.9&23.76&0.68\\
2892 & 5073 & & E & 6810.0 &23.6&-21.1&23.36&\nodata\\
\enddata
\tablecomments{Col. (1): NGC~number where available; Col. (2): {\em Upsalla General Catalog} (UGC) number; Col. (3): Alternative names; Col. (4): From the {\em NASA Extragalactic Database} (NED); Col. (5): Measured from the stellar kinematics (from NED); Col. (6): Parsecs per unbinned STIS pixel; Col. (7): Absolute blue magnitude from \citet{CB88}; Col. (8): Radio Luminosity from \citet{CB88}; Col. (9): Dust disk axis ratio from \citet{gijs99}, numbers in parentheses indicate a dust lane where the width:length ratio is given instead. All physical parameters assume ${\rm H_0 = 70~km s^{-1} Mpc^{-1}}$}
\end{deluxetable}

\clearpage

\begin{deluxetable}{llcccr}
\tablewidth{0pc}
\tablecaption{HST-STIS G750M observing log for this program.\label{obsets}}
\tablehead{\colhead{Galaxy} & \colhead{Visit ID} & \colhead{Aperture} & \colhead{Offsets} & \colhead{Exposure times}&\colhead{Brightest Row} \\
\colhead{}& \colhead{}&\colhead{} & \colhead{(\arcsec)} & \colhead{(seconds)}
\\ \colhead{(1)}&\colhead{(2)}&\colhead{(3)}&\colhead{(4)}&\colhead{(5)}&\colhead{(6)}}
\startdata
NGC~193 & {\tt o5ee01} & 52x0.2 & -0.20, 0.00, +0.20 & 1700, 1300, 1560&300 \\
NGC~315 &{\tt o5ee02} & 52x0.1 & -0.10, 0.00, +0.10 & 1082, 1000, 1098&598 \\
NGC~383 &{\tt o5ee03} & 52x0.1 & -0.10, 0.00, +0.10 & 1082, 1000, 1098&599 \\
NGC~541 &{\tt o5ee04} & 52x0.2 & -0.20, 0.00, +0.20 & 1442, 1300, 1500&300 \\
NGC~741 &{\tt o5ee05} & 52x0.2 & -0.20, 0.00, +0.20 & 1667, 1145, 1500&300 \\
UGC~1841 &{\tt o5ee06} & 52x0.2 & -0.20, 0.00, +0.20 & 1870, 1300, 1560&300 \\
NGC~2329 &{\tt o5ee07} & 52x0.2 & -0.20, 0.00, +0.20 & 1967, 1245, 1560&300 \\
NGC~2892 &{\tt o5ee09} & 52x0.2 & -0.20, 0.00, +0.20 & 2037, 1300, 1560&299 \\
\enddata
\tablecomments{Col. (1): NGC/UGC~Number; Col. (2): Root name for the HST visit and data sets; Col. (3): HST/STIS aperture used; Col. (4): Slit positions with offsets perpendicular to the slit direction, given relative to the central slit position. UGC~7115 was observed in only one position. Further data not included here is available for NGC~4374 (M84) and NGC~4486 (M87); Col. (5): Exposure times in each offset position. For each position two observations were obtained separated by 0\farcs2028 along the direction of the slit, the combined exposure time is shown; Col. (6): Row number in the final {\tt x2d} file of the brightest row.}
\end{deluxetable}

\clearpage

\begin{deluxetable}{llcccccc}
\rotate
\tablewidth{0pc}
\tablecaption{HST/STIS Instrumental properties for the configurations used. \label{stisets}}
\tablehead{
\colhead{Grating} & \colhead{Aperture}& \colhead{Slit Width} & \colhead{Binning} & \colhead{Central $\lambda$}  & \colhead{Wavelength Covered} & \colhead{$\lambda$ Scale} & \colhead{Spatial Scale} \\
\colhead{} &\colhead{} & \colhead{(\arcsec)} &\colhead{}& \colhead{(\AA)}   & \colhead{(\AA\ ... \AA )} & \colhead{(\AA / pixel)} & \colhead{(\arcsec / pixel)}
\\ \colhead{(1)}&\colhead{(2)}&\colhead{(3)}&\colhead{(4)}&\colhead{(5)}&\colhead{(6)}&\colhead{(7)}&\colhead{(8)}
}
\startdata
G750M & 52x0.1 & 0.1 & $2 \times 1$ & 6768 & 6482 ... 7054 & 1.108 & 0.05 \\
G750M & 52x0.2 & 0.2 & $2 \times 2$ & 6768 & 6482 ... 7054 & 1.108 & 0.1\phn \\
G750M & 52x0.2 (M84) & 0.2 & $1 \times 1$ & 6581 & 6295 ... 6867 & 0.554 & 0.05\\
G750M & 52x0.2 (M87) & 0.2 & $1 \times 1$ & 6581 & 6295 ... 6867 & 0.554 & 0.05\\
\enddata
\tablecomments{Col. (1): HST/STIS Grating Used; Col. (2): HST/STIS Aperture Used; Col. (3): The slit width of the chosen aperture; Col. (4): The degree of on chip binning used given as {\tt binaxis1}$ \times\ ${\tt binaxis2} (wavelength by spatial binning); Col. (5): The set central wavelength; Col. (6): The minimum and maximum wavelengths on the CCD chip at that central wavelength position; Col. (7) The wavelength scale per binned pixel; Col. (8) The spatial scale per binned pixel. Further details of HST/STIS configurations can be found in Kimble, et al. (1998).}
\end{deluxetable}

\clearpage
\begin{deluxetable}{lrrrrcr}
\tablewidth{0pc}
\tablecaption{Position angles of various axes. \label{pas}}
\tablehead{
\colhead{Galaxy} & \colhead{Major Axis}& \colhead{Radio Axis}&\colhead{Dust Axis}&\colhead{STIS Slits}&\colhead{Target}&\colhead{$\Delta PA$}\\
\colhead{}&\colhead{($^\circ N\hookrightarrow E$)} &\colhead{($^\circ N\hookrightarrow E$)} & \colhead{($^\circ N\hookrightarrow E$)}&\colhead{($^\circ N\hookrightarrow E$)}&\colhead{}&\colhead{($^\circ$)}\\
\colhead{(1)}&\colhead{(2)}&\colhead{(3)}&\colhead{(4)}&\colhead{(5)}&\colhead{(6)}&\colhead{(7)}
}
\startdata
NGC~193 & 58 & 104 & 0 & 57.9&$\parallel$ M.Ax.&0.1\\
NGC~315 & 39 & 130 & 40 &  -143.8&$\parallel$ M.Ax.&2.8\\
NGC~383 & 144 & 162 & 138 &  144.1&$\parallel$ M.Ax.&0.1\\
NGC~541 & 143 & 76 & \nodata &  -85.6&\nodata&\nodata\\
NGC~741 & 92 & \nodata  & \nodata & -107.4&$\parallel$ M.Ax.$^\dagger $&19.4\\
UGC~1841 & 91 & 50 & \nodata & 90.0 &$\parallel$ M.Ax.&1.0\\
NGC~2329 & 171 & 150 & 174  & 170.8&$\parallel$ M.Ax.&0.2\\
NGC~2892 & 164 & 52 & \nodata & 142.0&$\parallel$ M.Ax.$^\dagger$&22.0\\
\enddata
\tablecomments{Col. (1): NGC/UGC~Name; Col. (2): Position angles of the major axes just outside the central dust distributions; Col. (3): Position angles of the radio jet axes on arcsecond scales or smaller; Col. (4): Position angles of the central dust distributions; Col. (5): Position angles of the STIS slits on the nucleus; Col. (6): The target orientation of the STIS slits ($\parallel$ M.Ax.: Parallel to the Major Axis; $\parallel$ Dust: Parallel to the Dust Axis; $\perp$ Jet: Perpendicular to the radio jet; $^\dagger$ Additional flexibility allowed to reduce scheduling constraints); Col. (7): Offset in degrees of the STIS slits from the target orientation.}
\tablerefs{Data for Columns 3 - 5 taken from Verdoes Kleijn, et al. (1999)}
\end{deluxetable}
\clearpage

\begin{deluxetable}{lccrr}
\tablewidth{0pc}
\tablecaption{Spectral Lines in the region of H$\alpha$. \label{medlinelist}}
\tablehead{
\colhead{Ion} & \colhead{Ref.} & \colhead{Transition} & \colhead{$\lambda_{\rm air}$} &\colhead{$\lambda_{\rm vac}$} \\
 \colhead{} & \colhead{} & \colhead{} & \colhead{(\AA )} & \colhead{(\AA )} \\
\colhead{(1)}&\colhead{(2)}&\colhead{(3)}&\colhead{(4)}&\colhead{(5)}}
\startdata
{[\ion{N}{2}]}& (a) &{${\rm ^{3}P_{1} - ^{1}D_{2}}$} & 6548.05 & 6549.86 \\
H$\alpha$ &(b)&& 6562.80 & 6564.61\\
{[\ion{N}{2}]}&(a)& {${\rm ^{3}P_{2} - ^{1}D_{2}}$} & 6583.39 & 6585.21 \\
{[\ion{S}{2}]}&(a)& {${\rm ^{4}S_{3/2} - ^{2}D_{5/2}}$} & 6716.44 & 6718.29 \\
{[\ion{S}{2}]}&(a)& {${\rm ^{4}S_{3/2} - ^{2}D_{3/2}}$} & 6730.81 & 6732.67 \\
\enddata
\tablecomments{Col. (1): Ion responsible for the line; Col. (2): Wavelength reference; Col. (3): Transition configuration terms; Col. (4): Air wavelength; Col. (5): Vacuum wavelength. Note that the value for H$\alpha$ is computed from the intensity weighted mean of fine structure lines at $\lambda_{air}=6562.7247$ and $\lambda_{air}=6562.8516$ with relative intensities of 120 and 180 respectively.}
\tablerefs{(a) Wallerstein, et al. (2001); (b) Reader \& Corliss (1998)}
\end{deluxetable}

\clearpage\begin{deluxetable}{cccccccc}
\rotate
\tablecolumns{8}
\tablewidth{0pc}
\tablecaption{NGC 193: Measured Parameters. \label{ngc193_p1}}
\tablehead{\colhead{Row} & \colhead{Y-Offset} & \colhead{$v_{\rm r}$} & \colhead{$\sigma_{gas}$} & \colhead{F$(\rm{H}\alpha / 10^{-16}$)} & \colhead{\underline{F(\ion{N}{2}$_{6585}$)}} & \colhead{\underline{F(\ion{S}{2}$_{total}$)}}&\colhead{$R^2$}
\\ \colhead{} & \colhead{(\arcsec)} & \colhead{(\kms)} & \colhead{(\kms)} &  \colhead{\small{(${\rm erg~s^{-1} cm^{-2} Hz^{-1}}$)}} & \colhead{F(H$\alpha$)} & \colhead{F(H$\alpha$)} &\colhead{}}
\startdata
\cutinhead{Slit 1: X-Offset -0.2\arcsec}
 2&-0.9128&$ 4334 \pm\    30$&$  135 \pm\    30$&$   6.6 \pm\    2.9$&$ 1.7 \pm\  0.9$&$ 1.2 \pm\  0.7$&3.79\\
 3&-0.8114&$ 4270 \pm\    30$&$  121 \pm\    32$&$   6.9 \pm\    3.3$&$ 1.5 \pm\  1.0$&$ 0.2 \pm\  0.3$&3.17\\
 4&-0.7099&$ 4323 \pm\    22$&$  124 \pm\    23$&$  10.6 \pm\    4.0$&$ 1.5 \pm\  0.7$&$ 0.1 \pm\  0.2$&3.90\\
 5&-0.6085&$ 4390 \pm\     7$&$   40 \pm\     7$&$   7.8 \pm\    2.3$&$ 1.4 \pm\  0.6$&$ 0.1 \pm\  0.2$&3.23\\
 6&-0.5071&$ 4602 \pm\    42$&$  233 \pm\    41$&$  12.9 \pm\    4.5$&$ 0.0 \pm\  0.3$&$ 1.8 \pm\  0.8$&2.83\\
 7&-0.4057&$ 4371 \pm\    10$&$   67 \pm\    10$&$   7.4 \pm\    2.3$&$ 1.5 \pm\  0.6$&$ 1.2 \pm\  0.5$&3.67\\
 8&-0.3043&$ 4402 \pm\     9$&$   48 \pm\     9$&$   6.2 \pm\    2.2$&$ 1.1 \pm\  0.5$&$ 0.9 \pm\  0.5$&3.46\\
 9&-0.2028&$ 4369 \pm\    23$&$  210 \pm\    18$&$  24.7 \pm\    4.4$&$ 1.2 \pm\  0.3$&$ 1.3 \pm\  0.3$&3.28\\
\enddata
\end{deluxetable}

\iffullout
\clearpage\begin{deluxetable}{cccccccc}
\rotate
\tablecolumns{8}
\tablewidth{0pc}
\tablecaption{NGC 315: Measured Parameters. \label{ngc315_p1}}
\tablehead{\colhead{Row} & \colhead{Y-Offset} & \colhead{$v_{\rm r}$} & \colhead{$\sigma_{gas}$} & \colhead{F$(\rm{H}\alpha / 10^{-16}$)} & \colhead{\underline{F(\ion{N}{2}$_{6585}$)}} & \colhead{\underline{F(\ion{S}{2}$_{total}$)}}&\colhead{$R^2$}
\\ \colhead{} & \colhead{(\arcsec)} & \colhead{(\kms)} & \colhead{(\kms)} &  \colhead{\small{(${\rm erg~s^{-1} cm^{-2} Hz^{-1}}$)}} & \colhead{F(H$\alpha$)} & \colhead{F(H$\alpha$)} &\colhead{}}
\startdata
\cutinhead{Slit 1: X-Offset -0.1\arcsec}
 1&-0.5071&$ 4827 \pm\    22$&$   70 \pm\    21$&$  39.1 \pm\   17.8$&$ 0.0 \pm\  0.3$&$ 0.5 \pm\  0.4$&1.67\\
 7&-0.2028&$ 4884 \pm\    34$&$  161 \pm\    31$&$  52.2 \pm\   19.8$&$ 1.2 \pm\  0.6$&$ 1.2 \pm\  0.6$&1.90\\
 8&-0.1521&$ 4943 \pm\    33$&$  175 \pm\    29$&$  33.0 \pm\   17.3$&$ 3.0 \pm\  1.7$&$ 2.7 \pm\  1.6$&1.93\\
 9&-0.1014&$ 4866 \pm\    32$&$  193 \pm\    27$&$  62.0 \pm\   19.8$&$ 2.4 \pm\  0.9$&$ 1.0 \pm\  0.4$&1.65\\
10&-0.0507&$ 4905 \pm\    37$&$  272 \pm\    32$&$  88.4 \pm\   25.3$&$ 2.4 \pm\  0.8$&$ 0.9 \pm\  0.4$&1.84\\
11& 0.0000&$ 4917 \pm\    28$&$  302 \pm\    24$&$ 151.2 \pm\   29.4$&$ 2.5 \pm\  0.6$&$ 1.3 \pm\  0.3$&1.83\\
12& 0.0507&$ 4814 \pm\    48$&$  485 \pm\    44$&$  60.7 \pm\   45.0$&$ 8.3 \pm\  6.3$&$ 3.0 \pm\  2.3$&1.90\\
13& 0.1014&$ 4882 \pm\    36$&$  337 \pm\    35$&$  61.7 \pm\   28.1$&$ 5.4 \pm\  2.6$&$ 2.2 \pm\  1.1$&1.88\\
\enddata
\end{deluxetable}

\clearpage\begin{deluxetable}{cccccccc}
\rotate
\tablecolumns{8}
\tablewidth{0pc}
\tablecaption{NGC 383: Measured Parameters. \label{ngc383_p1}}
\tablehead{\colhead{Row} & \colhead{Y-Offset} & \colhead{$v_{\rm r}$} & \colhead{$\sigma_{gas}$} & \colhead{F$(\rm{H}\alpha / 10^{-16}$)} & \colhead{\underline{F(\ion{N}{2}$_{6585}$)}} & \colhead{\underline{F(\ion{S}{2}$_{total}$)}}&\colhead{$R^2$}
\\ \colhead{} & \colhead{(\arcsec)} & \colhead{(\kms)} & \colhead{(\kms)} &  \colhead{\small{(${\rm erg~s^{-1} cm^{-2} Hz^{-1}}$)}} & \colhead{F(H$\alpha$)} & \colhead{F(H$\alpha$)} &\colhead{}}
\startdata
\cutinhead{Slit 1: X-Offset -0.1\arcsec}
 2&-0.4564&$ 4953 \pm\     7$&$   37 \pm\     7$&$   9.5 \pm\    5.5$&$ 3.4 \pm\  2.2$&$ 3.6 \pm\  2.4$&1.40\\
 3&-0.4057&$ 4880 \pm\    21$&$  105 \pm\    21$&$  21.6 \pm\   10.0$&$ 2.4 \pm\  1.3$&$ 1.6 \pm\  1.1$&1.58\\
 4&-0.3550&$ 4923 \pm\    41$&$  184 \pm\    38$&$  29.1 \pm\   13.5$&$ 2.4 \pm\  1.3$&$ 0.1 \pm\  0.7$&1.62\\
 5&-0.3043&$ 4915 \pm\    30$&$  163 \pm\    28$&$  42.5 \pm\   14.2$&$ 1.9 \pm\  0.8$&$ 0.1 \pm\  0.4$&1.86\\
 6&-0.2535&$ 4952 \pm\    24$&$  201 \pm\    20$&$  95.3 \pm\   17.3$&$ 1.3 \pm\  0.3$&$ 0.6 \pm\  0.3$&1.91\\
 7&-0.2028&$ 4945 \pm\    18$&$  205 \pm\    15$&$ 116.8 \pm\   16.9$&$ 1.5 \pm\  0.3$&$ 1.1 \pm\  0.3$&1.48\\
 8&-0.1521&$ 4957 \pm\    13$&$  173 \pm\    11$&$ 112.3 \pm\   14.9$&$ 1.7 \pm\  0.3$&$ 1.0 \pm\  0.2$&1.68\\
 9&-0.1014&$ 5046 \pm\    16$&$  248 \pm\    12$&$ 156.8 \pm\   18.2$&$ 2.1 \pm\  0.3$&$ 0.9 \pm\  0.2$&1.64\\
\enddata
\end{deluxetable}

\clearpage\begin{deluxetable}{cccccccc}
\rotate
\tablecolumns{8}
\tablewidth{0pc}
\tablecaption{NGC 541: Measured Parameters. \label{ngc541_p1}}
\tablehead{\colhead{Row} & \colhead{Y-Offset} & \colhead{$v_{\rm r}$} & \colhead{$\sigma_{gas}$} & \colhead{F$(\rm{H}\alpha / 10^{-16}$)} & \colhead{\underline{F(\ion{N}{2}$_{6585}$)}} & \colhead{\underline{F(\ion{S}{2}$_{total}$)}}&\colhead{$R^2$}
\\ \colhead{} & \colhead{(\arcsec)} & \colhead{(\kms)} & \colhead{(\kms)} &  \colhead{\small{(${\rm erg~s^{-1} cm^{-2} Hz^{-1}}$)}} & \colhead{F(H$\alpha$)} & \colhead{F(H$\alpha$)} &\colhead{}}
\startdata
\cutinhead{Slit 1: X-Offset -0.2\arcsec}
 2&-0.9128&$ 5600 \pm\    41$&$  188 \pm\    35$&$  11.4 \pm\    4.5$&$ 1.6 \pm\  0.8$&$ 1.1 \pm\  0.6$&2.73\\
 5&-0.6085&$ 5516 \pm\     4$&$   51 \pm\     4$&$  29.4 \pm\    4.0$&$ 0.4 \pm\  0.1$&$ 0.2 \pm\  0.1$&3.11\\
 6&-0.5071&$ 5528 \pm\     6$&$   60 \pm\     6$&$  26.7 \pm\    4.1$&$ 0.5 \pm\  0.2$&$ 0.2 \pm\  0.1$&3.27\\
 7&-0.4057&$ 5535 \pm\     7$&$   54 \pm\     6$&$  21.5 \pm\    3.9$&$ 0.7 \pm\  0.2$&$ 0.4 \pm\  0.1$&2.86\\
 8&-0.3043&$ 5195 \pm\     7$&$   43 \pm\     7$&$   3.2 \pm\    1.8$&$ 4.5 \pm\  2.8$&$ 1.9 \pm\  1.3$&2.36\\
 9&-0.2028&$ 5220 \pm\    23$&$  142 \pm\    22$&$   4.7 \pm\    3.2$&$ 5.6 \pm\  4.1$&$ 4.3 \pm\  3.1$&2.66\\
10&-0.1014&$ 5297 \pm\    19$&$  140 \pm\    18$&$  11.6 \pm\    4.1$&$ 2.8 \pm\  1.1$&$ 1.2 \pm\  0.6$&2.34\\
11& 0.0000&$ 5389 \pm\    15$&$  168 \pm\    14$&$  17.6 \pm\    4.3$&$ 4.1 \pm\  1.1$&$ 0.0 \pm\  0.2$&2.61\\
\enddata
\end{deluxetable}

\clearpage\begin{deluxetable}{cccccccc}
\rotate
\tablecolumns{8}
\tablewidth{0pc}
\tablecaption{NGC 741: Measured Parameters. \label{ngc741_p1}}
\tablehead{\colhead{Row} & \colhead{Y-Offset} & \colhead{$v_{\rm r}$} & \colhead{$\sigma_{gas}$} & \colhead{F$(\rm{H}\alpha / 10^{-16}$)} & \colhead{\underline{F(\ion{N}{2}$_{6585}$)}} & \colhead{\underline{F(\ion{S}{2}$_{total}$)}}&\colhead{$R^2$}
\\ \colhead{} & \colhead{(\arcsec)} & \colhead{(\kms)} & \colhead{(\kms)} &  \colhead{\small{(${\rm erg~s^{-1} cm^{-2} Hz^{-1}}$)}} & \colhead{F(H$\alpha$)} & \colhead{F(H$\alpha$)} &\colhead{}}
\startdata
\cutinhead{Slit 1: X-Offset -0.2\arcsec}
 2&-0.9128&$ 5337 \pm\    33$&$   86 \pm\    31$&$   5.6 \pm\    3.5$&$ 0.0 \pm\  0.3$&$ 1.2 \pm\  0.9$&2.41\\
 3&-0.8114&$ 5309 \pm\    28$&$  125 \pm\    28$&$  15.0 \pm\    5.3$&$ 0.1 \pm\  0.2$&$ 0.6 \pm\  0.3$&2.69\\
 6&-0.5071&$ 5303 \pm\    32$&$  190 \pm\    29$&$   7.1 \pm\    4.2$&$ 4.1 \pm\  2.6$&$ 1.3 \pm\  1.0$&3.00\\
 7&-0.4057&$ 5465 \pm\    23$&$  100 \pm\    22$&$   7.5 \pm\    3.8$&$ 1.8 \pm\  1.1$&$ 0.8 \pm\  0.6$&3.08\\
12& 0.1014&$ 5686 \pm\    20$&$  138 \pm\    19$&$  11.9 \pm\    3.9$&$ 1.7 \pm\  0.7$&$ 1.9 \pm\  0.7$&2.83\\
13& 0.2028&$ 5616 \pm\    41$&$  192 \pm\    37$&$   9.5 \pm\    4.7$&$ 2.4 \pm\  1.4$&$ 0.6 \pm\  0.5$&3.50\\
18& 0.7099&$ 5252 \pm\    42$&$  181 \pm\    41$&$   9.6 \pm\    4.2$&$ 0.1 \pm\  0.3$&$ 1.5 \pm\  0.8$&3.33\\
19& 0.8114&$ 5317 \pm\    10$&$   37 \pm\    10$&$  10.5 \pm\    3.9$&$ 0.3 \pm\  0.2$&$ 0.0 \pm\  0.1$&2.94\\
\enddata
\end{deluxetable}

\clearpage\begin{deluxetable}{cccccccc}
\rotate
\tablecolumns{8}
\tablewidth{0pc}
\tablecaption{UGC 01841: Measured Parameters. \label{ugc01841_p1}}
\tablehead{\colhead{Row} & \colhead{Y-Offset} & \colhead{$v_{\rm r}$} & \colhead{$\sigma_{gas}$} & \colhead{F$(\rm{H}\alpha / 10^{-16}$)} & \colhead{\underline{F(\ion{N}{2}$_{6585}$)}} & \colhead{\underline{F(\ion{S}{2}$_{total}$)}}&\colhead{$R^2$}
\\ \colhead{} & \colhead{(\arcsec)} & \colhead{(\kms)} & \colhead{(\kms)} &  \colhead{\small{(${\rm erg~s^{-1} cm^{-2} Hz^{-1}}$)}} & \colhead{F(H$\alpha$)} & \colhead{F(H$\alpha$)} &\colhead{}}
\startdata
\cutinhead{Slit 1: X-Offset -0.2\arcsec}
 1&-1.0142&$ 6626 \pm\    13$&$   87 \pm\    13$&$  10.6 \pm\    2.6$&$ 1.0 \pm\  0.3$&$ 0.4 \pm\  0.2$&4.21\\
 2&-0.9128&$ 6276 \pm\    27$&$  172 \pm\    25$&$   5.5 \pm\    2.4$&$ 3.5 \pm\  1.7$&$ 1.3 \pm\  0.9$&4.16\\
 5&-0.6085&$ 6444 \pm\     7$&$   60 \pm\     7$&$  13.1 \pm\    2.6$&$ 0.8 \pm\  0.2$&$ 0.3 \pm\  0.2$&4.90\\
 6&-0.5071&$ 6258 \pm\    47$&$  289 \pm\    35$&$  22.2 \pm\    4.6$&$ 1.0 \pm\  0.3$&$ 1.1 \pm\  0.3$&4.45\\
 7&-0.4057&$ 6416 \pm\     9$&$  115 \pm\     9$&$   6.9 \pm\    2.1$&$ 4.6 \pm\  1.5$&$ 1.5 \pm\  0.6$&4.76\\
 8&-0.3043&$ 6411 \pm\     7$&$  137 \pm\     7$&$   9.3 \pm\    2.2$&$ 5.8 \pm\  1.5$&$ 2.9 \pm\  0.8$&4.34\\
 9&-0.2028&$ 6379 \pm\    13$&$  234 \pm\    10$&$  48.3 \pm\    4.0$&$ 1.5 \pm\  0.2$&$ 0.5 \pm\  0.1$&3.53\\
10&-0.1014&$ 6155 \pm\    29$&$  574 \pm\    25$&$  28.0 \pm\   11.2$&$ 7.2 \pm\  2.9$&$ 3.0 \pm\  1.3$&4.58\\
\enddata
\end{deluxetable}

\clearpage\begin{deluxetable}{cccccccc}
\rotate
\tablecolumns{8}
\tablewidth{0pc}
\tablecaption{NGC 2329: Measured Parameters. \label{ngc2329_p1}}
\tablehead{\colhead{Row} & \colhead{Y-Offset} & \colhead{$v_{\rm r}$} & \colhead{$\sigma_{gas}$} & \colhead{F$(\rm{H}\alpha / 10^{-16}$)} & \colhead{\underline{F(\ion{N}{2}$_{6585}$)}} & \colhead{\underline{F(\ion{S}{2}$_{total}$)}}&\colhead{$R^2$}
\\ \colhead{} & \colhead{(\arcsec)} & \colhead{(\kms)} & \colhead{(\kms)} &  \colhead{\small{(${\rm erg~s^{-1} cm^{-2} Hz^{-1}}$)}} & \colhead{F(H$\alpha$)} & \colhead{F(H$\alpha$)} &\colhead{}}
\startdata
\cutinhead{Slit 1: X-Offset -0.2\arcsec}
 2&-0.9128&$ 6017 \pm\     7$&$   52 \pm\     7$&$  13.0 \pm\    2.6$&$ 0.4 \pm\  0.1$&$ 0.3 \pm\  0.2$&3.27\\
 3&-0.8114&$ 5709 \pm\    49$&$  164 \pm\    48$&$   4.6 \pm\    2.7$&$ 1.7 \pm\  1.2$&$ 1.7 \pm\  1.4$&2.73\\
 4&-0.7099&$ 5752 \pm\    52$&$  284 \pm\    40$&$  28.8 \pm\    5.7$&$ 0.4 \pm\  0.2$&$ 0.2 \pm\  0.2$&3.24\\
 5&-0.6085&$ 5973 \pm\    10$&$   67 \pm\    11$&$   8.9 \pm\    2.3$&$ 1.2 \pm\  0.4$&$ 0.5 \pm\  0.3$&3.19\\
 6&-0.5071&$ 5989 \pm\    13$&$   59 \pm\    12$&$   8.8 \pm\    3.1$&$ 0.5 \pm\  0.3$&$ 0.8 \pm\  0.4$&2.98\\
 7&-0.4057&$ 5493 \pm\    26$&$  112 \pm\    26$&$   7.4 \pm\    2.7$&$ 1.2 \pm\  0.6$&$ 0.6 \pm\  0.5$&3.33\\
 8&-0.3043&$ 5816 \pm\    39$&$  223 \pm\    30$&$  12.7 \pm\    3.8$&$ 2.1 \pm\  0.7$&$ 0.2 \pm\  0.4$&3.04\\
 9&-0.2028&$ 5810 \pm\    20$&$  154 \pm\    20$&$   9.5 \pm\    3.2$&$ 3.3 \pm\  1.3$&$ 0.2 \pm\  0.5$&2.53\\
\enddata
\end{deluxetable}

\clearpage\begin{deluxetable}{cccccccc}
\rotate
\tablecolumns{8}
\tablewidth{0pc}
\tablecaption{NGC 2892: Measured Parameters. \label{ngc2892_p1}}
\tablehead{\colhead{Row} & \colhead{Y-Offset} & \colhead{$v_{\rm r}$} & \colhead{$\sigma_{gas}$} & \colhead{F$(\rm{H}\alpha / 10^{-16}$)} & \colhead{\underline{F(\ion{N}{2}$_{6585}$)}} & \colhead{\underline{F(\ion{S}{2}$_{total}$)}}&\colhead{$R^2$}
\\ \colhead{} & \colhead{(\arcsec)} & \colhead{(\kms)} & \colhead{(\kms)} &  \colhead{\small{(${\rm erg~s^{-1} cm^{-2} Hz^{-1}}$)}} & \colhead{F(H$\alpha$)} & \colhead{F(H$\alpha$)} &\colhead{}}
\startdata
\cutinhead{Slit 1: X-Offset -0.2\arcsec}
 1&-1.0142&$ 6843 \pm\     8$&$   25 \pm\     6$&$   1.4 \pm\    0.9$&$ 1.1 \pm\  0.9$&$ 4.9 \pm\  3.6$&3.48\\
 3&-0.8114&$ 6649 \pm\    13$&$   76 \pm\    12$&$  15.2 \pm\    3.6$&$ 0.1 \pm\  0.1$&$ 0.4 \pm\  0.2$&3.31\\
 4&-0.7099&$ 6895 \pm\    22$&$  143 \pm\    22$&$   7.9 \pm\    2.5$&$ 1.8 \pm\  0.7$&$ 1.2 \pm\  0.7$&3.43\\
 6&-0.5071&$ 6700 \pm\    20$&$   71 \pm\    20$&$   3.1 \pm\    1.9$&$ 2.0 \pm\  1.5$&$ 0.6 \pm\  1.0$&3.25\\
 8&-0.3043&$ 6905 \pm\    14$&$   81 \pm\    13$&$  10.1 \pm\    2.7$&$ 0.9 \pm\  0.3$&$ 0.3 \pm\  0.3$&3.19\\
 9&-0.2028&$ 6912 \pm\    23$&$   71 \pm\    23$&$   5.7 \pm\    2.7$&$ 0.5 \pm\  0.4$&$ 1.0 \pm\  0.7$&3.18\\
11& 0.0000&$ 6844 \pm\     9$&$   98 \pm\     9$&$  10.4 \pm\    2.7$&$ 2.5 \pm\  0.7$&$ 0.5 \pm\  0.4$&3.17\\
12& 0.1014&$ 6868 \pm\    12$&$   99 \pm\    12$&$   3.9 \pm\    2.9$&$ 5.8 \pm\  4.4$&$ 2.5 \pm\  2.1$&2.98\\
\enddata
\end{deluxetable}

\clearpage\begin{deluxetable}{cccccccc}
\rotate
\tablecolumns{8}
\tablewidth{0pc}
\tablecaption{NGC 3801: Measured Parameters. \label{ngc3801_p1}}
\tablehead{\colhead{Row} & \colhead{Y-Offset} & \colhead{$v_{\rm r}$} & \colhead{$\sigma_{gas}$} & \colhead{F$(\rm{H}\alpha / 10^{-16}$)} & \colhead{\underline{F(\ion{N}{2}$_{6585}$)}} & \colhead{\underline{F(\ion{S}{2}$_{total}$)}}&\colhead{$R^2$}
\\ \colhead{} & \colhead{(\arcsec)} & \colhead{(\kms)} & \colhead{(\kms)} &  \colhead{\small{(${\rm erg~s^{-1} cm^{-2} Hz^{-1}}$)}} & \colhead{F(H$\alpha$)} & \colhead{F(H$\alpha$)} &\colhead{}}
\startdata
\cutinhead{Slit 1: X-Offset -0.2\arcsec}
 2&-0.9128&$ 3372 \pm\    40$&$   89 \pm\    40$&$   3.7 \pm\    2.2$&$ 0.0 \pm\  0.3$&$ 0.2 \pm\  0.6$&3.00\\
 3&-0.8114&$ 3436 \pm\    16$&$   57 \pm\    16$&$   4.3 \pm\    1.7$&$ 0.4 \pm\  0.3$&$ 0.4 \pm\  0.5$&3.49\\
 4&-0.7099&$ 3468 \pm\     9$&$   46 \pm\     9$&$   3.4 \pm\    1.2$&$ 1.0 \pm\  0.5$&$ 1.7 \pm\  0.9$&3.59\\
 7&-0.4057&$ 3476 \pm\     9$&$   48 \pm\     9$&$   1.5 \pm\    1.0$&$ 1.0 \pm\  0.9$&$ 7.5 \pm\  5.2$&3.41\\
 9&-0.2028&$ 3325 \pm\    13$&$  161 \pm\    12$&$  25.4 \pm\    3.3$&$ 0.9 \pm\  0.2$&$ 0.8 \pm\  0.2$&3.78\\
10&-0.1014&$ 3383 \pm\    10$&$  143 \pm\     9$&$  20.8 \pm\    2.9$&$ 1.7 \pm\  0.3$&$ 0.8 \pm\  0.2$&3.66\\
11& 0.0000&$ 3378 \pm\    12$&$  142 \pm\    11$&$  15.6 \pm\    2.8$&$ 1.9 \pm\  0.4$&$ 1.3 \pm\  0.3$&3.01\\
12& 0.1014&$ 3345 \pm\    17$&$  148 \pm\    17$&$  14.5 \pm\    2.9$&$ 1.0 \pm\  0.3$&$ 1.1 \pm\  0.4$&3.50\\
\enddata
\end{deluxetable}

\clearpage\begin{deluxetable}{cccccccc}
\rotate
\tablecolumns{8}
\tablewidth{0pc}
\tablecaption{NGC 3862: Measured Parameters. \label{ngc3862_p1}}
\tablehead{\colhead{Row} & \colhead{Y-Offset} & \colhead{$v_{\rm r}$} & \colhead{$\sigma_{gas}$} & \colhead{F$(\rm{H}\alpha / 10^{-16}$)} & \colhead{\underline{F(\ion{N}{2}$_{6585}$)}} & \colhead{\underline{F(\ion{S}{2}$_{total}$)}}&\colhead{$R^2$}
\\ \colhead{} & \colhead{(\arcsec)} & \colhead{(\kms)} & \colhead{(\kms)} &  \colhead{\small{(${\rm erg~s^{-1} cm^{-2} Hz^{-1}}$)}} & \colhead{F(H$\alpha$)} & \colhead{F(H$\alpha$)} &\colhead{}}
\startdata
\cutinhead{Slit 1: X-Offset -0.2\arcsec}
 1&-1.0142&$ 6244 \pm\    13$&$   48 \pm\    13$&$   9.5 \pm\    4.4$&$ 0.6 \pm\  0.4$&$ 1.1 \pm\  0.8$&2.10\\
 5&-0.6085&$ 6452 \pm\     9$&$   56 \pm\     8$&$  23.3 \pm\    5.4$&$ 0.4 \pm\  0.2$&$ 0.1 \pm\  0.2$&2.22\\
 6&-0.5071&$ 6464 \pm\     5$&$   43 \pm\     5$&$  20.7 \pm\    4.2$&$ 0.7 \pm\  0.2$&$ 1.0 \pm\  0.3$&1.94\\
 7&-0.4057&$ 6450 \pm\     5$&$   43 \pm\     4$&$  34.1 \pm\    5.5$&$ 0.2 \pm\  0.1$&$ 0.1 \pm\  0.1$&2.25\\
 8&-0.3043&$ 6451 \pm\     9$&$   78 \pm\     9$&$  23.7 \pm\    4.9$&$ 1.3 \pm\  0.4$&$ 0.4 \pm\  0.3$&2.07\\
 9&-0.2028&$ 6457 \pm\     8$&$   80 \pm\     8$&$  20.3 \pm\    4.9$&$ 2.0 \pm\  0.6$&$ 0.7 \pm\  0.4$&2.30\\
10&-0.1014&$ 6486 \pm\    13$&$  149 \pm\    13$&$  10.2 \pm\    5.1$&$ 7.7 \pm\  4.0$&$ 1.9 \pm\  1.4$&2.36\\
11& 0.0000&$ 6468 \pm\    13$&$  214 \pm\    12$&$  42.7 \pm\    7.7$&$ 3.1 \pm\  0.6$&$ 1.2 \pm\  0.4$&2.41\\
\enddata
\end{deluxetable}

\clearpage\begin{deluxetable}{cccccccc}
\rotate
\tablecolumns{8}
\tablewidth{0pc}
\tablecaption{UGC 7115: Measured Parameters. \label{ugc7115-cor_p1}}
\tablehead{\colhead{Row} & \colhead{Y-Offset} & \colhead{$v_{\rm r}$} & \colhead{$\sigma_{gas}$} & \colhead{F$(\rm{H}\alpha / 10^{-16}$)} & \colhead{\underline{F(\ion{N}{2}$_{6585}$)}} & \colhead{\underline{F(\ion{S}{2}$_{total}$)}}&\colhead{$R^2$}
\\ \colhead{} & \colhead{(\arcsec)} & \colhead{(\kms)} & \colhead{(\kms)} &  \colhead{\small{(${\rm erg~s^{-1} cm^{-2} Hz^{-1}}$)}} & \colhead{F(H$\alpha$)} & \colhead{F(H$\alpha$)} &\colhead{}}
\startdata
\cutinhead{Slit 0: X-Offset 0.0\arcsec}
 5&-0.6085&$ 6792 \pm\    17$&$   63 \pm\    18$&$   5.7 \pm\    2.7$&$ 1.0 \pm\  0.7$&$ 0.6 \pm\  0.5$&2.59\\
 6&-0.5071&$ 6886 \pm\     9$&$   49 \pm\     8$&$   8.4 \pm\    2.5$&$ 1.0 \pm\  0.4$&$ 0.5 \pm\  0.3$&2.90\\
 7&-0.4057&$ 6863 \pm\     6$&$   57 \pm\     6$&$  15.8 \pm\    3.0$&$ 0.8 \pm\  0.2$&$ 0.7 \pm\  0.2$&2.65\\
 8&-0.3043&$ 6897 \pm\     7$&$   64 \pm\     7$&$  15.1 \pm\    2.9$&$ 1.2 \pm\  0.3$&$ 0.6 \pm\  0.2$&2.84\\
 9&-0.2028&$ 6914 \pm\     9$&$  107 \pm\     9$&$  17.1 \pm\    3.8$&$ 2.3 \pm\  0.6$&$ 0.9 \pm\  0.3$&2.53\\
10&-0.1014&$ 6963 \pm\    10$&$  271 \pm\     9$&$  74.7 \pm\    6.9$&$ 2.6 \pm\  0.3$&$ 1.1 \pm\  0.1$&2.48\\
11& 0.0000&$ 6801 \pm\     9$&$  496 \pm\     7$&$ 372.8 \pm\   19.3$&$ 2.4 \pm\  0.1$&$ 1.4 \pm\  0.1$&2.79\\
12& 0.1014&$ 6625 \pm\    13$&$  545 \pm\    10$&$ 138.7 \pm\   20.7$&$ 5.6 \pm\  0.8$&$ 2.3 \pm\  0.4$&3.86\\
13& 0.2028&$ 6669 \pm\     8$&$  224 \pm\     7$&$  62.4 \pm\    5.0$&$ 2.3 \pm\  0.2$&$ 0.9 \pm\  0.1$&2.92\\
15& 0.4057&$ 6639 \pm\     8$&$   69 \pm\     8$&$   3.1 \pm\    1.9$&$ 5.6 \pm\  3.7$&$ 4.7 \pm\  3.1$&2.78\\
16& 0.5071&$ 6661 \pm\     5$&$   40 \pm\     6$&$   5.7 \pm\    1.8$&$ 1.9 \pm\  0.7$&$ 0.7 \pm\  0.4$&2.42\\
\enddata
\end{deluxetable}

\clearpage\begin{deluxetable}{cccccccc}
\rotate
\tablecolumns{8}
\tablewidth{0pc}
\tablecaption{NGC 4261: Measured Parameters. \label{ngc4261_p1}}
\tablehead{\colhead{Row} & \colhead{Y-Offset} & \colhead{$v_{\rm r}$} & \colhead{$\sigma_{gas}$} & \colhead{F$(\rm{H}\alpha / 10^{-16}$)} & \colhead{\underline{F(\ion{N}{2}$_{6585}$)}} & \colhead{\underline{F(\ion{S}{2}$_{total}$)}}&\colhead{$R^2$}
\\ \colhead{} & \colhead{(\arcsec)} & \colhead{(\kms)} & \colhead{(\kms)} &  \colhead{\small{(${\rm erg~s^{-1} cm^{-2} Hz^{-1}}$)}} & \colhead{F(H$\alpha$)} & \colhead{F(H$\alpha$)} &\colhead{}}
\startdata
\cutinhead{Slit 1: X-Offset -0.1\arcsec}
 1&-0.5071&$ 2442 \pm\    17$&$  126 \pm\    16$&$  77.3 \pm\   19.6$&$ 1.6 \pm\  0.5$&$ 1.0 \pm\  0.4$&1.62\\
 2&-0.4564&$ 2324 \pm\    17$&$  149 \pm\    17$&$  68.2 \pm\   18.7$&$ 2.3 \pm\  0.7$&$ 1.2 \pm\  0.5$&1.62\\
 3&-0.4057&$ 2388 \pm\    14$&$  114 \pm\    13$&$  43.0 \pm\   15.6$&$ 3.3 \pm\  1.3$&$ 1.8 \pm\  0.8$&1.54\\
 4&-0.3550&$ 2365 \pm\    16$&$  186 \pm\    15$&$  94.8 \pm\   20.8$&$ 2.7 \pm\  0.7$&$ 1.5 \pm\  0.4$&1.45\\
 5&-0.3043&$ 2357 \pm\    11$&$  193 \pm\    10$&$ 174.7 \pm\   23.0$&$ 2.5 \pm\  0.4$&$ 0.7 \pm\  0.2$&1.83\\
 6&-0.2535&$ 2377 \pm\     7$&$  195 \pm\     6$&$ 223.3 \pm\   23.1$&$ 3.3 \pm\  0.4$&$ 1.4 \pm\  0.2$&1.64\\
 7&-0.2028&$ 2393 \pm\     6$&$  237 \pm\     5$&$ 381.4 \pm\   27.7$&$ 3.4 \pm\  0.3$&$ 1.4 \pm\  0.1$&1.91\\
 8&-0.1521&$ 2360 \pm\     7$&$  343 \pm\     7$&$ 864.7 \pm\   48.2$&$ 2.8 \pm\  0.2$&$ 1.0 \pm\  0.1$&2.24\\
\enddata
\end{deluxetable}

\clearpage\begin{deluxetable}{cccccccc}
\rotate
\tablecolumns{8}
\tablewidth{0pc}
\tablecaption{NGC 4335: Measured Parameters. \label{ngc4335_p1}}
\tablehead{\colhead{Row} & \colhead{Y-Offset} & \colhead{$v_{\rm r}$} & \colhead{$\sigma_{gas}$} & \colhead{F$(\rm{H}\alpha / 10^{-16}$)} & \colhead{\underline{F(\ion{N}{2}$_{6585}$)}} & \colhead{\underline{F(\ion{S}{2}$_{total}$)}}&\colhead{$R^2$}
\\ \colhead{} & \colhead{(\arcsec)} & \colhead{(\kms)} & \colhead{(\kms)} &  \colhead{\small{(${\rm erg~s^{-1} cm^{-2} Hz^{-1}}$)}} & \colhead{F(H$\alpha$)} & \colhead{F(H$\alpha$)} &\colhead{}}
\startdata
\cutinhead{Slit 1: X-Offset -0.2\arcsec}
 1&-1.0142&$ 4955 \pm\    12$&$   94 \pm\    12$&$   5.4 \pm\    1.6$&$ 2.3 \pm\  0.8$&$ 0.8 \pm\  0.5$&2.80\\
 3&-0.8114&$ 4943 \pm\     7$&$   92 \pm\     7$&$  13.0 \pm\    1.9$&$ 1.8 \pm\  0.3$&$ 0.8 \pm\  0.3$&2.81\\
 4&-0.7099&$ 4915 \pm\     9$&$  107 \pm\     9$&$  14.3 \pm\    2.2$&$ 1.7 \pm\  0.3$&$ 1.0 \pm\  0.3$&2.37\\
 5&-0.6085&$ 4928 \pm\     7$&$   84 \pm\     7$&$   8.5 \pm\    1.9$&$ 2.9 \pm\  0.7$&$ 1.3 \pm\  0.5$&2.29\\
 6&-0.5071&$ 4967 \pm\     6$&$   83 \pm\     6$&$  13.7 \pm\    2.2$&$ 2.0 \pm\  0.4$&$ 0.3 \pm\  0.2$&2.15\\
 7&-0.4057&$ 4947 \pm\     5$&$   85 \pm\     5$&$  20.8 \pm\    2.6$&$ 1.7 \pm\  0.3$&$ 0.8 \pm\  0.2$&1.84\\
 8&-0.3043&$ 4904 \pm\     8$&$  111 \pm\     8$&$  19.2 \pm\    3.1$&$ 2.2 \pm\  0.4$&$ 1.0 \pm\  0.3$&2.20\\
 9&-0.2028&$ 4807 \pm\    10$&$  160 \pm\     9$&$  29.5 \pm\    4.0$&$ 2.3 \pm\  0.4$&$ 1.0 \pm\  0.3$&2.43\\
\enddata
\end{deluxetable}

\clearpage\begin{deluxetable}{cccccccc}
\rotate
\tablecolumns{8}
\tablewidth{0pc}
\tablecaption{M84 : Measured Parameters. \label{m84_p1}}
\tablehead{\colhead{Row} & \colhead{Y-Offset} & \colhead{$v_{\rm r}$} & \colhead{$\sigma_{gas}$} & \colhead{F$(\rm{H}\alpha / 10^{-16}$)} & \colhead{\underline{F(\ion{N}{2}$_{6585}$)}} & \colhead{\underline{F(\ion{S}{2}$_{total}$)}}&\colhead{$R^2$}
\\ \colhead{} & \colhead{(\arcsec)} & \colhead{(\kms)} & \colhead{(\kms)} &  \colhead{\small{(${\rm erg~s^{-1} cm^{-2} Hz^{-1}}$)}} & \colhead{F(H$\alpha$)} & \colhead{F(H$\alpha$)} &\colhead{}}
\startdata
\cutinhead{Slit 1: X-Offset -0.2\arcsec}
 1&-0.5071&$ 1088 \pm\     7$&$   89 \pm\     8$&$  19.3 \pm\    3.2$&$ 1.9 \pm\  0.4$&$ 0.9 \pm\  0.3$&1.33\\
 2&-0.4564&$ 1071 \pm\     6$&$   88 \pm\     6$&$  14.4 \pm\    2.8$&$ 3.2 \pm\  0.7$&$ 1.6 \pm\  0.4$&1.28\\
 3&-0.4057&$ 1064 \pm\     7$&$   95 \pm\     7$&$  21.7 \pm\    3.3$&$ 2.2 \pm\  0.4$&$ 1.0 \pm\  0.2$&1.36\\
 4&-0.3550&$ 1061 \pm\     6$&$  101 \pm\     6$&$  31.6 \pm\    3.6$&$ 1.9 \pm\  0.3$&$ 1.1 \pm\  0.2$&1.22\\
 5&-0.3043&$ 1068 \pm\     5$&$  118 \pm\     5$&$  45.4 \pm\    4.0$&$ 1.9 \pm\  0.2$&$ 0.9 \pm\  0.1$&1.34\\
 6&-0.2535&$ 1106 \pm\     4$&$  126 \pm\     4$&$  53.1 \pm\    3.9$&$ 2.5 \pm\  0.2$&$ 1.2 \pm\  0.1$&1.39\\
 7&-0.2028&$ 1102 \pm\     3$&$  115 \pm\     3$&$  67.7 \pm\    4.0$&$ 2.2 \pm\  0.2$&$ 1.1 \pm\  0.1$&1.42\\
 8&-0.1521&$ 1102 \pm\     2$&$  102 \pm\     2$&$  65.0 \pm\    3.6$&$ 2.3 \pm\  0.2$&$ 1.5 \pm\  0.1$&1.54\\
\enddata
\end{deluxetable}

\clearpage\begin{deluxetable}{cccccccc}
\rotate
\tablecolumns{8}
\tablewidth{0pc}
\tablecaption{NGC 4486: Measured Parameters. \label{ngc4486_p1}}
\tablehead{\colhead{Row} & \colhead{Y-Offset} & \colhead{$v_{\rm r}$} & \colhead{$\sigma_{gas}$} & \colhead{F$(\rm{H}\alpha / 10^{-16}$)} & \colhead{\underline{F(\ion{N}{2}$_{6585}$)}} & \colhead{\underline{F(\ion{S}{2}$_{total}$)}}&\colhead{$R^2$}
\\ \colhead{} & \colhead{(\arcsec)} & \colhead{(\kms)} & \colhead{(\kms)} &  \colhead{\small{(${\rm erg~s^{-1} cm^{-2} Hz^{-1}}$)}} & \colhead{F(H$\alpha$)} & \colhead{F(H$\alpha$)} &\colhead{}}
\startdata
\cutinhead{Slit 1: X-Offset -0.2\arcsec}
 1&-0.5071&$ 1535 \pm\    14$&$  374 \pm\    12$&$ 113.0 \pm\   17.5$&$ 3.9 \pm\  0.6$&$ 2.4 \pm\  0.4$&1.93\\
 2&-0.4564&$ 1634 \pm\    12$&$  376 \pm\    10$&$ 252.0 \pm\   19.7$&$ 2.0 \pm\  0.2$&$ 1.4 \pm\  0.1$&2.10\\
 3&-0.4057&$ 1775 \pm\     5$&$  266 \pm\     4$&$ 353.6 \pm\   14.4$&$ 1.7 \pm\  0.1$&$ 1.3 \pm\  0.1$&2.18\\
 4&-0.3550&$ 1765 \pm\     4$&$  256 \pm\     3$&$ 352.3 \pm\   14.1$&$ 2.0 \pm\  0.1$&$ 1.4 \pm\  0.1$&1.89\\
 5&-0.3043&$ 1755 \pm\     5$&$  274 \pm\     4$&$ 332.3 \pm\   14.4$&$ 2.2 \pm\  0.1$&$ 1.5 \pm\  0.1$&1.90\\
 6&-0.2535&$ 1788 \pm\     3$&$  274 \pm\     2$&$ 477.4 \pm\   15.6$&$ 2.5 \pm\  0.1$&$ 1.6 \pm\  0.1$&2.33\\
 7&-0.2028&$ 1850 \pm\     2$&$  235 \pm\     1$&$ 794.4 \pm\   16.2$&$ 2.4 \pm\  0.1$&$ 1.5 \pm\  0.0$&2.68\\
 8&-0.1521&$ 1842 \pm\     1$&$  206 \pm\     1$&$1109.7 \pm\   17.9$&$ 2.4 \pm\  0.0$&$ 1.5 \pm\  0.0$&2.81\\
\enddata
\end{deluxetable}

\clearpage\begin{deluxetable}{cccccccc}
\rotate
\tablecolumns{8}
\tablewidth{0pc}
\tablecaption{NGC 5127: Measured Parameters. \label{ngc5127_p1}}
\tablehead{\colhead{Row} & \colhead{Y-Offset} & \colhead{$v_{\rm r}$} & \colhead{$\sigma_{gas}$} & \colhead{F$(\rm{H}\alpha / 10^{-16}$)} & \colhead{\underline{F(\ion{N}{2}$_{6585}$)}} & \colhead{\underline{F(\ion{S}{2}$_{total}$)}}&\colhead{$R^2$}
\\ \colhead{} & \colhead{(\arcsec)} & \colhead{(\kms)} & \colhead{(\kms)} &  \colhead{\small{(${\rm erg~s^{-1} cm^{-2} Hz^{-1}}$)}} & \colhead{F(H$\alpha$)} & \colhead{F(H$\alpha$)} &\colhead{}}
\startdata
\cutinhead{Slit 1: X-Offset -0.2\arcsec}
 5&-0.6085&$ 4931 \pm\    17$&$   52 \pm\    17$&$   4.4 \pm\    2.3$&$ 0.4 \pm\  0.5$&$ 1.3 \pm\  0.9$&3.61\\
 8&-0.3043&$ 4851 \pm\    10$&$  125 \pm\    10$&$  11.4 \pm\    2.8$&$ 2.5 \pm\  0.7$&$ 2.7 \pm\  0.7$&3.72\\
 9&-0.2028&$ 4873 \pm\     9$&$   91 \pm\     9$&$   5.4 \pm\    2.2$&$ 4.3 \pm\  1.9$&$ 3.5 \pm\  1.5$&3.01\\
10&-0.1014&$ 4799 \pm\     8$&$   61 \pm\     8$&$   6.7 \pm\    2.1$&$ 1.5 \pm\  0.6$&$ 1.8 \pm\  0.7$&3.31\\
11& 0.0000&$ 4919 \pm\    22$&$  125 \pm\    22$&$   7.6 \pm\    2.9$&$ 1.8 \pm\  0.8$&$ 1.2 \pm\  0.6$&2.99\\
12& 0.1014&$ 4711 \pm\    27$&$  161 \pm\    26$&$  11.1 \pm\    3.5$&$ 1.4 \pm\  0.6$&$ 1.0 \pm\  0.4$&3.59\\
13& 0.2028&$ 4705 \pm\    10$&$   68 \pm\    10$&$   9.6 \pm\    2.5$&$ 0.6 \pm\  0.3$&$ 1.4 \pm\  0.5$&3.65\\
14& 0.3043&$ 4663 \pm\    36$&$  219 \pm\    31$&$  14.0 \pm\    4.1$&$ 1.0 \pm\  0.4$&$ 1.3 \pm\  0.5$&3.24\\
\enddata
\end{deluxetable}

\clearpage\begin{deluxetable}{cccccccc}
\rotate
\tablecolumns{8}
\tablewidth{0pc}
\tablecaption{NGC 5141: Measured Parameters. \label{ngc5141_p1}}
\tablehead{\colhead{Row} & \colhead{Y-Offset} & \colhead{$v_{\rm r}$} & \colhead{$\sigma_{gas}$} & \colhead{F$(\rm{H}\alpha / 10^{-16}$)} & \colhead{\underline{F(\ion{N}{2}$_{6585}$)}} & \colhead{\underline{F(\ion{S}{2}$_{total}$)}}&\colhead{$R^2$}
\\ \colhead{} & \colhead{(\arcsec)} & \colhead{(\kms)} & \colhead{(\kms)} &  \colhead{\small{(${\rm erg~s^{-1} cm^{-2} Hz^{-1}}$)}} & \colhead{F(H$\alpha$)} & \colhead{F(H$\alpha$)} &\colhead{}}
\startdata
\cutinhead{Slit 1: X-Offset -0.2\arcsec}
 1&-1.0142&$ 5498 \pm\    15$&$  128 \pm\    15$&$   6.0 \pm\    2.4$&$ 3.9 \pm\  1.7$&$ 1.4 \pm\  0.9$&2.78\\
 4&-0.7099&$ 5335 \pm\    26$&$  160 \pm\    25$&$   5.2 \pm\    2.6$&$ 4.2 \pm\  2.3$&$ 0.5 \pm\  0.9$&2.97\\
 5&-0.6085&$ 5494 \pm\    10$&$   87 \pm\    10$&$   8.9 \pm\    2.3$&$ 2.3 \pm\  0.7$&$ 0.8 \pm\  0.4$&2.27\\
 6&-0.5071&$ 5389 \pm\    20$&$  179 \pm\    18$&$   7.1 \pm\    2.8$&$ 5.3 \pm\  2.2$&$ 0.7 \pm\  0.8$&2.26\\
 7&-0.4057&$ 5456 \pm\    23$&$  198 \pm\    21$&$  10.8 \pm\    3.3$&$ 3.4 \pm\  1.2$&$ 0.6 \pm\  0.6$&2.35\\
 8&-0.3043&$ 5383 \pm\    12$&$  158 \pm\    11$&$  12.6 \pm\    3.0$&$ 4.1 \pm\  1.1$&$ 2.1 \pm\  0.7$&2.67\\
 9&-0.2028&$ 5295 \pm\     7$&$  141 \pm\     7$&$  25.5 \pm\    3.3$&$ 3.0 \pm\  0.5$&$ 2.3 \pm\  0.4$&2.21\\
10&-0.1014&$ 5265 \pm\     3$&$  131 \pm\     3$&$  50.9 \pm\    3.6$&$ 3.2 \pm\  0.3$&$ 1.5 \pm\  0.2$&2.35\\
\enddata
\end{deluxetable}

\clearpage\begin{deluxetable}{cccccccc}
\rotate
\tablecolumns{8}
\tablewidth{0pc}
\tablecaption{NGC 5490: Measured Parameters. \label{ngc5490_p1}}
\tablehead{\colhead{Row} & \colhead{Y-Offset} & \colhead{$v_{\rm r}$} & \colhead{$\sigma_{gas}$} & \colhead{F$(\rm{H}\alpha / 10^{-16}$)} & \colhead{\underline{F(\ion{N}{2}$_{6585}$)}} & \colhead{\underline{F(\ion{S}{2}$_{total}$)}}&\colhead{$R^2$}
\\ \colhead{} & \colhead{(\arcsec)} & \colhead{(\kms)} & \colhead{(\kms)} &  \colhead{\small{(${\rm erg~s^{-1} cm^{-2} Hz^{-1}}$)}} & \colhead{F(H$\alpha$)} & \colhead{F(H$\alpha$)} &\colhead{}}
\startdata
\cutinhead{Slit 1: X-Offset -0.2\arcsec}
 3&-0.8114&$ 5164 \pm\    33$&$   76 \pm\    33$&$   5.8 \pm\    3.5$&$ 0.5 \pm\  0.5$&$ 0.3 \pm\  0.6$&2.42\\
 5&-0.6085&$ 5104 \pm\     7$&$   33 \pm\    10$&$   9.8 \pm\    4.1$&$ 0.1 \pm\  0.2$&$ 0.6 \pm\  0.4$&2.05\\
10&-0.1014&$ 5008 \pm\    16$&$  162 \pm\    15$&$  16.1 \pm\    4.4$&$ 3.4 \pm\  1.0$&$ 0.8 \pm\  0.5$&2.02\\
11& 0.0000&$ 4999 \pm\    10$&$  205 \pm\     9$&$  41.3 \pm\    5.4$&$ 3.3 \pm\  0.5$&$ 0.8 \pm\  0.2$&2.29\\
12& 0.1014&$ 5154 \pm\    11$&$  261 \pm\     9$&$  79.9 \pm\    7.0$&$ 2.6 \pm\  0.3$&$ 1.0 \pm\  0.2$&2.29\\
13& 0.2028&$ 5220 \pm\     9$&$  160 \pm\     8$&$  29.9 \pm\    4.4$&$ 3.6 \pm\  0.6$&$ 1.5 \pm\  0.3$&1.94\\
14& 0.3043&$ 5204 \pm\    18$&$  145 \pm\    18$&$  24.3 \pm\    5.3$&$ 1.3 \pm\  0.4$&$ 0.8 \pm\  0.3$&2.00\\
16& 0.5071&$ 5213 \pm\    17$&$   63 \pm\    17$&$   7.9 \pm\    3.3$&$ 1.0 \pm\  0.6$&$ 0.5 \pm\  0.5$&2.07\\
\enddata
\end{deluxetable}

\clearpage\begin{deluxetable}{cccccccc}
\rotate
\tablecolumns{8}
\tablewidth{0pc}
\tablecaption{NGC 7052: Measured Parameters. \label{ngc7052_p1}}
\tablehead{\colhead{Row} & \colhead{Y-Offset} & \colhead{$v_{\rm r}$} & \colhead{$\sigma_{gas}$} & \colhead{F$(\rm{H}\alpha / 10^{-16}$)} & \colhead{\underline{F(\ion{N}{2}$_{6585}$)}} & \colhead{\underline{F(\ion{S}{2}$_{total}$)}}&\colhead{$R^2$}
\\ \colhead{} & \colhead{(\arcsec)} & \colhead{(\kms)} & \colhead{(\kms)} &  \colhead{\small{(${\rm erg~s^{-1} cm^{-2} Hz^{-1}}$)}} & \colhead{F(H$\alpha$)} & \colhead{F(H$\alpha$)} &\colhead{}}
\startdata
\cutinhead{Slit 1: X-Offset -0.1\arcsec}
 2&-0.4564&$ 4310 \pm\    26$&$  153 \pm\    23$&$  35.8 \pm\   10.0$&$ 1.9 \pm\  0.7$&$ 0.3 \pm\  0.3$&2.33\\
 5&-0.3043&$ 4466 \pm\    26$&$  186 \pm\    24$&$  28.5 \pm\    9.9$&$ 3.3 \pm\  1.3$&$ 0.6 \pm\  0.5$&1.94\\
 6&-0.2535&$ 4527 \pm\    24$&$  179 \pm\    22$&$  25.7 \pm\    9.8$&$ 4.0 \pm\  1.7$&$ 0.7 \pm\  0.5$&2.75\\
 7&-0.2028&$ 4461 \pm\    26$&$  241 \pm\    22$&$  50.7 \pm\   12.0$&$ 2.9 \pm\  0.8$&$ 1.0 \pm\  0.4$&2.97\\
 8&-0.1521&$ 4481 \pm\    27$&$  281 \pm\    21$&$  86.0 \pm\   15.0$&$ 2.3 \pm\  0.5$&$ 1.6 \pm\  0.4$&2.15\\
 9&-0.1014&$ 4487 \pm\    15$&$  285 \pm\    12$&$ 167.2 \pm\   17.1$&$ 2.4 \pm\  0.3$&$ 1.1 \pm\  0.2$&2.62\\
10&-0.0507&$ 4463 \pm\    15$&$  413 \pm\    14$&$ 289.4 \pm\   29.2$&$ 3.0 \pm\  0.3$&$ 1.1 \pm\  0.2$&2.31\\
11& 0.0000&$ 4541 \pm\    17$&$  456 \pm\    16$&$ 354.7 \pm\   35.1$&$ 2.7 \pm\  0.3$&$ 0.8 \pm\  0.1$&2.94\\
\enddata
\end{deluxetable}

\clearpage\begin{deluxetable}{cccccccc}
\rotate
\tablecolumns{8}
\tablewidth{0pc}
\tablecaption{UGC 12064: Measured Parameters. \label{ugc12064_p1}}
\tablehead{\colhead{Row} & \colhead{Y-Offset} & \colhead{$v_{\rm r}$} & \colhead{$\sigma_{gas}$} & \colhead{F$(\rm{H}\alpha / 10^{-16}$)} & \colhead{\underline{F(\ion{N}{2}$_{6585}$)}} & \colhead{\underline{F(\ion{S}{2}$_{total}$)}}&\colhead{$R^2$}
\\ \colhead{} & \colhead{(\arcsec)} & \colhead{(\kms)} & \colhead{(\kms)} &  \colhead{\small{(${\rm erg~s^{-1} cm^{-2} Hz^{-1}}$)}} & \colhead{F(H$\alpha$)} & \colhead{F(H$\alpha$)} &\colhead{}}
\startdata
\cutinhead{Slit 1: X-Offset -0.1\arcsec}
 2&-0.4564&$ 4876 \pm\    17$&$  136 \pm\    17$&$  39.5 \pm\   10.9$&$ 1.8 \pm\  0.6$&$ 1.3 \pm\  0.5$&2.29\\
 3&-0.4057&$ 4971 \pm\    18$&$  107 \pm\    18$&$  47.6 \pm\   11.6$&$ 0.4 \pm\  0.2$&$ 0.7 \pm\  0.3$&1.88\\
 4&-0.3550&$ 4962 \pm\    19$&$   73 \pm\    19$&$  29.8 \pm\   10.6$&$ 0.1 \pm\  0.2$&$ 0.2 \pm\  0.3$&2.06\\
 5&-0.3043&$ 4880 \pm\    25$&$  127 \pm\    25$&$  37.2 \pm\   11.3$&$ 1.0 \pm\  0.4$&$ 0.1 \pm\  0.3$&2.24\\
 6&-0.2535&$ 4868 \pm\    18$&$  134 \pm\    16$&$  44.1 \pm\   10.1$&$ 1.7 \pm\  0.5$&$ 0.0 \pm\  0.3$&2.11\\
 7&-0.2028&$ 4940 \pm\    13$&$   47 \pm\    12$&$  17.8 \pm\    7.1$&$ 0.5 \pm\  0.4$&$ 0.9 \pm\  0.6$&1.77\\
 8&-0.1521&$ 4957 \pm\     8$&$   46 \pm\     8$&$  32.3 \pm\    8.5$&$ 0.4 \pm\  0.2$&$ 0.7 \pm\  0.3$&2.44\\
 9&-0.1014&$ 5015 \pm\    21$&$  191 \pm\    18$&$  69.0 \pm\   13.2$&$ 1.6 \pm\  0.4$&$ 0.7 \pm\  0.3$&2.36\\
\enddata
\end{deluxetable}

\clearpage\begin{deluxetable}{cccccccc}
\rotate
\tablecolumns{8}
\tablewidth{0pc}
\tablecaption{NGC 7626: Measured Parameters. \label{ngc7626_p1}}
\tablehead{\colhead{Row} & \colhead{Y-Offset} & \colhead{$v_{\rm r}$} & \colhead{$\sigma_{gas}$} & \colhead{F$(\rm{H}\alpha / 10^{-16}$)} & \colhead{\underline{F(\ion{N}{2}$_{6585}$)}} & \colhead{\underline{F(\ion{S}{2}$_{total}$)}}&\colhead{$R^2$}
\\ \colhead{} & \colhead{(\arcsec)} & \colhead{(\kms)} & \colhead{(\kms)} &  \colhead{\small{(${\rm erg~s^{-1} cm^{-2} Hz^{-1}}$)}} & \colhead{F(H$\alpha$)} & \colhead{F(H$\alpha$)} &\colhead{}}
\startdata
\cutinhead{Slit 1: X-Offset -0.2\arcsec}
 8&-0.3043&$ 3554 \pm\    30$&$  210 \pm\    25$&$  17.5 \pm\    4.1$&$ 1.6 \pm\  0.5$&$ 0.5 \pm\  0.3$&2.25\\
 9&-0.2028&$ 3583 \pm\    22$&$  230 \pm\    18$&$  22.5 \pm\    4.2$&$ 2.3 \pm\  0.5$&$ 0.4 \pm\  0.3$&2.17\\
10&-0.1014&$ 3457 \pm\    13$&$  200 \pm\    12$&$  34.4 \pm\    4.1$&$ 1.9 \pm\  0.3$&$ 0.9 \pm\  0.2$&2.46\\
11& 0.0000&$ 3378 \pm\     8$&$  152 \pm\     8$&$  14.0 \pm\    3.1$&$ 5.1 \pm\  1.2$&$ 1.4 \pm\  0.5$&2.32\\
12& 0.1014&$ 3332 \pm\    15$&$  185 \pm\    14$&$   8.6 \pm\    3.4$&$ 6.5 \pm\  2.6$&$ 1.3 \pm\  0.8$&2.51\\
13& 0.2028&$ 3331 \pm\    18$&$  163 \pm\    17$&$  20.5 \pm\    4.0$&$ 1.5 \pm\  0.4$&$ 0.4 \pm\  0.3$&2.14\\
14& 0.3043&$ 3295 \pm\    27$&$  155 \pm\    27$&$  13.4 \pm\    4.1$&$ 1.3 \pm\  0.5$&$ 0.4 \pm\  0.4$&2.25\\
16& 0.5071&$ 3289 \pm\    16$&$   63 \pm\    16$&$   7.9 \pm\    2.8$&$ 0.0 \pm\  0.2$&$ 0.2 \pm\  0.3$&2.89\\
\enddata
\end{deluxetable}

\fi

\clearpage
\begin{deluxetable}{ccccccc}
\tablewidth{0pc}
\tablecaption{Effect of making various fit parameters free. \label{hafree}}
\tablehead{\colhead{Fit description}&\colhead{$v_{H\alpha}$}&\colhead{$v_{[NII]}$}&\colhead{$\sigma_{H\alpha}$}&\colhead{$\sigma_{[NII]}$}&\colhead{\underline{F([\ion{N}{2}]$_{6585}$)}}&\colhead{$R^2$} \\ \colhead{}&\colhead{(\kms)}&\colhead{(\kms)}&\colhead{(\kms)}&\colhead{(\kms)}&\colhead{F([\ion{N}{2}]$_{6549}$)}&\colhead{}
\\ \colhead{(1)}&\colhead{(2)}&\colhead{(3)}&\colhead{(4)}&\colhead{(5)}&\colhead{(6)}&\colhead{(7)}}
\startdata
initial parameters & \multicolumn{2}{c}{4620 $\pm$\ 4} & \multicolumn{2}{c}{249 $\pm$\ 3}& 3.0 & 3.21\\
$v_{H\alpha}$ free & 4616 $\pm$\ 9 & 4620 $\pm$\ 3 & \multicolumn{2}{c}{249 $\pm$\ 3}& 3.0 & 3.22\\
$\sigma_{H\alpha}$ free &\multicolumn{2}{c}{4619 $\pm$\ 3}& 228 $\pm$\ 9& 256 $\pm$\ 4& 3.0 &3.21\\
$v_{H\alpha} + \sigma_{H\alpha}$ free & 4618 $\pm$\ 9&4619 $\pm$\ 4&228 $\pm$\ 9&256 $\pm$\ 4 & 3.0 & 3.22\\
{[\ion{N}{2}]} ratio free & \multicolumn{2}{c}{4622 $\pm$\ 3}& \multicolumn{2}{c}{248 $\pm$\ 3}& 2.32 $\pm$\ 0.09 & 3.04\\
\enddata
\tablecomments{The measurements shown were made using the central spectral row in the central slit observed for NGC~4335. Col. (1): Parameters made free, Initially a set of lines with one velocity and one velocity dispersion was used, with the ratio of the two [\ion{N}{2}] lines fixed to 3.0. The two [\ion{S}{2}] lines are fixed in velocity and width to the [\ion{N}{2}] lines in every case; Col. (2-3): The radial velocities of the H$\alpha$ and [\ion{N}{2}] lines; Col. (3-4): The velocity dispersions of the H$\alpha$ and [\ion{N}{2}] lines; Col. (5): The ratio of the [\ion{N}{2}]$_{6585}$ and [\ion{N}{2}]$_{6549}$ lines; Col. (6): The reduced $\chi^2$ value of the resulting fit.}
\end{deluxetable}

\clearpage
\begin{deluxetable}{lrrcrc}
\tablewidth{0pc}
\tablecaption{Presence of a Nuclear Broad Line.\label{blrtab}}
\tablehead{
\colhead{Galaxy} & \colhead{$\Delta\overline{|residual|}$} & \colhead{\underline{$R^2_{\rm N}-R^2_{\rm A}$}} & \colhead{By eye} &\colhead{Score} &\colhead{Broad Cpt.}\\
\colhead{} & \colhead{(\%)} & \colhead{$R^2_{\rm N}$} & & & \colhead{}\\ \colhead{(1)}&\colhead{(2)}&\colhead{(3)}&\colhead{(4)}&\colhead{(5)}&\colhead{(6)}}
\startdata
NGC~193  &  9.6 &  0.19 & $\checkmark$ &3 & Yes \\
NGC~315  & 25.0 &  0.46 & $\checkmark$ &3 & Yes \\
NGC~383  &  8.7 &  0.16 & $\circ$ &2 &Yes \\
NGC~541  &  4.6 &  0.12 & $\checkmark$ &1& No \\
NGC~741  &  1.9 &  0.01 & $\circ$ &0& No \\
UGC~1841 & 14.7 &  0.35 & $\checkmark$ &3& Yes \\
NGC~2329 &  9.4 &  0.32 & $\circ$ &2& * \\
NGC~2892 &  2.8 &  0.09 & $\circ$ &0& No \\
\enddata 
\tablecomments{Col. (1): NGC/UGC Identification; Col. (2): Difference in the mean of the absolute values of the residuals from the fits including and excluding the additional component; Col. (3): Relative change in reduced $\chi^2$ value between fits with the narrow lines alone (N) and with the additional component (A); Col. (4): Could an improvement in the fit be judged by eye?; Col. (5): Additional component `score', one point available from each of columns 2 to 6, see text for criteria; Col. (6) Is an additional broad component useful in the fit based on these results? An asterisk (*) indicates the component appears to represent a non-flat continuum.}
\end{deluxetable}

\clearpage
\begin{deluxetable}{lrrrrrr}
\tablewidth{0pc}
\tablecaption{Fits to the central pixel for each galaxy, including broad lines. Kinematics.\label{ffbblfits}}
\tablehead{\colhead{Galaxy} & \colhead{Row} & \colhead{$v_{\rm r}$(NL)}&\colhead{$v_{\rm r}$(BL)} & \colhead{$\sigma_{gas}$(NL)}& \colhead{$\sigma_{gas}$(BL)} & \colhead{$R^2$}
\\ \colhead{} & \colhead{} & \colhead{(\kms)}& \colhead{(\kms)}& \colhead{(\kms)}& \colhead{(\kms)} &\colhead{} \\
\colhead{(1)}&\colhead{(2)}&\colhead{(3)}&\colhead{(4)}&\colhead{(5)}&\colhead{(6)}&\colhead{(7)}
}
\startdata
NGC~193   & 11 & 4417 $\pm$\ \phn 7 & 5128 $\pm$\ 36 & 277 $\pm$\ \phn 6 & 967 $\pm$\ \phn 21 & 3.15\\
NGC~315   & 11 & 4830 $\pm$\ \phn 9 & 5738 $\pm$\ 55 & 406 $\pm$\ \phn 9 & 1790 $\pm$ \phn 67 & 2.06\\
NGC~383   & 11 & 5278 $\pm$\ 54 & 5568 $\pm$\ 29 & 669 $\pm$\ 39 & 1367 $\pm$\ \phn 79 &2.25\\
UGC~1841  & 11 & 6409 $\pm$\ \phn 8 & 6585 $\pm$\ 20 & 436 $\pm$\ \phn 7 & 1254 $\pm$\ \phn 35 &3.11\\
NGC~4261$^\dagger$ & 11 & 2056 $\pm$\ \phn 9 & 2703 $\pm$\ 68 & 366 $\pm$\ 12 & 1079 $\pm$\ \phn 55 &1.69\\
NGC~4335  & 11 & 4618 $\pm$\ \phn 3 & 4893 $\pm$\ 37 & 188 $\pm$\ \phn 4 & 1299 $\pm$\ \phn 51 &1.93\\
NGC~4486 & 11 & 1417 $\pm$\ 14 & 1408 $\pm$\ 28 & 534 $\pm$\ 12 & 1479 $\pm$\ \phn 42 & 6.01\\
NGC~5141 & 11 & 5229 $\pm$\ \phn 3 & 5356 $\pm$\ 86 & 121 $\pm$\ \phn 4 & 1223 $\pm$\ \phn 96 &2.01\\
\enddata
\tablecomments{\dag For NGC~4261 slit 1 (offset -0.1 arcsec) was used; Col. (1): NGC/UGC~Number; Col. (2): Central pixel row number referenced from the earlier data tables for each galaxy; Col. (3): The line of sight velocities of the narrow line components; Col. (4): The line of sight velocities of the broad components (where present); Col. (5): The line of sight velocity dispersions of the narrow line components; Col. (6): The line of sight velocity dispersions of the broad components (where present); Col. (7): The reduced $\chi^{2}$ value of the fit. The parameters were measured by fitting a models with 5 narrow lines, fixed to each other in velocity and velocity dispersion and one free broad component, except as noted above. Errors quoted are the formal errors on the various parameters from the fit.}
\end{deluxetable}

\clearpage
\begin{deluxetable}{lrrrrr}
\tablewidth{0pc}
\tablecaption{Fits to the central pixel for each galaxy, including broad lines. Fluxes.\label{ffbblflux}}
\tablehead{\colhead{Galaxy} & \colhead{F(H$\alpha_{Narrow}$)} &\colhead{\underline{F([\ion{N}{2}]$_{6585}$)}} & \colhead{\underline{F([\ion{S}{2}]$_{total}$)}}&\colhead{\underline{F(Broad Line)}}&\colhead{\underline{F([\ion{S}{2}]$_{6733}$)}}
\\ \colhead{} & \colhead{(${\rm erg s^{-1} cm^{-2} Hz^{-1}}$)} &\colhead{F(H$\alpha_{Narrow}$)} &\colhead{F(H$\alpha_{Narrow}$)} & \colhead{F(H$\alpha_{Narrow}$)}&\colhead{F([\ion{S}{2}]$_{6718}$)}
\\ \colhead{(1)}&\colhead{(2)}&\colhead{(3)}&\colhead{(4)}&\colhead{(5)}&\colhead{(6)}}
\startdata
NGC 193&    1.33 $\pm$ 0.23& 4.21 $\pm$ \phn 0.77& 3.86 $\pm$ \phn 0.71& 7.60 $\pm$ \phn 1.46& 1.18 $\pm$ 0.07\\
NGC 315&    9.28 $\pm$ 0.77& 4.72 $\pm$ \phn 0.42& 1.81 $\pm$ \phn 0.21& 4.40 $\pm$ \phn 0.51& 1.80 $\pm$ 0.14\\
NGC 383&   27.72 $\pm$ 3.65& 0.33 $\pm$ \phn 0.18& 0.56 $\pm$ \phn 0.63& 2.14 $\pm$ \phn 0.50& 0.11 $\pm$ 0.12\\
UGC 1841&   7.71 $\pm$ 0.46& 2.41 $\pm$ \phn 0.16& 1.39 $\pm$ \phn 0.11& 2.57 $\pm$ \phn 0.28& 1.09 $\pm$ 0.06\\
NGC 4261&   4.36 $\pm$ 1.53& 6.76 $\pm$ \phn 2.42& 3.37 $\pm$ \phn 1.23& 7.22 $\pm$ \phn 2.93& 1.55 $\pm$ 0.15\\
NGC 4335&   1.33 $\pm$ 0.13& 4.46 $\pm$ \phn 0.44& 2.63 $\pm$ \phn 0.32& 7.54 $\pm$ \phn 0.87& 1.04 $\pm$ 0.08\\
NGC 4486&  44.96 $\pm$ 1.53& 0.43 $\pm$ \phn 0.04& 0.59 $\pm$ \phn 0.60& 1.37 $\pm$ \phn 0.11& 0.03 $\pm$ 0.03\\
NGC 5141&   0.91 $\pm$ 0.07& 2.95 $\pm$ \phn 0.25& 1.38 $\pm$ \phn 0.22& 3.11 $\pm$ \phn 0.45& 0.97 $\pm$ 0.14\\
\enddata
\tablecomments{Col. (1): NGC/UGC~Number; Col. (2): H$\alpha$ narrow line flux; Col. (3): Ratio of the broad line flux to the H$\alpha$ narrow line flux; Col. (4) Ratio of the [\ion{N}{2}]$_{6585}$ narrow line flux to the H$\alpha$ Narrow line flux; Col. (5) Ratio of the total [\ion{S}{2}] flux to the H$\alpha$ Narrow line flux; Col. (6) Ratio of the [\ion{S}{2}]$_{6733}$ narrow line to the [\ion{S}{2}]$_{6718}$ narrow line. The parameters were measured by fitting a models with 5 narrow lines, fixed to each other in velocity and velocity dispersion and one free broad component, except as noted above. Errors quoted are the formal errors on the various parameters from the fit.}
\end{deluxetable}%

\end{document}